\documentclass[a4paper,11pt]{article}

\usepackage{amsmath}
\usepackage{amssymb}
\usepackage[applemac]{inputenc}
\usepackage{latexsym}
\usepackage{graphicx}
\usepackage{color}
\usepackage[ german, frenchb, english]{babel}

\usepackage{amsfonts}
\input cyracc.def

\newtheorem{thm}{Theorem}[section]
\newtheorem{prop}[thm]{Proposition}
\newtheorem{rem}[thm]{Remark}
\newtheorem{lem}[thm]{Lemma}
\newtheorem{cor}[thm]{Corollary}

\newtheorem{defe}[thm]{Definition}

\def\proofp{{\hspace{-0.5 cm} $\Box$ \textbf{Proof of Proposition }}}
\def\proofc{{\hspace{-0.5 cm} $\Box$ \textbf{Proof of Corollary
      }}}
\def\prooft{{\hspace{-0.5 cm} $\blacksquare$
\textbf{Proof of Theorem }}}
\def\proofl{{\hspace{-0.5 cm} $\vartriangle$ \textbf{Proof of Lemma  }}}
\def\dem{{\hspace{-0.5 cm} $\Box$ \textbf{Proof of Proposition }}}

\def\demthm{{\hspace{-0.5 cm} $\blacksquare$
\textbf{Proof of Theorem  }}}

\def\cqfd{{\hfill $\Box$}}
\def\cqfdt{{\hfill $\blacksquare$}}
\def\cq{{\hfill $\vartriangle$}}

\def\C{{\mathbb{C}}}
\def\R{{\mathbb{R}}}

\def\Z{{\mathbb{Z}}}
\def\N{{\mathbb{N}}}

\def\K{{\mathbb{K}}}

\def\Id{{\textrm{ Id} }}
\def\Tr{{\textrm{ Tr }}}

\def\ker{\textrm{ Ker }}

\def\del{\partial}

\def\U{{\mathcal{U}}}

\def\M{{\mathcal{M}}}
\def\m{{\mathcal{M}_{k}}}

\def\Ws{{\mathcal{W}_{k}^{s_{1}}}}
\def\wss{{\mathcal{W}_{k}^{s_{3}}}}
\def\W{{\mathcal{W}_{k}}}
\newcommand{\Ran}{{\rm Ran}}

\def\w{{\omega}}

\newcommand{\gm}{{\rm g}}

\def\g{{\mathfrak{g}}}

\def\a{{\mathfrak{a}}}
\def\b{{\mathfrak{b}}}

\def\~?{{\textbf{\Large{~?}}}}
\def\.{{$\centerdot$}}

\begin{document}

\title{{Hyperk\"ahler structures and \\infinite-dimensional Grassmannians }}
\author{Alice Barbara TUMPACH\footnote{EPFL, Lausanne, Switzerland.
   This
  work was partially supported by
  the University of Paris VII, the University of
  Paris XI, and the \'Ecole Polytechnique, Palaiseau, France.}}
\date{November 2005}
\maketitle

\abstract In this paper, we describe an example of a hyperk\"ahler
quotient  of a Banach manifold by a Banach Lie group. Although the
initial manifold is not diffeomorphic to a Hilbert manifold (not
even to a manifold modelled on a reflexive Banach space), the
quotient space obtained  is a Hilbert manifold, which can be
furthermore identified either with the cotangent space of a
connected component $Gr_{res}^{j}$, ($j \in \Z$), of the
restricted Grassmannian or with a natural complexification of this
connected component, thus proving that these two manifolds are
isomorphic hyperk\"ahler manifolds. Moreover, K\"ahler potentials
associated with the natural complex structure of the cotangent
space of $Gr_{res}^{j}$ and with the natural complex structure of
the complexification of $Gr_{res}^{j}$ are computed using
Kostant-Souriau's theory of prequantization.

\begin{center}
\textbf{R\'esum\'e}
\end{center}
Dans cet article, nous pr\'esentons un exemple de quotient
hyperk\"ahl\'erien d'une vari\'et\'e banachique par un groupe de
Lie banachique. Bien que la vari\'et\'e initiale ne soit pas
diff\'eomorphe \`a une vari\'et\'e hilbertienne (ni m\^eme \`a une
vari\'et\'e  model\'ee sur un espace de Banach r\'eflexif),
l'espace quotient obtenu est une vari\'et\'e hilbertienne, qui
peut \^etre identifi\'ee, selon la structure complexe
distingu\'ee, soit \`a l'espace cotangent d'une composante connexe
$Gr_{res}^{j}$ ($j \in \Z$) de la grassmannienne restreinte, soit
\`a une complexification naturelle de cette m\^eme composante
connexe.
 De plus, les potentiels k\"ahl\'eriens associ\'es respectivement
\`a la structure complexe naturelle de l'espace cotangent de
$Gr_{res}^{j}$ et \`a la structure complexe naturelle de la
complexification  de $Gr_{res}^{j}$ sont calcul\'es \`a l'aide de
la th\'eorie de pr\'equantisation de Kostant-Souriau.

\tableofcontents

\section{Introduction}

The restricted Grassmannian in a Hilbert manifold closely related
to Loop groups (see [38], [41]) and equations of the KdV type (see
[39], [37], [40], [43]). It is also a Hermitian symmetric space.
Therefore it is natural to ask (see below) whether its
complexification or its cotangent space admit hyperk\"ahler
structures.

Recall that a K{\"a}hler manifold of finite dimension is a
Riemannian manifold endowed with a complex structure that is
parallel for the Levi-Civita connection, or equivalently, a
manifold equipped with a closed symplectic real form, called the
K{\"a}hler form, and a compatible integrable complex structure. A
hyperk{\"a}hler manifold of finite dimension is a manifold endowed
with a Riemannian  metric $\textrm{g}$ and three complex
structures $I$, $J$, $K$ such that~: $ I J K = -1, $ and
$\textrm{g}$ is K{\"a}hler with respect to each complex structure.
Hence a hyperk{\"a}hler manifold admits three K{\"a}hler forms
$\omega_{1}$, $\omega_{2}$, et $\omega_{3}$, and the choice of one
complex structure, for instance  $I$, allows to build a
holomorphic symplectic form, namely $\Omega = \omega_{2} + i
\omega_{3}$.

D. Kaledin and B. Feix  have found independently in
 \cite{Kal} and \cite{Fei} that, given a finite-dimensional manifold $N$
  endowed with a K{\"a}hler metric $\textrm{g}_{N}$,
there exists a hyperk{\"a}hler metric $\textrm{g}$ defined on a
 neighborhood of the zero section of the cotangent space $T^{*}N$ of
 $N$, compatible with the natural holomorphic symplectic structure
 of the cotangent space, and
 such that the restriction of $\textrm{g}$ to $N$ is $\textrm{g}_{N}$.
In addition, $\textrm{g}$ is unique if one requires
$S^{1}$-invariance. D. Kaledin
 uses for his proof the theory of Hodge manifolds, whereas B. Feix uses
 twistor spaces. As far as we know, this result has not been extended
 to the infinite-dimensional Banach setting. Moreover this existence
 result does not lead to an explicit expression of the metric and
 examples of explicit hyperk{\"a}hler
 metrics are rare.

O. Biquard and P. Gauduchon provided in \cite{BG1} a construction
of hyperk{\"a}hler metrics on cotangent bundles of Hermitian
symmetric spaces, and in \cite{BG2} hyperk{\"a}hler metrics on
coadjoint orbits of symmetric type of a complex semi-simple Lie
group. Furthermore they established formulas for K{\"a}hler
potentials that allow explicit expressions of these metrics. In
\cite{BG3} the same authors identified these hyperk{\"a}hler
structures, showing that the cotangent space and the complexified
coadjoint orbit are, in the case of Hermitian symmetric spaces of
finite dimension, two aspects of the same hyperk{\"a}hler object,
which  appear in accordance with the chosen complex structure
within the $2$-sphere of complex structures.

In the case of the cotangent space of the Grassmannian  $Gr(p, n)$
of subspaces of dimension $p$  in
 $\C^{n}$, the aforementioned  hyperk\"ahler structure can be
obtained by a hyperk\"ahler quotient, and the corres\-ponding
K\"ahler potential can be computed  via the theory of
Kostant-Souriau's prequantization. It is the study of this
particular example of Hermitian  symmetric space that leads to the
theory of O. Biquard and P. Gauduchon as developed in \cite{BG1}.
  In the present work, we show that each connected component
  $Gr_{res}^{j}$ ($j \in \Z$) of
the restricted Grassmannian  $Gr_{res}$ introduced by Pressley and
Segal in \cite{PS}, gives an example of an infinite-dimensional
 Hermitian  symmetric space whose cotangent space  can be obtained
by an infinite-dimensional hyperk\"ahler quotient of a Banach
manifold  by a Banach Lie group. Moreover, we show that the
resulting  quotient space can also be identified with a natural
complexification $\mathcal{O}^{\C}_{j}$ of $Gr_{res}^{j}$, also
called complexified orbit, consisting of pairs $(P, Q)$ of
elements of $Gr_{res}^{j}$ such that $P \cap Q^{\perp} = \{0\}$.
In this way, the study of this particular example provides a first
step towards the generalization of the aforementioned results of
O. Biquard and P. Gauduchon to the infinite-dimensional setting.
The full generalization of these results has been carried out in
\cite{Tum3} by the construction of hyperk\"ahler metrics on
complexifications of Hermitian-symmetric affine coadjoint orbits
of semi-simple $L^{*}$-groups of compact type, and by the
identification of these complexifications  with the cotangent
spaces of the orbits under consideration. This generalization is
based on Mostow's Decomposition Theorem  (see \cite{Tum2}) and on
the notion of strongly orthogonal roots of a $L^{*}$-algebra (see
the Appendix in \cite{Tum3}).

The theory of symplectic quotients was initiated by J.\,E. Marsden
and A. Weinstein in~\cite{MW1}. In finite dimension it was used in
particular to construct new examples of symplectic manifolds.
In~\cite{MR} J.\,E. Marsden and T. Ratiu applied this theory to
infinite-dimensional manifolds and obtained new developments of
V.\,I. Arnold's idea that fluid motion equations are the equations
of geodesics on a suitable infinite-dimensional Lie group. Another
example of infinite-dimensional symplectic reduction is given by
J.\,E. Marsden and A. Weinstein in~\cite{MW3} in relation with the
Maxwell-Vlasov  equation. (For an overview of applications of the
symplectic reduction see~\cite{MW2}.) K{\"a}hler and hyperk\"ahler
reductions are refinements of this theory. An infinite-dimensional
version based on the study of  Nahm's  equations was used by
P.\,B. Kronheimer in~\cite{Kro1} and~\cite{Kro2} in order to
construct hyperk\"ahler structures on maximal semi-simple and
nilpotent coadjoint orbits of  semi-simple complex
(finite-dimensional) Lie groups. These results were generalized to
all orbits by O.~Biquard in~\cite{Biq} and A.\,G. Kovalev in
\cite{Kov}. In [24], the study of the same Nahm's equations but
with different boundary conditions leads P.B. Kronheimer to prove
that the cotangent space of any complex Lie group carries a
hyper\"ahler structure.

In~\cite{Kac}, V.\,G. Kac classifies infinite-dimensional
Lie groups and algebras into four (overlapping) categories:\\
1) Groups of diffeomorphisms of manifolds and the
corresponding Lie algebras of vector fields;\\
2) Lie groups (resp. Lie algebras) of maps from a
finite-dimensional
manifold to a finite-dimensional Lie group (resp. Lie algebra);\\
3) Classical Lie groups and algebras of operators on Hilbert and
Banach spaces;\\
4) Kac-Moody algebras.\\
The examples of infinite-dimensional reduction mentioned
above concern only the first two classes of groups. In this
work we construct an example of hyperk\"ahler
reduction involving the third class of groups.

The structure of the paper is as follows. In Section \ref{p2}, we
introduce the necessary background on K{\"a}hler and hyperk\"ahler
quotients of Banach manifolds as well as the theory of K\"ahler
potential on a quotient induced by
 Kostant-Souriau's theory of prequantization.
In the infinite-dimensional setting, the definition of a
hyperk{\"a}hler manifold has to be specified so as to avoid
problems such as the possible non-existence of a Levi-Civita
connection for weak Riemannian  metrics. The conditions needed to
get a smooth K\"ahler structure on a K\"ahler quotient of a Banach
manifold by a Banach Lie group have to be strengthened in
comparison to the finite-dimensional case. In this Section, a
basic definition of the notion of stable manifold associated with
a level set and an holomorphic action of a complex Banach Lie
group on a K\"ahler manifold is given. It is the most adapted to
our purpose, but can be related to more sophisticated definitions
as the one appearing in the Hilbert-Mumford Geometric Invariant
Theory (see \cite{MFK}) or the one appearing in Donaldson's work
(see \cite{Don} for an exposition of the circle of ideas around
this notion). In Subsection   \ref{stabman}, the existence of a
\emph{smooth} projection of the stable manifold to the level set
is proved and used in the identification of  a smooth
 K\"ahler quotient with  the complex quotient of the
associated stable manifold by the complexification of the group.
In Subsection \ref{potquo}, we give a survey of the theory of
K\"ahler potential on a K\"ahler quotient which includes a natural
generalization to the Banach setting of the formula for the
K\"ahler potential proved by O. Biquard and P. Gauduchon in
Theorem  3.1 in \cite{BG1}.

In Section  \ref{p4}, we construct a smooth hyperk\"ahler quotient
of the tangent bundle $T\mathcal{M}_{k}$  of a flat non-reflexive
Banach space $\mathcal{M}_{k}$ (indexed by $k \in \R^{*}$) by a
Banach Lie group $G$. The key point in the proof of this result is
the existence of a $G$-equivariant slice of the tangent space to
the level set, which is orthogonal to the $G$-orbits and allows
one to define a structure of smooth Riemannian  manifold on the
quotient. As far as we know, the general procedure for finding
closed complements to  closed subspaces of a Banach space recently
developed by D.~Belti\c{t}\u{a} and B.~Prunaru in \cite{BP} does
not apply in our cases, so that the existence of closed
complements has to be worked out by hand. For this purpose, the
properties of Schatten ideals (see [44]) are extensively used.

In Section  \ref{p5}, we show that the quotient space obtained in
Section  \ref{p4} can be identified with  the cotangent bundle of
a connected component $Gr_{res}^{j}$ of the restricted
Grassmannian , which is therefore endowed with a (1-parameter
family of strong) hyperk\"ahler structure(s). To prove this
identification, we  use the stable manifold associated with one of
the complex structures of the quotient space and the general
results of Section  \ref{p2}. At the end of Section  \ref{p5}, we
compute the K\"ahler potential associated with this complex
structure using the theory explained in Subsection \ref{potquo},
and we give an expression of this potential using the curvature of
$Gr_{res}$. The formulas obtained by this method are analogous to
the ones proved by O. Biquard and P. Gauduchon in the
finite-dimensional setting. By restriction to the zero section,
the theorems proved in Section \ref{p5} realize each connected
component of the restricted Grassmannian  as a K\"ahler quotient
and provide
 the expression of the K\"ahler potential of the restricted Grassmannian
 induced by
Pl\"ucker's embedding (for a description of Pl\"ucker embedding,
see [38] and [45]). The realization of the restricted Grassmannian
as a
 symplectic quotient  was
independently obtained  by T. Wurzbacher in unplublished work and
explained in numerous talks (see \cite{Wur2}).

In Section \ref{p6}, we show that the hyperk\"ahler quotient
constructed in   Section  \ref{p4} can also be identified with a
natural complexification $\mathcal{O}_{j}^{\C}$ of $Gr_{res}^{j}$.
For this purpose, we use various equivalent definitions of the
complexified orbit $\mathcal{O}_{j}^{\C}$. To give an explicit
formula of the K\"ahler potential associated with this complex
structure, we use
 an invariant of the $G^{\C}$-orbits. The
 expression of the potential as a function of the curvature of
 $Gr_{res}$ obtained by this method
  is again analogous to the
 one given by O. Biquard and P. Gauduchon in \cite{BG1}.
An equivalent expression, in terms of  characteristic angles of a
pair of subspaces $(P, Q) \in \mathcal{O}_{j}^{\C}$ is also given.

\section{Background on K\"ahler and hyperk\"ahler quotients
of Banach manifolds} \label{p2}

\subsection{K\"ahler quotient}


Let $\M$ be a smooth Banach manifold over the field $\K = \R$ or $\C$, endowed
 with a smooth
action of a Banach Lie group $G$ (over $\K$), whose Lie algebra
will be denoted by $\g$. For a Banach space $B$ over $\K$, we will
denote by $B'$ the topological dual space of $B$, i.e. the Banach
space of continuous linear applications from
 $B$ to $\K$.

\begin{defe}{\rm
A weak symplectic form $\omega$ on $\mathcal{M}$ is a closed smooth $2$-form
on $\mathcal{M}$ such that for all $x$ in $\mathcal{M}$ the map
$$
\begin{array}{llll}
\varphi_{x}: & T_{x}\mathcal{M} & \rightarrow & T_{x}'\mathcal{M}\\
& X & \mapsto & i_{X}\omega
\end{array}
$$
is an injection.}
\end{defe}

\begin{defe} {\rm
A moment map for a $G$-action on a weakly symplectic Banach
manifold $\M$ is a map $ \mu~: ~ \M ~ \rightarrow ~ \g',$
satisfying $$ d\mu_{x}(\a) =  i_{X^{\a}} \, \omega, $$
 for all $x$
in $\mathcal{M}$ and for all $\a$ in $\g$, where $X^{\a}$ denotes
the vector field on $\M$ generated by the infinitesimal action of
$\a \in \g$. The $G$-action is called {\it
  Hamiltonian } if there exists a $G$-equivariant moment map $\mu$,
i.e. a moment map satisfying the following condition :
$$
\mu(g \cdot x) = \textrm{Ad}^{*}(g)\left(\mu(x)\right).
$$
}\end{defe}

\begin{defe}{\rm
A regular value of the moment map is an element $\xi \in \g'$ such
that, for every $x$ in the level set ${\mu}^{-1}(\xi)$, the map
$d\mu_{x}~:T_{x}\M \rightarrow  \g' $ is surjective and its kernel
admits a closed complement in $T_{x}\M$. }\end{defe}

\begin{rem}{\rm
If $\xi$ is a regular value of $\mu$, then ${\mu}^{-1}(\xi)$ is a
submanifold of $\M$. If $\mu$ is $G$-equivariant and if $\xi$ is
an $\textrm{Ad}^{*}(G)$-invariant element of $\g'$, then the
manifold ${\mu}^{-1}(\xi)$ is globally $G$-stable, and one can
consider the quotient space ${\mu}^{-1}(\xi)/G$. }
\end{rem}
In the following, we consider an Hamiltonian  action of $G$ on
$\M$ and a regular $\textrm{Ad}^{*}(G)$-invariant element $\xi$ of
$\g'$. We recall some classical results on the topology and
geometry of the quotient space. Propositions \ref{haus},
\ref{propre} and \ref{exisquo} are respectively Proposition  3
chap.III \S 4.2 in \cite{Bou1}, Proposition  6 chap.III \S 4.3 in
\cite{Bou1} and Proposition  10 chap.III \S 1.5 in \cite{Bou2}, up
to notational changes.

\begin{prop}[\cite{Bou1}]\label{haus}
If  $G$ acts properly on a manifold $\mathcal{N}$, then the
quotient space $\mathcal{N}/G$ endowed with the quotient topology
is Hausdorff.
\end{prop}

\begin{prop}[\cite{Bou1}]\label{propre}
If  $G$ acts freely on $\mathcal{N}$, the action of $G$ is proper
if and only if the graph $\mathcal{C}$ of the equivalence relation
defined by $G$ is closed in $\mathcal{N} \times \mathcal{N}$ and
the canonical application from $\mathcal{C}$ to $G$ is continuous.
\end{prop}

\begin{prop}[\cite{Bou2}]\label{exisquo}
Assume that $G$ acts freely and properly on $\mathcal{N}$. If, for
every $x \in \mathcal{N}$,  the tangent space $T_{x}G\!\cdot\!x$
to the orbit $G\!\cdot\!x$ at $x$  is closed in $T_{x}\mathcal{N}$
and admits a closed complement, then the quotient space
$\mathcal{N}/G$ has a unique structure of  Banach manifold such
that the projection $\pi~: \mathcal{N} \rightarrow \mathcal{N}/G$
is a submersion.
\end{prop}

\begin{rem}\label{toutou}{\rm
Let $b$ be a continuous bilinear  form on a Banach vector space
$B$. Suppose that $b$ realizes an injection of $B$ into its
topological dual $B'$ by $\tilde{b}(X)~:= b(X, \cdot ~)$, for $X
\in B$. For any linear subspace $A$ of $B$, we have the inclusion
$$
\bar{A} \subset \left({A}^{\perp_{b}}\right)^{\perp_{b}}
$$
but not necessarily the equality. The equality means that any
continuous linear form vanishing on ${A}^{\perp_{b}}$ is of the
form $\tilde{b}(X)$ for some $X \in A$, which is a particular
property of the subspace~$A$. Along the same lines, if $b$ is a
positive definite symmetric bilinear form on $B$, we have~:
$$
A \cap A^{\perp_{b}} = \{ 0 \},
$$
but in general we do not have~:
\begin{equation}\label{compl}
B = A \oplus A^{\perp_{{b}}}
\end{equation}
even if $A$ is closed, since the right hand side may not be
closed. Furthermore J. Lindenstrauss and L. Tzafriri have proved
in \cite{LT} that a Banach space in which every closed subspace is
complemented is isomorphic to a Hilbert space. This implies in
particular that for a non-reflexive Banach space $B$ endowed with
a weak Riemannian metric, equality (\ref{compl}) is certainly not
fulfilled by every closed subspace $A$. }
\end{rem}

\begin{prop}\label{weaksymp}
If ${\mu}^{-1}(\xi)/G$ has a Banach manifold structure such that
the quotient map is a submersion, and if $G$ acts by
symplectomorphisms, the condition
$$
T_{x}G\!\cdot\!x =
\left({(T_{x}G\!\cdot\!x)}^{\perp_{\omega}}\right)^{\perp_{\omega}}
~~{\rm for ~all }~~x \in {\mu}^{-1}(\xi)
$$
 implies that ${\mu}^{-1}(\xi)/G$ is a weakly
symplectic manifold.
\end{prop}

\proofp \ref{weaksymp}:\\
Denote by $\pi$ the quotient map $\pi : {\mu}^{-1}(\xi)
\rightarrow {\mu}^{-1}(\xi)/G$. Let us show that the following
expression
\begin{equation}\label{wred}
\omega^{\textrm{red}}_{[x]}(X, Y)~:= \omega_{x}(\tilde{X},
\tilde{Y}),
\end{equation}
where $X, Y$ are in $T_{[x]}({\mu}^{-1}(\xi)/G)$ and where
$\pi_{*}\tilde{X} = X$ and $\pi_{*}\tilde{Y} = Y$, defines a weak
symplectic structure on the quotient. Note that for all $x \in
{\mu}^{-1}(\xi)$, the tangent space $T_{x}({\mu}^{-1}(\xi))$ is
precisely the kernel of the differential $d{\mu}_{x}$, so that for
all $\a \in \g$, the $1$-form $i_{X^{\a}}\omega$ vanishes on
${\mu}^{-1}(\xi)$. This implies that the right-hand side of
(\ref{wred}) does not depend on the choice of $\tilde{X}$ and
$\tilde{Y}$. Since $G$ acts by symplectomorphisms, it does not
depend on the choice of the element $x$ in the class $[x]$ either.
It follows that ${\pi}^{*}\omega^{\textrm{red}} =
\omega_{|{\mu}^{-1}(\xi)}$. Since $\omega$ is closed, so is
$\omega^{\textrm{red}}$. The kernel of $\omega^{\textrm{red}}$ at
a point $[x]$ in the quotient space is~:
$$\pi_{*}\left(T_{x}\left({\mu}^{-1}(\xi)\right)^{\perp_{\omega}}\right).$$
Note that the tangent space $T_{x}G\!\cdot\!x$ to the $G$-orbit of
$x$ is spanned by $\{ X^{\a}(x), \a \in \g \}$, hence we have
$$
T_{x}\left({\mu}^{-1}(\xi)\right) =
{(T_{x}G\!\cdot\!x)}^{\perp_{\omega}},
$$
and
$$
\left(T_{x}\left({\mu}^{-1}(\xi)\right)\right)^{\perp_{\omega}} =
\left({(T_{x}G\!\cdot\!x)}^{\perp_{\omega}}\right)^{\perp_{\omega}}.
$$
For $\omega^{\textrm{red}}$ to be symplectic, one needs
$T_{x}\left({\mu}^{-1}(\xi)\right)^{\perp_{\omega}} =
T_{x}G\!\cdot\!x$, which is precisely the hypothesis. \cqfd
\\
\\
 Recall the following definition~:

\begin{defe}\label{oiu} {\rm A $G$-equivariant slice of the manifold ${\mu}^{-1}(\xi)$
is a subbundle $H$ of the tangent bundle
$T\left({\mu}^{-1}(\xi)\right)$ such that, for every $x$ in
${\mu}^{-1}(\xi)$, $H_{x}$ is a closed complement
 to the tangent space $T_{x}G\!\cdot\!x$ of the $G$-orbit  $G\!\cdot\!x$, and such that
$$
H_{g \cdot x} = g_{*}H_{x},
$$
for all $x$ in ${\mu}^{-1}(\xi)$ and for all $g$ in $G$.}
\end{defe}

\begin{rem}{\rm
Suppose that the manifold $\M$ is endowed with a weakly Riemannian
$G$-invariant metric $\mathrm{g}$. Then the existence of a
$G$-invariant slice $H$ of ${\mu}^{-1}(\xi)$ allows one to define
a weakly Riemannian  metric $\mathrm{g}^{\textrm{red}}$ on the
quotient ${\mu}^{-1}(\xi)/G$, as follows. For every $x \in
{\mu}^{-1}(\xi)$, we set
$$
\begin{array}{lcll}
\mathrm{g}^{\textrm{red}}_{[x]}~:&
T_{[x]}\left({\mu}^{-1}(\xi)/G\right) \times
T_{[x]}\left({\mu}^{-1}(\xi)/G\right)
                   & \rightarrow & \R \\
              &   (~X~,~ Y~) & \mapsto & \mathrm{g}_{x}(\tilde{X}, \tilde{Y}),
\end{array}
$$
where $\tilde{X}$ and $\tilde{Y}$ are the unique elements of
$H_{x}$ such that $\pi_{*}(\tilde{X}) = X$ and $\pi_{*}(\tilde{Y})
= Y$. }\end{rem}

\begin{defe}{\rm
A weak K\"ahler manifold is a Banach manifold $\M$ endowed with a
weak symplectic form $\omega$ and a weak Riemannian  metric
$\textrm{g}$ (i.e. such that at every $x$ in $\M$, $\textrm{g}$
defines an injection of $T_{x}\M$ into its dual), satisfying the
following compatibility condition~:
\begin{itemize}
\item[(C)] the endomorphism $I$ of the tangent bundle of $\M$
defined by $\mathrm{g}(IX, Y) = \omega(X, Y)$ satisfies ${I}^{2} =
-1$
 and the Nijenhuis tensor $N$ of $I$ vanishes.
\end{itemize}
Recall that the Nijenhuis tensor has the following expression at
$x$ in $\M$~:
$$
N_{x}(X, Y)  := [X, Y] + I[X, IY] + I[IX, Y] -
        [IX, IY] ,
$$
where  $X$ and $Y$ belong to  $T_{x}\M$.}
\end{defe}

\begin{thm}\label{basic}
Let $\M$ be a smooth K\"ahler Banach manifold endowed with a
 free and proper Hamiltonian action of a Banach Lie group $G$
preserving the K\"ahler structure. If, for every  $x$ in the
preimage ${\mu}^{-1}(\xi)$ of an $\textrm{Ad}^{*}(G)$-invariant
regular value of the moment map $\mu$, the tangent space
$T_{x}G\!\cdot\!x$ of the orbit $G\!\cdot\!x$ satisfies the direct
sum condition
$$
 {\rm (D) } \hspace{40pt} T_{x}G\!\cdot\!x ~\oplus ~{(T_{x}G\!\cdot\!x)}^{\perp_{\mathrm{g}}} ~=~ T_{x}\left({\mu}^{-1}(\xi)\right),
$$
 then the  quotient space $\M//G := \mu^{-1}(\xi)/G$ is  a
  smooth K\"ahler manifold.
\end{thm}

\prooft \ref{basic}:\\
Let us denote by $({\rm g}, \omega, I)$ the K\"ahler structure of
$\M$. For $x$ in ${\mu}^{-1}(\xi)$, the condition
$$
T_{x}G\!\cdot\!x \oplus {(T_{x}G\!\cdot\!x)}^{\perp_{\mathrm{g}}}
= T_{x}{\mu}^{-1}(\xi),
$$
implies that the tangent space of the orbit $G\!\cdot\!x$
satisfies the following property~:
\begin{equation}\label{ellela}
\left({(T_{x}G\!\cdot\!x)}^{\perp_{\mathrm{g}}}\right)^{\perp_{\mathrm{g}}}
= T_{x}G\!\cdot\!x
\end{equation}
(the converse may not be true). In particular, $T_{x}G\!\cdot\!x$
is closed and splits,  so one is able to define on the quotient
space a Banach manifold structure by use of Proposition
\ref{exisquo}. Now equality (\ref{ellela})  is equivalent to
$$
\left({(T_{x}G\!\cdot\!x)}^{\perp_{\omega}}\right)^{\perp_{\omega}}
= T_{x}G\!\cdot\!x,
$$
since $I$ is orthogonal with respect to $\textrm{g}$. So the
condition needed for the definition of the symplectic structure on
the quotient in Proposition  \ref{weaksymp} is fulfilled. Moreover
 the orthogonal
 $H_{x}:= {(T_{x}G\!\cdot\!x)}^{\perp_{\mathrm{g}}}$ of
$T_{x}G\!\cdot\!x$ in $T_{x}\left({\mu}^{-1}(\xi)\right)$ defines
a $G$-equivariant slice for the manifold ${\mu}^{-1}(\xi)$, and,
by Remark  \ref{oiu}, allows one to define a Riemannian  metric on
the quotient. It remains to define a compatible complex structure
$I^{\textrm{red}}$ on ${\mu}^{-1}(\xi)/G$. For this purpose, let
us remark that
$$
I H_{x} = I \left(T_{x}G\!\cdot\!x \right)^{\perp_{\textrm{g}}}
\subset T_{x}{\mu}^{-1}(\xi) =
\left(T_{x}G\!\cdot\!x\right)^{\perp_{\omega}},
$$
since $\omega(X^{\a}, IU) = \textrm{g}(I X^{\a}, IU) =
\textrm{g}(X^{\a}, U) = 0$ for all $\a$ in $\g$ and $U$  in
$H_{x}$. In addition, $ I H_{x}$ is orthogonal to
$T_{x}G\!\cdot\!x$ with respect to $\textrm{g}$ since
$\textrm{g}(IU, X^{\a}) = \omega(U, X^{\a}) = 0$ for $U \in
T_{x}\left({\mu}^{-1}(\xi)\right) =
\left(T_{x}G\!\cdot\!x\right)^{\perp_{\omega}}$. This implies that
 $H_{x}$
is stable under $I$ and one can  define a complex structure
$I^{\textrm{red}}$ on the quotient space by:
$$
\begin{array}{llll}
I^{\textrm{red}}~:& T_{[x]}\left({\mu}^{-1}(\xi)/G\right) &\rightarrow &  T_{[x]}\left({\mu}^{-1}(\xi)/G\right) \\
         &   X  & \mapsto &     {\pi}_{*}I_{x}\tilde{X},
\end{array}
$$
where $x $ is in $[x]$ and where $\tilde{X}$ is the unique element
of $H_{x}$ whose projection on
$T_{[x]}\left({\mu}^{-1}(\xi)/G\right)$ is $X$. Hence one has~:
$$
I^{\textrm{red}}\pi_{*}\tilde{X} = \pi_{*}I\tilde{X}.
$$
The application $\pi$ being a submersion, given two vector fields
$X$ and $Y$ on ${\mu}^{-1}(\xi)/G$, one has~:
$$
[X, Y] = \pi_{*}\left([\tilde{X}, \tilde{Y}]\right),
$$
where again $\tilde{X}$ satisfies $\tilde{X}(x) \in H_{x}$ and
$\pi_{*}(\tilde{X}) = X$, and similar conditions for $\tilde{Y}$.
Therefore the formal integrability condition  on $I$ implies the
formal integrability condition on $I_{\textrm{red}}$ since the
Nijenhuis tensor of $I^{\textrm{red}}$ has the following
expression:
$$
\begin{array}{ll}
N(X, Y) & := [X, Y] + I^{\textrm{red}}[X, I^{\textrm{red}}Y] +
I^{\textrm{red}}[I^{\textrm{red}}X, Y] -
        [I^{\textrm{red}}X, I^{\textrm{red}}Y]\\
        & = {\pi}_{*}[\tilde{X}, \tilde{Y}] + I^{\textrm{red}}
{\pi}_{*}[\tilde{X}, I\tilde{Y}] + I^{\textrm{red}}{\pi}_{*}[I
\tilde{X},
\tilde{Y}] - {\pi}_{*}[I\tilde{X}, I\tilde{Y}]  \\
& = {\pi}_{*}(N(\tilde{X}, \tilde{Y})).
\end{array}
$$
where $X, Y$ are in  $T_{[x]}\left({\mu}^{-1}(\xi)/G\right)$ and
$\tilde{X}, \tilde{Y}$ are as before. \cqfdt

\begin{rem} {\rm
In contrast to the finite-dimensional case illustrated by the
Newlander-Nirenberg Theorem  (see \cite{NN}),  the formal
integrability condition of an almost complex structure $I$ on  a
Banach manifold $\M$ given by a vanishing Nijenhuis tensor is not
sufficient for $\M$ to admit a system of holomorphic charts. An
example of a formally integrable complex structure on a real
Banach manifold which does not admit any open subset biholomorphic
to an open subset of a complex Banach manifold was recently
constructed by I. Patyi in \cite{Pat}. However, if $\M$ is a real
analytic manifold and $I$ a formally integrable analytic complex
structure, then $\M$ can be endowed with a holomorphic atlas (see
\cite{Pen}, and \cite{Bel} for the details of this result). Note
also that, in the Fr\'echet context, L. Lempert showed in
\cite{Lem} that the complex structure defined in \cite{MW4} by
J.\,E. Marsden and A. Weinstein on the space of knots  does not
lead to the existence of holomorphic charts, although this
structure was shown to be formally integrable by J.\,L. Brylinski
in \cite{Bry}. In the context of formally integrable complex
structures, we will call a map $f$ between two complex manifolds
$\left(M, I_{M}\right)$ and $\left(N, I_{N}\right)$
\emph{holomorphic} if $df \circ I_{M} = I_{N} \circ df$.}
\end{rem}

\subsection{Stable manifold}\label{stabman}

Let $(\M, \omega, \mathrm{g}, I)$  be a  smooth K\"ahler Banach
manifold endowed with a smooth Hamiltonian  action of a Banach Lie
group $G$ preserving the K\"ahler structure. Let $\xi$ be an
$\textrm{Ad}^{*}(G)$-invariant regular value of the moment map
$\mu$. Assume that there exists a complex Lie group $G^{\C}$ with
Lie algebra $g^{\C} := \g \oplus i \g$ which acts holomorphically
and smoothly on $\M$ extending the action of $G$, and that the
following assumption holds~:
\begin{itemize}
\item[(H)] for every $x$ in ${\mu}^{-1}(\xi)$, ~one has:
~$T_{x}\M~=~ T_{x}\left({\mu}^{-1}(\xi)\right) \oplus
I~T_{x}G\!\cdot\!x~$ as topological sum.
\end{itemize}
(Note that, by definition of the moment map, one has
$T_{x}\left({\mu}^{-1}(\xi)\right)~=~\left(I~T_{x}G\!\cdot\!x\right)^{\perp_{\textrm{g}}}$,
so  (H) states that the direct sum of $I~T_{x}G\!\cdot\!x$ and its
orthogonal is closed in $T_{x}\M$, which is not always the case as
mentioned in Remark  \ref{toutou}, but nevertheless a natural
assumption to make.)
 The action of $G^{\C}$ on $\M$ allows one to define a
notion of stable manifold associated with the level set
${\mu}^{-1}(\xi)$~:

\begin{defe}
The stable manifold $\M^{s}$ associated with the level set
${\mu}^{-1}(\xi)$ is defined by:
$$
\M^{s}~:= \{ x \in \M ~~|~~ \exists~ g \in G^{\C}, ~g \cdot x \in
{\mu}^{-1}(\xi) \}.
$$
\end{defe}

\begin{rem}{\rm
The assumption (H) implies that $\M^{s}$ is open in $\M$ since
$T_{x}\M^{s} = T_{x}\M$ for every element $x$ in
${\mu}^{-1}(\xi)$, hence, by translation by an element of $G^{
\C}$, for every $x$ in $\M^{s}$.}
\end{rem}

\begin{prop}\label{inter}
If $G^{\C}$ admits a polar decomposition $~G^{\C} = \exp
i\g\!\cdot\!G~$, then for every $x$ in ${\mu}^{-1}(\xi)$, one has:
$$
G^{\C}\!\cdot\!x~ \cap ~{\mu}^{-1}(\xi)~ = ~G\!\cdot\!x.
$$
\end{prop}

\proofp \ref{inter}:\\
The argument below has already been used by F. Kirwan (see [17],
[18], [19]). It goes like this. Clearly $G\!\cdot\!x~ \subset
~G^{\C}\!\cdot\!x~ \cap ~{\mu}^{-1}(\xi)$ since $\xi$ is
$\textrm{Ad}^{*}(G)$-invariant. Let us show that $
G^{\C}\!\cdot\!x~ \cap ~{\mu}^{-1}(\xi)~ \subset ~G\!\cdot\!x$.
Suppose that there exists $g \in G^{\C}$ such that $g \cdot x \in
{\mu}^{-1}(\xi)$. Since ${\mu}^{-1}(\xi)$ is $G$-invariant and
since $G^{\C} = \exp i\g\!\cdot\! G$,~ it is sufficient to
consider the case when $g = \exp i\a$~,~ $\a \in \g$.

Define the function $h~: \R \rightarrow \R$  by $h(t) =
\mu\left((\exp it \a) \cdot x\right)(\a)$. One has $h(0) = h(1) =
\xi(\a)$, hence there exists $t_{0} \in~ (0, 1)$ such that~:
$$
0 = h'(t_{0}) = d_{y}\mu(i\a \cdot y)(\a) = - \omega_{y}(i\a \cdot
y, \a \cdot y) = \|\a \cdot y\|^{2},
$$
where $y = \exp(it_{0}\a) \cdot x$. Hence $\a \cdot y = 0$ and
$\exp(i\a \R)$ fixes $y$, thus also $x$. It follows that~:
$$\exp(i\a \R) \cdot x ~\cap~ {\mu}^{-1}(\xi) ~=~ \{x\}.$$\cqfd
\\
From now on and till the end of Section  \ref{p2}, it will be
assumed that $G^{\C}$ admits a polar decomposition.

\begin{cor}\label{free}
If $G$ acts freely on ${\mu}^{-1}(\xi)$,  then  $G^{\C}$ acts
freely on $\M^{s}$.
\end{cor}

\proofc \ref{free}:\\
 Let
$x$ be an element of ${\mu}^{-1}(\xi)$ and let $g \in G^{\C}$ be
such that $g \cdot x =x$. Since $G^{\C} = \exp i\g\!\cdot\!G$,
there exists $u \in G$ and $\a \in \g$  such that $g =
\exp(i\a)u$, and one has $\exp (i\a)u \cdot x = x$. From the proof
of the previous Proposition , one has:
$$
\exp(i\a\R).(ux) ~\cap~ {\mu}^{-1}(\xi) ~=~ \{ux\}.
$$
It follows that $ux = x$, thus $u = e$, since $G$ acts freely on
${\mu}^{-1}(\xi)$. Now,  the condition $\exp(i\a\R) \cdot x = x$
implies that $\a$ fixes $x$, hence $\a = 0$.
 \cqfd

\begin{prop}\label{toctoc}
Assume that $G$ acts freely on ${\mu}^{-1}(\xi)$. Then, for every
$y$ in $\M^{s}$, there is a unique element $g(y)$ in $\exp i\g$
such that $g(y)$ maps $y$ to the level set. The resulting
application
$$
\begin{array}{llll}
g~:&  \M^{s} & \rightarrow & \exp i\g\\
& y & \mapsto & g(y)
\end{array}
$$
is smooth and  the projection $q$
$$
\begin{array}{llll}
q~:&  \M^{s} & \rightarrow & {\mu}^{-1}(\xi)\\
& y & \mapsto & g(y) \cdot y
\end{array}
$$
is smooth and $G$-equivariant.
\end{prop}

\proofp \ref{toctoc}~:

\. Let $y$ be in $\M^{s}$ and suppose that there exist two
elements $\a$ and $\b$ in $\g$ such that both $\exp i\a \cdot y$
and $\exp i\b \cdot y$ belong to ${\mu}^{-1}(\xi)$. Since $\exp
i\a \cdot y$ and $\exp i\b \cdot y$ are in the same
$G^{\C}$-orbit, by Proposition  \ref{inter}, there exists $u$ in
$G$ such that $\exp i\a \cdot y = u \cdot \exp i\b \cdot y$. Since
$G$ acts freely on the level set, Corollary  \ref{free} implies
that $G^{\C}$ acts freely on $\M^{s}$. It follows that $\exp i\a =
u \cdot \exp i\b$, hence, by uniqueness of the polar
decomposition, $u$ is the unit element of $G$ and $\exp i\a \,=\,
\exp i\b$. Therefore $g(y) = \exp i\a$ is well defined, and so is
the projection $q$.

\. Let us show that the application~:
$$
\begin{array}{llll}
g ~: & \M^{s} & \rightarrow & \exp{i\g} \\
& y & \mapsto & g(y)
\end{array}
$$
is smooth.  Since $G^{\C}$ acts smoothly on $\M^{s}$, it will
imply that the projection $q$ is smooth also. Consider the
following map~:
$$
\begin{array}{llll}
\phi~: & \exp i\g \times \M^{s} & \rightarrow & \M\\
& (~\exp i\a ~, ~ y~) & \mapsto & \exp i\a \cdot y,
\end{array}
$$
which maps $\exp i\g \times \M^{s}$ onto $\M^{s}$. (Recall that
$\exp i\g$ inherits a Banach manifold structure from its
identification with the homogeneous space $G^{\C}/G$.) We will
prove that $\phi$ is transversal to ${\mu}^{-1}(\xi)$ (see \S
5.11.6 and \S 5.11.7 of \cite{Bou3} for a definition of this
notion), so that the subset
$$
\phi^{-1}\left({\mu}^{-1}(\xi)\right) ~= ~ \{~(g(y)~,~y)~,~ y
~\in~ \M^{s} ~\}
$$
is a smooth submanifold of $\exp i\g \times \M^{s}$. The
smoothness of the application $g$ will therefore follow from the
smoothness of the projection $p_{1}~: \exp i\g \times \M^{s}
\rightarrow \exp i\g$ on the first factor. We will denote by
$R_{y}$ the right translation by $y$ on $G^{\C}$. The differential
of $\phi$ at a point $(\exp i\a, y)$ in $\exp i\g \times \M^{s}$
reads~:
$$
\left(d\phi \right)_{(\exp i\a, y)} \left((\left(R_{\exp
i\a}\right)_{*}(i\b), Z)\right) := i\b \cdot \left(\exp i\a
 \cdot y \right) \oplus \left( \exp i\a \right)_{*}(Z).
$$
Note that, for every element $(\exp i\a, y)$ in
$\phi^{-1}\left({\mu}^{-1}(\xi)\right)$, one has~:
$$
\left(d\phi \right)_{(\exp i\a, y)} \left( \{0\} \times
T_{y}\M^{s} \right) ~= ~ T_{x}\M^{s}
$$
where $x := \exp i\a \cdot y$, so that $\left(d\phi \right)_{(\exp
i\a, y)}$
 is surjective. It remains to show that the subspace
 $$
 \left(d\phi \right)_{(\exp i\a, y)}^{-1}
 \left(T_{x}({\mu}^{-1}(\xi))\right)
$$ is complemented.
For this purpose recall that, by  assumption (H), the tangent
space $T_{y}\M^{s}$ is isomorphic to $$~(\exp
(-i\a))_{*}\left(T_{x}\left({\mu}^{-1}(\xi)\right)\right) \oplus
~(\exp (-i\a))_{*}\left(i\g \cdot x\right),$$ so that the tangent
space
$$T_{(\exp i\a, y)} \left( \exp i\g \times \M^{s}\right)~=~ T_{\exp i\a}\left(\exp i\g \right)
\times T_{y}\M^{s}
$$
is isomorphic to $\g
\times T_{x}\left({\mu}^{-1}(\xi)\right) \times \g$ by the
following isomorphism~:
$$
\begin{array}{llll}
\jmath~: & \g \times T_{x}\left({\mu}^{-1}(\xi)\right) \times \g &
\rightarrow &   T_{\exp i\a}\left(\exp i\g \right)
\times T_{y}\M^{s} \\
& (~\b~, ~W ~,~ \mathfrak{c} ~) & \mapsto & \left(\left(R_{\exp
i\a}\right)_{*}(i\b) ~,~ (\exp (-i\a))_{*}\left(W\right) + (\exp
(-i\a))_{*}\left( i\mathfrak{c} \cdot x \right) \right).
\end{array}
$$
The element $\jmath(\b, W, \mathfrak{c})$ belongs to $
 \left(d\phi \right)_{(\exp i\a, y)}^{-1}
 \left(T_{x}({\mu}^{-1}(\xi))\right)
$  whenever $i\b \cdot x + W + i\mathfrak{c} \cdot x \in
T_{x}\left({\mu}^{-1}(\xi)\right)$. Since $G$ acts freely on $\M$,
 it follows that the subspace~:
$$
\left(d\phi \right)_{(\exp i\a, y)}^{-1}
 \left(T_{x}({\mu}^{-1}(\xi))\right) $$ equals $$ \{~ \jmath(\b, W, -\b)~,~\b \in \g, W \in
T_{x}\left({\mu}^{-1}(\xi)\right)~\},
$$
and
$$
\{~ \jmath(\b, 0, \b)~,~\b \in \g ~\}
$$
is a closed complement to it.

\. Let us check the $G$-equivariance of $q$.
 Since $\mu$ is $G$-equivariant and $\xi$ is $\textrm{Ad}^{*}(G)$-invariant,
one has~:
$$
\mu( u \cdot g(y) \cdot y ) = \textrm(Ad)^{*}(u)\left(\mu( g(y)
\cdot y) \right) = \textrm(Ad)^{*}(u)(\xi) = \xi
$$
for all $u$ in $G$, and $y$ in $\M^{s}$. We can write $g(y) = \exp
i\a$ for some $\a \in \g$. Now the equality $ u \cdot \exp i\a =
\exp \left( \textrm{Ad}(u)(i\a) \right) \cdot u$ and the
uniqueness of the element $g ( u \cdot y)$ satisfying $g(u \cdot
y)\cdot (u \cdot y)\in {\mu}^{-1}(\xi)$ proved above, imply that~:
$$
g( u \cdot y)~ = ~\exp \left( \textrm{Ad}(u)(i\a) \right).
$$
Hence $q$ satisfies the $G$-equivariant condition~:
$$
q( u \cdot y) ~=~ u \cdot q(y)
$$
for all $u \in G$ and $y \in \M^{s}$. \cqfd

\begin{prop}\label{propre2}
If $G$ acts freely and properly on ${\mu}^{-1}(\xi)$, then
$G^{\C}$ acts (freely and) properly on $\M^{s}$.
\end{prop}

\proofp \ref{propre2}:\\
By Proposition \ref{free}, $G^{\C}$ acts freely on $\M^{s}$. By
Proposition \ref{propre}, $G^{\C}$ acts properly on $\M^{s}$ if
and only if the graph $\tilde{\mathcal{C}}$ of the equivalence
relation defined by $G^{\C}$ is closed in $\M^{s} \times \M^{s}$
and the canonical map from $\tilde{\mathcal{C}}$ to $G^{\C}$ is
continuous.

Let us show that $\tilde{\mathcal{C}}$ is closed in $\M^{s} \times
\M^{s}$. Denote by $\mathcal{C}$ the graph of the equivalence
relation defined by the action of $G$ on ${\mu}^{-1}(\xi)$. Let
$\{\left(y_{n}, v_{n} \cdot y_{n}\right)\}_{n \in \N}$ be a
sequence in $\tilde{\mathcal{C}}$, where $y_n\in \M^{s}$ and
$v_n\in G^{\C}$, which converges to an element $(y_{\infty},
z_{\infty})$ in $\M^{s} \times \M^{s}$. From Proposition
\ref{toctoc} and from the continuity of the projection $q$, the
sequence $\{\left(q(y_{n}), q(v_{n} \cdot y_{n}) \right)\}_{n \in
\N}$ belongs to $\mathcal{C}$ and converges to $\left(
q(y_{\infty}), q(z_{\infty}) \right)$. Since $\mathcal{C}$ is
closed in ${\mu}^{-1}(\xi) \times {\mu}^{-1}(\xi)$, it follows
that $q(z_{\infty}) = u_{\infty} \cdot q(y_{\infty})$ for some
$u_{\infty}$ in $G$. Hence $z_{\infty}  = g(z_{\infty})^{-1}
u_{\infty} g(y_{\infty}) \cdot y_{\infty}$. Thus
$\tilde{\mathcal{C}}$ is closed in $\M^{s} \times \M^{s}$.

Let us show that the canonical map $\tilde{\iota}$ from
$\tilde{\mathcal{C}}$ to $G^{\C}$ is continuous. Denote by $\iota$
the canonical map from $\mathcal{C}$ to $G$. One has~:
$$
\tilde{\iota}(y, z)~
\mapsto ~ g(z)^{-1} \circ \iota(q(y), q(z)) \circ g(y),
$$
and the continuity of $\tilde{\iota}$ follows from the continuity
of the applications $\iota$, $g$ and $q$. \cqfd
\\
\\
We conclude this Subsection with
\begin{thm}\label{joujou}
Let $\M$ be a Banach K\"ahler manifold endowed with a smooth, free
and proper Hamiltonian  action of  a Banach Lie group $G$, which
preserves the K\"ahler structure and extends to a smooth
holomorphic action of a complex Lie group $G^{\C} =  \exp
i\g\!\cdot\!G$. Let $\xi$ be an $\textrm{Ad}^{*}(G)$-invariant
regular value of the moment map $\mu$. If for every $x$ in
${\mu}^{-1}(\xi)$, the orthogonal
$\left(T_{x}G\!\cdot\!x\right)^{\perp_{\textrm{g}}}$ of
$T_{x}G\!\cdot\!x$ in $T_{x}\left({\mu}^{-1}(\xi)\right)$
satisfies the following direct sum condition
$$
{\rm ( D + H ) } \hspace{40pt}
 T_{x}\M ~=~ T_{x}G\!\cdot\!x ~\oplus~
(T_{x}G\!\cdot\!x)^{\perp_{\textrm{g}}} ~\oplus~
I\left(T_{x}G\!\cdot\!x\right),
$$
then the quotient space $\M^{s}/G^{\C}$ is a smooth complex
manifold isomorphic to the smooth K\"ahler quotient $\M//G :=
{\mu}^{-1}(\xi)/G$ as complex smooth manifold. Moreover the full
integrability of the complex structure on $\M$ implies the full
integrability of the complex structure on $\M^{s}/G^{\C}$, hence
on $\M//G$.
\end{thm}

\prooft \ref{joujou}~:\\
By Proposition \ref{free} and Proposition \ref{propre2}, $G^{\C}$
acts freely and properly on $\M^{s}$. The smoothness of the
application $g$ and the $G$-equivariant slice $H$ of
${\mu}^{-1}(\xi)$ given by $H_{x} :=
\left(T_{x}G\!\cdot\!x\right)^{\perp_{\textrm{g}}}$, allows one to
define a $G$-equivariant slice on $\M^{s}$, also denoted by $H$,
by the following formula~:
$$
H_{y} := g(y)_{*}^{-1}\left(H_{q(y)}\right) \subset T_{y}\M^{s},
$$
for all $y$ in $\M^{s}$. The following decomposition of the
tangent space $T_{y}\M^{s}$ holds~:
$$
T_{y}\M^{s} = H_{y} \oplus T_{y}\left(G^{\C}\!\cdot\!y\right).
$$
By Proposition \ref{exisquo}, it follows that $\M^{s}/G^{\C}$ has
a unique (real) Banach manifold structure such that the quotient
map is a submersion. Since for every element $y$ in $\M^{s}$,
$H_{q(y)}$ is invariant under the complex structure $I$ of $\M$,
and since the $G^{\C}$-action on $\M$ is holomorphic, for every
$y$ in $\M^{s}$ the subspace $H_{y}$ of $T_{y}\M^{s}$ is
$I$-invariant and, further, $\M^{s}/G^{\C}$ inherits a natural
complex structure. Moreover, since the complex structure of the
quotient $\M^{s}/G^{\C}$ comes from  the complex structure of
$\M$, the natural injection ${\mu}^{-1}(\xi) \hookrightarrow
\M^{s}$ induces a complex isomorphism between ${\mu}^{-1}(\xi)/G$
and $\M^{s}/G^{\C}$. Finally, $\M^{s}$ being an open subset of
$\M$ because of (H), the existence of holomorphic charts on $\M$
allow one to apply Proposition \ref{exisquo} to the holomorphic
quotient $\M^{s}/G^{\C}$, thus implies the existence of
holomorphic charts on $\M^{s}/G^{\C}$. \cqfdt

\subsection{Hyperk{\"a}hler quotient}\label{hypsub}

Let $\M$ be a Banach manifold endowed with a weak hyperk\"ahler
 metric $\mathrm{g}$, with K\"ahler forms
 $\omega_{1}$, $\omega_{2}$ and $\omega_{3}$, and corresponding complex structures
$I_{1}$, $I_{2}$, and $I_{3}$. Let  $G$ be a connected Banach Lie
group with Lie
 algebra $\g$,
acting freely on  $\M$ by hyperk{\"a}hler diffeomorphisms. We
assume in this Subsection that
 there exists  a $G$-{e}quivariant moment map  ${\mu}_{i}$
for each symplectic structure. Define  ${\mu}~: \M \rightarrow \g'
\otimes \R^{3}$ by $\mu = \mu_{1} \oplus \mu_{2} \oplus \mu_{3}$.
Let $\xi = (\xi_{1}, \xi_{2}, \xi_{3})$ be  a regular
$\mathrm{Ad}^{*}(G)$-invariant value
 of the moment map
$\mu$.

\begin{thm}\label{hyper}
If $G$ acts freely  and properly  on $\M$ and if, for every $x$ in
${\mu}^{-1}(\xi)$, the orthogonal
$\left(T_{x}G\!\cdot\!x\right)^{\perp_{\textrm{g}}}$ of
$T_{x}G\!\cdot\!x$ in $T_{x}\left({\mu}^{-1}(\xi)\right)$
satisfies the direct sum condition
$$
 {\rm (D) } \hspace{40pt} T_{x}G\!\cdot\!x ~\oplus ~{(T_{x}G\!\cdot\!x)}^{\perp_{\mathrm{g}}}
 ~=~ T_{x}\left({\mu}^{-1}(\xi)\right),
$$
then  the quotient space ${\mu}^{-1}(\xi)/G$ carries a structure
of  smooth Banach hyperk\"ahler manifold.
\end{thm}

\prooft \ref{hyper}:\\
This is a direct application of Theorem  \ref{basic} with respect
to each complex structure $I_{1}$, $I_{2}$, and $I_{3}$ of $\M$.
By the very definition of the complex structures and symplectic
forms on the reduced space, ${\mu}^{-1}(\xi)/G$ inherits a
hyperk\"ahler structure from the hyperk\"ahler structure of $\M$.
\cqfdt
\\
\\
Let us now assume  that  the orthogonal
$\left(T_{x}G\!\cdot\!x\right)^{\perp_{\textrm{g}}}$ of
$T_{x}G\!\cdot\!x$ in $T_{x}\left({\mu}^{-1}(\xi)\right)$
satisfies the  direct sum condition
\begin{itemize}
\item[(S)] for every $x$ in ${\mu}^{-1}(\xi)$, ~one has: $ T_{x}\M
~=~ T_{x}G\!\cdot\!x ~\oplus
~\left(T_{x}G\!\cdot\!x\right)^{\perp_{\textrm{g}}}~ \oplus ~I_{1}
T_{x}G\!\cdot\!x ~\oplus~ I_{2} T_{x}G\!\cdot\!x ~\oplus~ I_{3}
T_{x}G\!\cdot\!x $ as a topological direct sum.
\end{itemize}
For each complex structure $I_{\vec{k}}= k_1 \,I_1 + k_2\, I_2 +
k_3 \,I_3$ in the $2$-sphere of complex structures on $\M$,
indexed by $\vec{k} = (k_{1}, k_{2}, k_{3}) $ in $S^{2}$, let us
define an action $\cdot_{\vec{k}}$ of $i\g$ on $\M$ by~:
$$
i\a \cdot_{\vec{k}}x = I_{\vec{k}}(\a \cdot x),
$$
for all $ \a$  in $\g$ and  for all $x$ in $\M$. Assume that  for
a given  $\vec{k}$, the action $\cdot_{\vec{k}}$ integrates into
an $I_{\vec{k}}$-holomorphic action of $G^{\C}$ on $\M$. Let us
choose an orthogonal complex structure  to $I_{\vec{k}}$ denoted
by  $I_{\vec{l}}$, and define a third complex structure by
$I_{\vec{m}}:= I_{\vec{k}}.I_{\vec{l}}$, so that
$\left(I_{\vec{k}}\,,\,I_{\vec{l}}\,,\,I_{\vec{m}}\right)$
satisfies the quaternionic identities. Denote by ${\mu}_{\vec{k}}$
the combination ${\mu}_{\vec{k}} := k_1 \,\mu_1 + k_2\, \mu_2 +
k_3\, \mu_3$ and similarly  ${\mu}_{\vec{l}}$ and
${\mu}_{\vec{m}}$. Consider $\xi_{\vec{l}} := l_1 \,\xi_1 + l_2
\,\xi_2 + l_3 \,\xi_3$ and $\xi_{\vec{m}} := m_1\, \xi_1 + m_2\,
\xi_2 + m_3 \,\xi_3$, and similarly $\omega_{\vec{l}} := l_1
\,\omega_1 + l_2 \,\omega_2 + l_3 \,\omega_3$ and
$\omega_{\vec{m}} := m_1\, \omega_1 + m_2\, \omega_2 + m_3
\,\omega_3$.

\begin{prop}\label{gcrajout}
The map ${\mu}^{\C} := {\mu}_{\vec{l}} + i {\mu}_{\vec{m}}$ is an
$G^{\C}$-equivariant holomorphic moment map for the
$I_{\vec{k}}$-complex symplectic structure $\omega_{\vec{l}} + i
\omega_{\vec{m}}$.
\end{prop}

\proofp \ref{gcrajout}:\\
By the same algebraic arguments than the ones given in
\cite{HKLR}, section 3\,(D), the map $\mu^{\C}$ is holomorphic and
a moment map for $\omega_{\vec{l}} + i \omega_{\vec{m}}$. Since
$G$ is connected, the $G^{\C}$-equivariance of ${\mu}^{\C}$ is
will follow from
$$
\langle d{\mu}^{\C}(i\a.x), \b\rangle = \langle {\mu}^{\C}(x),
[i\a, \b]\rangle,
$$
for all $x$ in $\mathcal{M}$, and all $\a, \b \in \g$, where the
bracket denotes the duality pairing. But this is an easy
consequence of the $G$-equivariance of ${\mu}_{\vec{l}}$ and
${\mu}_{\vec{m}}$ and the following identities
$$
d{\mu}^{\C}(i\a.x) := d{\mu}^{\C}(I_{\vec{k}} X^{\a}) = i
d{\mu}^{\C}(X^{\a}) \textrm{~~~and~~~} {\mu}^{\C}(x)\left([i\a,
\b] \right) = i {\mu}^{\C}(x)\left( [\a, \b] \right),
$$
where $X^{\a}$ denotes the vector field generated by $\a \in \g$.
\cqfd

\begin{lem}\label{decs}
The stable manifold $\M^{s_{\vec{k}}}$ with respect to the complex
structure $I_{\vec{k}}$, associated with the level set
${\mu}^{-1}(\xi)$,  is a submanifold of the Banach manifold $\M$,
contained in the preimage by ${\mu}^{\C}$ of the
$G^{\C}$-coadjoint orbit of $\xi_{\vec{l}} + i \xi_{\vec{m}}$. In
particular, if ${\xi}_{\vec{l}} + i {\xi}_{\vec{m}}$ is in the
center of $\g^{\C}$, then $\M^{s_{\vec{k}}} \subset
{\mu}_{\vec{l}}^{-1}(\xi_{\vec{l}})
~\cap~{\mu}_{\vec{m}}^{-1}(\xi_{\vec{m}})$.
\end{lem}

\proofl \ref{decs}:\\
The fact that $\M^{s_{\vec{k}}}$ is included in
$\left({\mu}^{\C}\right)^{-1}\left(\textrm{Ad}^{*}\left(G^{\C}\right)\left(\xi_{\vec{l}}
+ i \xi_{\vec{m}}\right)\right)$ is a direct consequence of the
$G^{\mathbb{C}}$-equivariance of ${\mu}^{\C}$ (Proposition
\ref{gcrajout}). Let $x$ be an element in the level set
${\mu}^{-1}(\xi)$. Since $\xi$ is a regular value of $\mu$, the
level set ${\mu}^{-1}(\xi)$ is a Banach submanifold of
$\mathcal{M}$. Consider an adapted chart $\left(\mathcal{V},
E\times F, \varphi\right)$ of the submanifold ${\mu}^{-1}(\xi)$ at
$x$, where $\mathcal{V}$ is a neighborhood of $x$ in
$\mathcal{M}$, $E$ and $F$ are two Banach spaces, and $\varphi$ is
an homeomorphism from $\mathcal{V}$ onto a neighborhood of $0$ in
$E\times F$ such that
$\mathcal{U}_{1}:=\varphi\left({\mu}^{-1}(\xi) \cap
\mathcal{V}\right) \subset E$. By assumption (S), one has~:
$$
T_{x}\M ~=~ T_{x}G\!\cdot\!x ~\oplus~
\left(T_{x}G\!\cdot\!x\right)^{\perp_{\textrm{g}}} ~\oplus~
I_{\vec{k}} \left(T_{x}G\!\cdot\!x\right)~ \oplus ~I_{\vec{l}}
\left(T_{x}G\!\cdot\!x\right) ~\oplus~ I_{\vec{m}}
\left(T_{x}G\!\cdot\!x\right),
$$
hence $F$ is isomorphic (as Banach space) to $I_{\vec{k}}
\left(T_{x}G\!\cdot\!x\right)~ \oplus ~I_{\vec{l}}
\left(T_{x}G\!\cdot\!x\right) ~\oplus~ I_{\vec{m}}
\left(T_{x}G\!\cdot\!x\right)$. For $\a$ in the Lie algebra $\g$
of $G$, denote by $X^{\a}$ the vector field on $\mathcal{M}$
generated by $\a$. Since $G$ acts freely on $\mathcal{M}$, the map
$\g \rightarrow I_{\vec{k}}\left(T_{x}G\!\cdot\!x\right)$ which
assigns to $\a \in \g$  the vector $I_{\vec{k}}X^{\a}$ is a
continuous bijection of Banach spaces, hence an isomorphism, and
similarly for the indexes $\vec{l}$ and $\vec{m}$. Since by
hypothesis the $G$-action on $\mathcal{M}$ extend to a
$I_{\vec{k}}$-holomorphic action of $G^{\C}$ on $\mathcal{M}$, the
vector fields $I_{\vec{k}}X^{\a}$, for $\a\in \g$, are complete,
whereas the flows $t \mapsto f^{t}_{I_{\vec{l}}X^{\b} +
I_{\vec{m}}X^{\mathfrak{c}}}$ of the vector fields
$\left(I_{\vec{l}}X^{\b} + I_{\vec{m}}X^{\mathfrak{c}}\right)$,
with $\b$ and $\mathfrak{c}$ in $\g$, may not be globally defined.
Nevertheless, by the smooth dependance of the solutions of a
differential equation with respect to a parameter, there exists a
small neighborhood $\mathcal{U}_{2}$ of $0$ in $I_{\vec{l}}
\left(T_{x}G\!\cdot\!x\right) ~\oplus~ I_{\vec{m}}
\left(T_{x}G\!\cdot\!x\right) \simeq \g \oplus \g$ for which the
flows $f^{t}_{I_{\vec{l}}X^{\b} + I_{\vec{m}}X^{\mathfrak{c}}}$
are defined for $t\in [0, 1]$. Now consider the following map~:
$$
\begin{array}{llll}
\Psi~:& \mathcal{U}_1 \oplus \g \oplus \mathcal{U}_{2} &
\rightarrow
& \mathcal{M}\\
& \left(u, \a, (\b, \mathfrak{c})\right) & \mapsto &
\exp{ia}\cdot_{\vec{k}}f_{I_{\vec{l}}X^{\b} +
I_{\vec{m}}X^{\mathfrak{c}}}^{1}\left({\varphi}^{-1}(u)\right).
\end{array}
$$
Then $\Psi$ provides an adapted chart in the neighborhood of any
$y$ in the fiber $q_{\vec{k}}^{-1}(x)$ where $q_{\vec{k}}~:
\M^{s_{\vec{k}}} \rightarrow {\mu}^{-1}(\xi)$ is the projection
defined in Proposition~\ref{toctoc}. \cq
\\
\\
 With the notation above, Theorem
\ref{joujou} reads:

\begin{thm}\label{kiot}
If  $G$ acts freely and properly on $\M$, and if, for every $x$ in
${\mu}^{-1}(\xi)$, the orthogonal of $T_{x}G\!\cdot\!x$ in
$T_{x}{\mu}^{-1}(\xi)$ satisfies the direct sum conditions {\rm
(D)} and {\rm (S)}, then the quotient space
$\M^{s_{\vec{k}}}/G^{\C}$ is a smooth $I_{\vec{k}}$-complex
manifold which is
 diffeomorphic to
${\mu}^{-1}(\xi)/G$ as a $I_{\vec{k}}$-complex smooth
manifold.\hfill $\blacksquare$
\end{thm}

\subsection{K\"ahler potential on a K\"ahler  quotient}\label{potquo}

This Subsection   is a generalization to the Banach setting of
results
 obtained in the finite-dimensional case by O. Biquard
and P. Gauduchon in \cite{BG1} (see Theorem  3.1 there), and based
on an idea in \cite{HKLR}.  In this Subsection, we will again make
use of the setting  of Subsection   \ref{stabman}  and we will
suppose that the complex structure $I$ of $\M$ is formally
integrable.
\\
\\
Recall that the complex structure $I$ acts on $n$-differential
forms by $\eta \mapsto I \eta$ where
$$
(I\eta)_{x}(X_{1}, \dots, X_{n} )~:= (-1)^{n} \eta(IX_{1}, \dots,
IX_{n}).
$$
and where $x$ belongs to $\M$, and
 $ X_{1}, \dots, X_{n}$ are elements in $T_{x}\M$. This action
 allows one to define the corresponding differential operator $d^{c}$ by
$d^{c} := I d I^{-1}$. Note that $dd^{c} = 2i \del \bar{\del}$
where $\del$ and $\bar{\del}$ are the Dolbeault operators. Recall
also the following definitions~:

\begin{defe}{\rm
 A K\"ahler potential on  $\M$ is a function $K$ on $\M$
such that $\omega= d d^{c} K$}.
\end{defe}
In the following we will make the assumption that $\M$ admits a
 $G$-invariant globally defined K\"ahler potential, which will be the case
 in the next Section. Under this assumption,
the action of  $G$ is Hamiltonian  with respect to the moment map
$\mu$ defined by:
$$
\mu(x)(\a)~:=   d K_{x}(I X^{\a}),
$$
for all  $x$  in $\M$, and for all $\a$  in $\g$.

\begin{lem}\label{triv}
If $K$ is a globally defined K\"ahler potential on $(\M, \omega, I
)$, then the trivial bundle $L = \M \times \C$ endowed with the
Chern connection  $\nabla$ associated with the Hermitian  product
$h$ on $L$ given by~: $$h(\sigma(x), \sigma(x))~:= e^{-2 K(x)},$$
where $\sigma$ is the canonical section $\sigma(x) = (x, 1)$,
pr{e}quantifies $\M$ in the sense that $~R^{\nabla} = i \omega$.
\end{lem}

\proofl \ref{triv}:\\
 Given a non-vanishing section $\sigma$ such that $\bar{\partial}\sigma = 0$,
the curvature of the Chern connection has the following expression
$$
R^{\nabla} = \frac{1}{2i} d d^{c} \log h(\sigma, \sigma)
$$
since the $\bar{\del}$ operator on $L = \M \times \C$ satisfies
the formal integrability condition $d^{\bar{\partial}} \circ
d^{\bar{\partial}} = 0$. Hence the Hermitian  product $h$
satisfies~: $d d^{c} \log h(\sigma, \sigma) = -2 \omega = -2 d
d^{c} K$, i.e. $\log h = -2 K + \eta$, where $\eta$ is in the
kernel of $dd^{c}$ and can be chosen to be $0$. \cq
\\

The aim of this Subsection   is to compute a K\"ahler potential on
a smooth K\"ahler quotient $\M//G := {\mu}^{-1}(\xi)/G$, given a
globally defined K\"ahler potential on $\M$. For this purpose we
will construct a holomorphic line bundle $(\hat{L}, \hat{h})$ over
${\mu}^{-1}(\xi)/G$ which pr{e}quantifies  ${\mu}^{-1}(\xi)/G$.
The previous Lemma  establishes the link between the Hermitian
scalar product on a trivial line bundle and a globally defined
potential. If the pull-back of $(\hat{L}, \hat{h})$ to the stable
manifold  $\M^{s}$ (which is an open subset of $\M$ since we
assume hypothesis (H) satisfied) is a trivial Hermitian  line
bundle, then the pull-back of the Hermitian  scalar product
$\hat{h}$ induces a globally defined K\"ahler potential on
$\M^{s}$.

Consider the following action of the Lie algebra
 $\mathfrak{g}$ on  the vector space $\Gamma(L)$ of sections
 of the trivial line bundle $L$ :
$$
\mathfrak{a}. \sigma = -\nabla_{ X^{\mathfrak{a}}}\sigma - i
\mu^{\mathfrak{a}}\sigma  + i \xi(\a).\sigma ,
$$
for all $\sigma$  in $\Gamma(L)$ and $\mathfrak{a}$ in
$\mathfrak{g}.$ This action corresponds to the action of $\g$  on
the total space of $L$ which assigns to an element $\mathfrak{a}
\in \mathfrak{g}$ the vector field $\hat{X}^{\mathfrak{a}}$ over
$L$ whose value at a point $\zeta \in
  L$ over $x \in \M$ is~:
$$
\hat{X}^{\mathfrak{a}}(\zeta) = \tilde{X}^{\mathfrak{a}}(\zeta) + i
    \mu^{\mathfrak{a}}(x) .T(\zeta) -  i \xi(\a).T(\zeta),
$$
where $\tilde{X}^{\mathfrak{a}}(\zeta)$ denotes the horizontal lift at
  $\zeta$  of $X^{\mathfrak{a}}$ for the trivial connection, and where
  $T$ denotes the vertical vector field given by $T(\zeta) = \zeta$.
This action integrates into an action of the group $G$ if and only
if the following integrability condition is satisfied~:
\begin{itemize}
\item[(I)] $\mathfrak{a} \mapsto i \xi(\mathfrak{a})$ is the
differential of a group homomorphism from $G$ to $S^{1}$ at the
unit element $e$ of $G$.
\end{itemize}
Suppose it is the case and denote by  $\chi$ the
 homomorphism such that $(d\chi)_{e} = -i\xi$. The homomorphism
 $\chi$ extends to a homomorphism of
 $G^{\C}$ to $\C^{*}$,
which will be also denoted by $\chi$. The corresponding action of
 $G^{\C}$ on $L$ is given by~:
$$
g \cdot (x\,, \,z_{x}) ~=~ (g \cdot x\,,\, \chi(g)^{-1}.z_{x}),
$$
where $g$ is an element in $G^{\C}$.
 This induces an action of $G^{\C}$ on $\Gamma(L)$ by~:
$$
(g \cdot \sigma)(x)~:= g(\sigma(g^{-1} \cdot x)).
$$
where $\sigma$ belongs to $\Gamma(L)$,  $g$  is an element in
$G^{\C}$ and $x$ is in $\M$. In the remainder of this Subsection,
we will assume that the integrability condition (I) is satisfied.

\begin{defe} {\rm
Let $\hat{L}$ be the complex line bundle over ${\mu}^{-1}(\xi)/G$
obtained as the $G$-orbit space of the restriction of the trivial
line bundle $L$ to the level set ${\mu}^{-1}(\xi)$. The fiber of
$\hat{L}$ over an element $[x]$ in ${\mu}^{-1}(\xi)/G$ is~:
$$
\hat{L}([x]) = [(x\,,\, z_{x})],
$$
where $(x\,,\, z_{x}) \sim (g \cdot x\,,\, \chi(g)^{-1}. z_{x})$
for all $g \in G$. }
\end{defe}

\begin{defe} {\rm
Define an Hermitian  scalar product  $\hat{h}$ on $\hat{L}$ by~:
$$
\hat{h}(\hat{\sigma}_{1}, \hat{\sigma}_{2}) = h(\sigma_{1},
\sigma_{2}),
$$
where, for $i \in \{ 1,\,2\}$,  $\sigma_{i}$  is the $G$-invariant
section of $L_{|{\mu}^{-1}(\xi)}$ whose projection to
$\Gamma(\hat{L})$ is $\hat{\sigma}_{i}$. }
\end{defe}

\begin{prop}\label{connexionhat}
Let $\sigma$ be a $G$-invariant section of $L_{|{\mu}^{-1}(\xi)}$
whose projection to  $\Gamma(\hat{L})$ will be denoted by
$\hat{\sigma}$, and let $X$ be a vector field on
${\mu}^{-1}(\xi)/G$ whose horizontal lift with respect to an
arbitrary  $G$-invariant connection $\tilde{\nabla}$ on the bundle
${\mu}^{-1}(\xi) \rightarrow {\mu}^{-1}(\xi)/G$ will be denoted by
$\tilde{X}$. Then  the identity
$$
\hat{\nabla}_{X} \hat{\sigma}~:= \nabla_{\tilde{X}} \sigma
$$
where $\nabla$ denotes the Chern connection  on $L$, defines a
connection $\hat{\nabla}$ on $\hat{L}$, which is independent of
$\tilde{\nabla}$ and for which $\hat{h}$ is parallel.
\end{prop}

\proofp \ref{connexionhat}~:\\
Note that~: $\hat{\nabla}_{X} \hat{h} = \nabla_{\tilde{X}} h = 0$,
since $\nabla$ preserves $h$. Let  $\tilde{X}_{1}$ be the
horizontal lift of  $X$ with respect to another $G$-invariant
connection. One has~: $\tilde{X}_{1} = \tilde{X} + X^{\a}$, with
$\a \in \g$. If $\sigma$ is $G$-invariant, then~:
$$
\nabla_{X^{\a}} \sigma = -\a.\sigma - i {\mu}^{\a}\sigma + i
\xi(\a).\sigma = 0.
$$
Consequently $\nabla_{\tilde{X}} \sigma =
\nabla_{\tilde{X}_{1}}\sigma$, hence $\hat{\nabla}$ is independent
of the connection  $\tilde{\nabla}$. To check that
$\nabla_{\tilde{X}} \sigma$ is $G$-invariant, note that for all
$\a \in \g$, one has~:
$$
\begin{array}{ll}
\a.\nabla_{\tilde{X}}\sigma & = -
\nabla_{{X}^{\a}}\nabla_{\tilde{X}}\sigma - i
{\mu}^{\a}\nabla_{\tilde{X}}\sigma +
i\xi(\a)\nabla_{\tilde{X}}\sigma\\
& = - \nabla_{\tilde{X}}\nabla_{X^{\a}}\sigma - \nabla_{[X^{\a},
  \tilde{X}]}\sigma - R_{X^{\a}, \tilde{X}}\sigma\\
& = -i \omega(X^{\a}, \tilde{X})\sigma = 0.
\end{array}
$$
\cqfd

\begin{prop}\label{preq1}
The Hermitian  line bundle $(\hat{L}, \hat{h}, \hat{\nabla})$
pr{e}quantifies the quotient ${\mu}^{-1}(\xi)/G$ in the sense that
$~R^{\hat{\nabla}} = i \omega^{\textrm{red}}$.
\end{prop}

\proofp \ref{preq1}~:\\
For any vector fields $X$ and $Y$ on ${\mu}^{-1}(\xi)/G$, and for
every section $ \hat{\sigma}$ of $\hat{L}$ given by a
$G$-invariant section $\sigma$ of $L_{|{\mu}^{-1}(\xi)}$, one
has~:
$$
\begin{array}{ll}
R^{\hat{\nabla}}_{X,Y} \hat{\sigma} & =
\nabla_{\tilde{X}}\nabla_{\tilde{Y}} \sigma - \nabla_{\tilde{Y}}
\nabla_{\tilde{X}}\sigma - \nabla_{\widetilde{[X, Y]}}\sigma\\
& = i \omega(\tilde{X},\tilde{Y})\sigma +
\nabla_{[\tilde{X},\tilde{Y}]-\widetilde{[X, Y]}}\sigma \\
& = i \omega^{\textrm{red}}(X, Y)\sigma,
\end{array}
$$
since $[\tilde{X},\tilde{Y}]-\widetilde{[X, Y]}$ is tangent to the
$G$-orbit and since, for every $\mathfrak{a} \in \mathfrak{g}$,
the identity
$$
\mathfrak{a}. \sigma = -\nabla_{ X^{\mathfrak{a}}}\sigma - i
\mu^{\mathfrak{a}}\sigma  + i \xi(\mathfrak{a}) .\sigma = 0
$$
implies that the covariant derivative of  $\sigma$ with respect to
a vertical vector vanishes. \cqfd

\begin{cor}\label{holocor}
The complex structure of the Hermitian  line bundle $(\hat{L},
\hat{h})$ is formally integrable and the connection $\hat{\nabla}$
is the Chern connection.
\end{cor}

\proofc \ref{holocor}~:\\
The curvature of  $\hat{L}$ being of type $(1, 1)$, the operator
$\bar{\partial}~:= \hat{\nabla}^{0.1}$ defines a formally
integrable complex structure on $\hat{L}$.  Moreover,
$\hat{\nabla}$ is $\C$-linear and preserves $\hat{h}$, thus it is
the associated Chern connection. \cqfd
\\
\\
To proceed, suppose that we are under the assumptions of Theorem
\ref{joujou}, so that we have a submersion:
$$
\begin{array}{llll}
p~: & \M^{s} & \rightarrow & {\mu}^{-1}(\xi)/G
\end{array}
$$
given by the identification of  ${\mu}^{-1}(\xi)/G$ with
$\M^{s}/G^{\C}$. Note that the differential of $p$ satisfies~: $dp
\circ I = I^{\textrm{red}} \circ dp$, where $I$ and
$I^{\textrm{red}}$ are the complex structures on $\M^s$ and
${\mu}^{-1}(\xi)/G$ respectively. From Proposition \ref{toctoc} it
follows that there exists a map
$$
\begin{array}{llll}
g~: & \M^{s} & \rightarrow \exp{i \g} &
\end{array}
$$
and a smooth projection~:
$$
\begin{array}{llll}
q~: & \M^{s} & \rightarrow {\mu}^{-1}(\xi), &
    \end{array}
$$
satisfying  $g(x)\!\cdot\!x = q(x)$ for all $ x$ in  $\M^{s}$.

\begin{prop}\label{rapport1}
The 2-form $ p^{*}\omega^{\textrm{red}}$ is the curvature of the
line bundle $(L_{|\M^{s}}, \bar{h})$, where $\bar{h}$ is defined
by:
$$
\bar{h}(\zeta, \zeta) = h\left(\,g(x)\!\cdot\!\zeta\,,\,
g(x)\!\cdot\!\zeta\right).
$$
for $x$ in $\M^s$ and $\zeta$ in $L_x$.
\end{prop}

\proofp \ref{rapport1}~:\\
Since $dp$ satisfies $dp \circ I = I^{\textrm{red}} \circ dp$, the
pull-back by $p$ of the Chern connection associated with $\hat{h}$
is the Chern connection of $p^{*}\hat{L}$ with respect to
$p^{*}\hat{h}$. Thus $R^{p^{*}\hat{\nabla}} = ip^{*}
\omega^{\textrm{red}}$. In addition, the fiber over  $x \in
\M^{s}$ of the bundle $p^{*}\hat{L}$ is~:
$$
(p^{*}\hat{L})_{x} = \hat{L}_{p(x)} = [L_{q(x)}]_{[q(x)]}.
$$
Since the element  $q(x)$ in the class $[q(x)]$ is distinguished,
the fiber  $(x, [L_{q(x)}])$ can be identified with the fiber $(x,
L_{q(x)})$. One can therefore define an isomorphism $\Phi$ of
complex bundles by~:
$$
\begin{array}{llll}
\Phi~: & L_{|\M^{s}} & \longrightarrow & p^{*}\hat{L}\\
       & \zeta \in L_{x} & \longmapsto & g(x)\!\cdot\!\zeta \in \left(p^{*}\hat{L}\right)_{x}.
\end{array}
$$
Clearly $\Phi^{*}\hat{h} = \bar{h}$. Now let  $\bar{\nabla}$ be
the Chern connection of the trivial Hermitian   bundle  $(L_{|
\M^{s}}, \bar{h})$. Its curvature is~:
$$
R^{\bar{\nabla}} =  \Phi^{-1} \circ R^{p^{*}\hat{\nabla}} \circ
\Phi.
$$
Since the bundle is a complex line bundle, one has~:
$$
R^{\bar{\nabla}} = R^{p^{*}\hat{\nabla}} = ip^{*}
\omega^{\textrm{red}}.
$$
\cqfd

\begin{thm}\label{potentiel}
The 2-form $p^{*}\omega^{\textrm{red}}$ on $\M^{s}$ satisfies
$~ip^{*}\omega^{\textrm{red}} = d d^{c} \hat{K},~ $ where for
every $x \in \M^{s}$,
$$
\hat{K}(x)~:= ~K(g(x)\cdot x) ~+~  \frac{1}{2}  \log ~
|\,\chi\left(g(x)\right)\,|^{2}.
$$
\end{thm}

\prooft \ref{potentiel}~:\\
The curvature of the Chern connection of the Hermitian  line
bundle $(L_{| \M^{s}}, \bar{h})$  is given by:
$$
R = \frac{1}{2i} d d^{c} \log~ \bar{h}\left(\sigma, \sigma\right),
$$
where $\sigma$ is the canonical section. Moreover:
$$
\bar{h}(\sigma, \sigma) = h\left(g(x)\!\cdot\!\sigma\,,\,
g(x)\!\cdot\!\sigma\right) ~=~ |\,\chi(g(x))\,|^{-2} h(\sigma,
\sigma).
$$
Thus:
$$
p^{*} \omega^{\textrm{red}}(x)  ~=~ -\frac{1}{2} d d^{c} \log~
|\,\chi(g(x))\,|^{-2} h(\sigma, \sigma)
   = d d^{c} K(\,g(x) \cdot x\,) ~+~ \frac{1}{2} d d^{c} \log ~ |\,\chi(g(x))\,|^{2}.
$$
\cqfdt
\\

\section{An Example of hyperk\"ahler quotient of a Banach mani\-fold
  by a Banach Lie group}\label{p4}

\subsection{Notation}\label{nota}

Let us summarize  the notation used in the remainder of this
paper.

$H$ will stand for a separable Hilbert space endowed with an
orthogonal decomposition $H = H_{+} \oplus H_{-}$ into two closed
infinite-dimensional subspaces. The orthogonal projection from $H$
onto $H_{+}$ (resp. $H_{-}$) will be denoted by $p_{+}$ (resp.
$p_{-}$). The restriction of $p_{\pm}$ to a subspace $P$ will
generally be denoted by $pr_{\pm}$ and the space $P$ be specified.

Given two Banach spaces $E$ and $F$, the set of Fredholm operators
from $E$ to $F$ will be denoted by $\textrm{Fred}(E, F)$, the
Hilbert space of Hilbert-Schmidt operators from $E$ to $F$ will be
denoted by $L^{2}(E, F)$, the Banach space of trace class
operators from $E$ to $F$ by $L^{1}(E, F)$, and the Banach space
of bounded operators from $E$ to $F$ by $B(E, F)$. The argument
$F$  in the previous operator spaces will be omitted when  $F =
E$.

The set of self-adjoint trace class operators on a complex Hilbert
space $E$ will be denoted by  $\mathcal{S}^{1}(E)$, and the set of
skew-Hermitian trace class operators on $E$ by
$\mathcal{A}^{1}(E)$. Similarly, $\mathcal{S}^{2}(E)$ will stand
for self-adjoint Hilbert-Schmidt operators on $E$, and
$\mathcal{A}^{2}(E)$ for skew-Hermitian Hilbert-Schmidt operators
on $E$.

 The unitary group of $H$ will be denoted by
$\U(H)$ and the identity map of a Hilbert space $E$ by $\Id_{E}$.
$\textrm{Ran}(A)$ will stand for the range of an operator $A$ and
$\textrm{Ker}(A)$ for its kernel.

\subsection{Introduction}

The restricted Grassmannian  $Gr_{res}$ of $H$, studied for
instance  in \cite{PS} and \cite{Wur1}, is defined as follows:
$$
\begin{array}{llll}
Gr_{res}(H) = \{~ P {\rm ~closed ~subspace ~of~}  H {\rm ~ such
 ~that ~}
                   &  pr_{+}~:  P & \rightarrow  & H_{+} \in \textrm{
                   Fred}(P, H_{+})\\
                   &  pr_{-}~:  P & \rightarrow  & H_{-} \in L^{2}(P, H_{-})
                 ~\},
\end{array}
$$
where $pr_{\pm}$ denotes the orthogonal projection from $P$ to
$H_{\pm}$. The space $Gr_{res}$ is a Hilbert manifold and a
homogeneous space under the restricted unitary group:
$$
\U_{res} = \left\{ \begin{array}{l}
         \left( \begin{array}{cc} U_{+} & U_{-+} \\
                                              U_{+-} & U_{-}
                            \end{array} \right) \in \U(H)~~ |~~
U_{-+} \in L^{2}(H_{-}, H_{+}), ~U_{+-} \in L^{2}(H_{+}, H_{-})
 \end{array} \right\}.
$$
Note that the stabilizer of  $H_{+}$ is $\U(H_{+}) \times
\U(H_{-})$. The space $Gr_{res}$  is a strong K\"ahler manifold
whose K\"ahler structure is invariant under $ \U_{res}$. The
expressions of the metric $\textrm{g}_{Gr}$,  the complex
structure $I_{Gr}$ and  the symplectic form $\w_{Gr}$ at the
tangent space of $Gr_{res}$ at $H_{+}$ are~:
$$
\begin{array}{l}
\textrm{g}_{Gr}(X, Y) =  \Re \Tr X^{*}Y \\
I_{Gr} Y  = i Y \\
\w_{Gr}(X, Y) = \textrm{g}_{Gr}(iX, Y) =   \Im \Tr X^{*}Y ,
\end{array}
$$
where $X$ and $Y$ belong to the tangent space $ T_{H_{+}}Gr_{res}$
 which can be identified with  $L^{2}(H_{+}, H_{-})$.
 Two elements $P_{1}$ and $P_{2}$ of
$Gr_{res}$ are in the same connected component $Gr_{res}^{j}$, ($j
\in \Z$), if and only if the projections $pr_{+}^{1}~: P_1
\rightarrow H_{+}$ and $pr_{+}^{2}~: P_2 \rightarrow  H_{+}$ have
the same index $j$. In particular $Gr_{res}^{0}$ denotes the
connected component of $Gr_{res}$ containing $H_{+}$.

The aim of this Section  is to construct a hyperk\"ahler quotient
whose quotient space will be identified with the cotangent space
of $Gr_{res}^{0}$ in Section  \ref{p5}. This will make
$T'Gr_{res}^{0}$ into a strong hyperk\"ahler manifold. In Section
\ref{p6}, the same quotient space will be identified with a
complexified orbit of $Gr_{res}^{0}$, which will therefore carry a
hyperk\"ahler structure in its own right. Since all the
constructions that follow can be carried out substituting an
arbitrary element of another connected component $Gr_{res}^{j}$
for $H_{+}$ and its orthogonal for $H_{-}$, it will follow in
particular that the cotangent space of the whole restricted
Grassmannian  is strongly hyperk\"ahler and that it can be
identified with the union of the complexifications of all
connected components $Gr_{res}^{j}$. The latter union is nothing
but the orbit of $p_{+}$ under the action by conjugation of the
(non-connected) group $GL_{res}$ which is the complexification of
$\U_{res}$. In other words, $T'Gr_{res} \simeq
GL_{res}\!\cdot\!p_{+}$.

\subsection{A weak hyperk\"ahler affine space $T\m$}

For $k \in \R^{*}$, let $\mathcal{\M}_{k}$ be the following affine
Banach space~:
$$
\m ~:= \left\{ ~x =  \left( \begin{array}{c} x_{+}  \\
                                               x_{-}
                            \end{array} \right) \in B(H_{+},
                            H)~~|~~x_+ - \Id_{H_{+}} \in
                            L^{1}(H_{+}),~ x_{-} \in L^{2}(H_{+},
                            H_{-}) \right\}.
$$
  Then $\m$ is
modelled over the Banach space  $L^{1}(H_{+}, H_{+}) \times
L^{2}(H_{+}, H_{-})$. The tangent space of $\m$
$$
T\m = \left\{ ~(x, X)   \in ~\m \times B(H_{+}, H)~~ |~~
                 p_{+} \circ X \in L^{1}(H_{+}),~
                p_{-} \circ X \in L^{2}(H_{+},H_{-}) ~\right\},
$$
injects into the continuous cotangent space $T'\m$ of $\m$ via the
application~:
$$
(x, X) \mapsto (x, (Y \mapsto \Tr X^{*}Y)). $$ Thus $T\m$ inherits
a structure of  weak complex symplectic manifold with symplectic
form $\Omega$ given by
$$
\Omega \left( (Z_{1}, T_{1})~;(Z_{2}, T_{2})\right) =
\Tr(T_{1}^{*}Z_{2}) -  \Tr(T_{2}^{*}Z_{1}),
$$
where $(Z_{1}, T_{1})$ and $(Z_{2}, T_{2})$ belong to $T_{(x,
X)}(T\m)$. We will denote by $\w_{2}$ and $\w_{3}$ the real
symplectic forms given respectively by the real and imaginary
parts of $\Omega$. Besides, from the natural inclusion of
$L^{1}(H_{+})$ into  $L^{2}(H_{+})$, it follows that $T\m$ admits
a natural weak Riemannian  metric whose expression is
$$
\textrm{g}_{(x, X)}\left( (Z_{1}, T_{1})~;(Z_{2}, T_{2})\right) =
      \Re \Tr Z_{1}^{*}Z_{2} + \Re \Tr T_{1}^{*}T_{2}
$$
where $(Z_{1}, T_{1})$ and $(Z_{2}, T_{2})$ belong to $T_{(x,
X)}(T\m)$. The complex symplectic form $\Omega$ and the Riemannian
metric $\textrm{g}$ give rise to a hyperk\"ahler structure on
$T\m$ with complex structures~:
$$
\begin{array}{ll}
I_{1}(Z, T) & = (iZ, -iT)\\
I_{2}(Z, T) & = (T, -Z)\\
I_{3}(Z, T) & = (iT, iZ).
\end{array}
$$
The real symplectic form associated with $I_{1}$ is given by
$$
\begin{array}{ll}
\omega_{1}\left( (Z_{1}, T_{1})~; (Z_{2}, T_{2})\right) & =
\textrm{g}_{(x, X)}\left( (I_{1}(Z_{1}, T_{1})~;(Z_{2}, T_{2})\right)\\
& =
  \Im \Tr Z_{1}^{*}Z_{2} - \Im \Tr T_{1}^{*}T_{2}.
\end{array}
$$

\subsection{Tri-Hamiltonian  action of a unitary group $G$}

Let $G$ be the following Banach Lie group of unitary operators~:
$$
G ~:= \U(H_{+}) ~\cap~ \{~ \Id + L^{1}(H_{+}) ~\}.
$$
The Lie algebra $\g$ of $G$ is the Lie algebra of skew-Hermitian
operators of trace class. We will denote by $G^{\C}$ the
complexification of $G$~:
$$
G^{\C}~:= GL(H_{+}) ~\cap~ \{~ \Id + L^{1}(H_{+}) ~\},
$$
and by $\g^{\C}$ its complex Lie algebra $\g \oplus i\g$. The
group $G$ acts on $T\m$ by
$$
u \cdot (x, X) := (x \circ u^{-1},\, X \circ u^{-1}),
$$
for all $u$ in $G$ and for all $(x, X)$ in $T\m$. This action
extends to an $I_{1}$-holomorphic action of  $G^{\C}$  by
$$
g \cdot (x, X) := (x \circ g^{-1},\, X \circ g^{*}),
$$
for all $g$ in $G^{\C}$ and for all $(x, X)$ in $T\m$.

\begin{prop}\label{rien}
The action of $G^{\C}$ on $T\m$ is Hamiltonian  with respect to
the complex symplectic form $\Omega$, with moment map $\mu^{\C}$~:
$$
\begin{array}{llll}
{\mu}^{\C}~: & T\m & \longrightarrow &  (\g^{\C})'\\
             & (x, X) & \longmapsto & \left( \a \mapsto \Tr \left( X^{*}x \a \right)
             \right).
     \end{array}
$$
\end{prop}

\proofp  \ref{rien}:\\
Let us check that $\mu^{\C}$ satisfies:
$$
\langle d{\mu}^{\C}_{(x, X)}\left((Z, T)\right), \a \rangle =
i_{\a \cdot (x, X)}\Omega \left((Z, T)\right),
$$
for all $\left(Z, T \right)$ in $T_{(x, X)}\m$ and  for all $\a$
in $\g^{\C}$, where $\langle ~,~ \rangle$ denotes the duality
pairing and where $\a \cdot (x, X) = (- x \circ \a,\, X \circ
\a^{*})$ is the vector induced by the infinitesimal action of $\a$
on $(x, X)$. One has~:
$$
\begin{array}{ll}
\langle d{\mu}^{\C}_{(x, X)}\left((Z, T)\right), \a \rangle & =
\Tr \left( (X^{*}Z + T^{*}x) \a \right) =  \Tr(X^{*}Z \a + T^{*}x
\a) \\ & = \Tr ( \a X^{*} Z ) - \Tr (T^{*} (-x \circ
\a))  \textrm{ ~~~~~~~~since } \a \in L^{1}(H_{+})\\
& = \Tr \left((X \circ \a^{*})^{*} Z \right) - \Tr (T^{*}(-x \circ
\a))  \\
& = ~i_{(-x \circ \a, X \circ \a^{*})} \Omega\left( (Z, T)
\right).
\end{array}
$$
\cqfd

 It follows from Proposition  \ref{rien} that the real
symplectic forms $\w_{2}$ and $\w_{3}$ are Hamiltonian  with
respect to the real moment maps~:
$$
\begin{array}{llll}
\mu_{2} = \Re({\mu}^{\C}): & T\mathcal{M}_{k} & \rightarrow & \g'\\
                         & (x, X) & \mapsto &
                         \left( \a \mapsto \frac{1}{2} \Tr \left( X^{*}x - x^{*}X \right) \a \right)\\
& &  & \\
\mu_{3} = \Im({\mu}^{\C}): &  T\mathcal{M}_{k} & \rightarrow & \g'\\
                         & (x, X) & \mapsto &
                         \left( \a \mapsto -\frac{i}{2} \Tr \left( X^{*}x + x^{*}X \right) \a \right)\\
\end{array}
$$

\begin{prop}\label{rien2}
The action of $G$ on $T\m$ is  Hamiltonian  with respect to the
real symplectic form $\w_{1}$ with moment map~:
$$
\begin{array}{llll}
{\mu}_{1}~: & T\m  & \longrightarrow &  \g'\\
            & (x, X) & \longmapsto &
            \left( \a \mapsto -\frac{i}{2} \Tr \left(  x^{*} x  -  X^{*} X \right) \a
            \right).
\end{array}
$$
\end{prop}

\proofp \ref{rien2}~:\\
One has to check  that $\mu_{1}$ satisfies~:
$$
\langle \left(d{\mu}_{1}\right)_{(x, X)}\left((Z, T)\right), \a
\rangle = i_{(-x \circ \a, -X \circ \a)}\omega_1 \left((Z,
T)\right),
$$
for all $\left(Z, T \right)$ in $T_{(x, X)}\m$ and all $\a $ in
$\g$. One has~:
$$
\begin{array}{ll}
\langle \left(d{\mu}_{1}\right)_{(x, X)}\left((Z, T)\right), \a
\rangle & = -\frac{i}{2} \Tr \left( Z^* x \a + x^* Z \a \right) +
\frac{i}{2} \Tr \left(  T^* X \a + X^*T \a \right)\\
& = ~\frac{i}{2} \Tr \left( Z^{*} (-x \circ \a) - \a x^{*} Z
\right) - \frac{i}{2} \Tr \left(
 T^{*} (-X \circ \a) - \a X^{*} T \right)\\
 & = ~\frac{i}{2} \Tr \left(Z^{*} (-x \circ \a) - (x \circ
 \a^{*})^{*} Z \right) - \frac{i}{2} \Tr \left( T^{*} (-X \circ
 \a) - (X \circ \a^{*})^{*} T \right)\\
 & = ~\frac{i}{2} \Tr \left( Z^{*}(-x \circ \a) - (-x \circ \a)^{*}
 Z \right) - \frac{i}{2} \Tr \left( T^{*}(-X \circ \a) - (-X \circ
 \a)^{*} T \right)\\
 & = ~\Im \Tr (-x \circ \a)^{*}Z - \Im \Tr (-X \circ \a)^{*} T\\
 & = ~i_{(-x \circ \a, -X \circ \a)}\omega_1 \left((Z,
T)\right).
\end{array}
$$
\cqfd \\
In the following, we will denote by $\mu$ the $\mathfrak{g}'
\otimes \R^{3}$-valued moment map defined by:
$$
\begin{array}{llll}
{\mu}~: & T\m & \rightarrow & \mathfrak{g}' \otimes \R^{3} \\
        & (x, X) & \mapsto & (\,{\mu}_{1}(x, X)~,~ \mu_{2}(x, X)~,~ \mu_{3}(x,
        X)\,).
\end{array}
$$
In the next Subsection   we will consider the
$\textrm{Ad}^{*}(G)$-invariant value $\xi_{k} :=
\left(-\frac{i}{2}k^{2} \Tr,\, 0,\, 0\right)$ of the moment map
$\mu$ and the level set:
$$
\mathcal{W}_{k}:= {\mu}^{-1}\left(\xi_{k}\right).
$$

\subsection{Smooth Banach manifold structure on the level set $\W$}

This Subsection   is devoted to the proof of the following Theorem
~:
\begin{thm}\label{w}
$\W := {\mu}^{-1}\left(\left(-\frac{i}{2}k^{2} \Tr,\, 0,\,
0\right)\right)$ is a smooth Riemannian  submanifold of $T\m$.
\end{thm}
We will prove  that $\xi_{k} = \left(-\frac{i}{2}k^{2} \Tr,\, 0,\,
0\right)$ is a regular value of the moment map $\mu$. For this
purpose we will need the following fact, which will be useful in
other parts of the paper, so that we single it out here~:

\begin{lem}\label{technical}
Let $B$ be a Banach space which injects continuously into a
Hilbert space $H$. Let $F$ be a closed subspace of $B$ and
$\bar{F}$  its closure in $H$. If the orthogonal projection of $H$
onto $\bar{F}$ maps $B$ onto $F$, then $F$ admits a closed
complement in $B$ which is ${\bar{F}}^{\perp} ~\cap~B$.
\end{lem}

\proofl \ref{technical}:\\
Since the orthogonal projection $p$ of $H$ onto $\bar{F}$ maps $B$
onto $F$, $B$ is the algebraic sum of $F$ and ${\bar{F}}^{\perp}
~\cap~B$. Since $B$ injects continuously into $H$ and $p$ is
continuous, the projection from $B$ onto $F$ with respect to
${\bar{F}}^{\perp} ~\cap~B$ is continuous. Hence $$B ~=~ F \oplus
({\bar{F}}^{\perp} ~\cap~B)$$ as a topological sum. \cq
\\
\\
\prooft \ref{w}:\\
Consider the following smooth map of Banach manifolds
$$
\begin{array}{llll}
\mathcal{F}~: & T\m & \longrightarrow & L^{1}(H_{+}) \times
\mathcal{S}^{1}(H_{+})\\
& (x, X) & \longmapsto & (X^{*}x,~ x^{*}x - X^{*}X),
\end{array}
$$
  We have:
$$
\begin{array}{llll}
d_{(x, X)}\mathcal{F}~: & T_{(x, X)}T\m & \longrightarrow & L^{1}(H_{+}) \times
\mathcal{S}_{1}(H_{+})\\
& (Z, T) & \longmapsto & (X^{*}Z + T^{*}x,~ x^{*}Z + Z^{*}x -
X^{*}T - T^{*}X).
\end{array}
$$
The level set $\W$ is the preimage of $(0, k^{2})$ under
$\mathcal{F}$. To prove that $\W$ is a smooth Banach submanifold
of $T\m$ it is sufficient to prove that
 the differential of $\mathcal{F}$ at a
point $(x, X)$ of $\W$ is surjective and that its kernel splits.

\. For this purpose consider the following decomposition of
$H_{+}$:
\begin{equation}\label{H+}
H_{+} = \ker X \oplus \left({\ker X}\right)^{\perp}.
\end{equation}
The operator $x$ is a  Fredholm operator hence it has closed
range, and $X$ is a compact operator. The equality $X^{*}x = 0$
implies that the range  $\textrm{Ran}\,x$ of $x$ is orthogonal to
the range  $\textrm{Ran}\,X$ of $X$. From the continuity of the
orthogonal projection of $H$ onto $\textrm{Ran}\,x$ it follows
that: $\textrm{Ran}\,x \perp \overline{\textrm{Ran}\,X}$. Let us
introduce the following decomposition of $H$:
\begin{equation}\label{H}
H = \overline{\textrm{Ran }X} \oplus \left(\textrm{ Ran
}X\right)^{\perp} \cap \left(\textrm{ Ran
  }x\right)^{\perp} \oplus \textrm{ Ran }x_{|\ker X} \oplus
\textrm{ Ran }x_{|{\ker X}^{\perp}}.
\end{equation}
With respect to the decompositions (\ref{H+}) and (\ref{H}) of
$H_{+}$ and $H$ into closed subspaces,  $x$ and $X$ have the
following expressions~:
$$
\begin{array}{ll}
x = \left( \begin{array}{cc} 0 & 0 \\ 0 & 0 \\ x_{31} & 0 \\ 0 & x_{42}
  \end{array} \right)
&
X = \left( \begin{array}{cc} 0 & X_{12} \\ 0 & 0 \\ 0 & 0 \\ 0 & 0
  \end{array} \right),
\end{array}
$$
where $x_{31}$ and $x_{42}$ are continuous bijections, thus
isomorphisms, and where  $X_{12}$ is 1-1 but not onto. Let $(Z,
T)$ be a tangent vector to $T\m$ at $(x, X)$, and denote by
$(Z_{ij})_{ 1 \leq i \leq 4,\, 1 \leq j \leq 2 }$ and $(T_{ij})_{
1 \leq i \leq 4,\, 1 \leq j \leq 2 }$ the block decompositions of
$Z$ and $T$ with respect to the direct sums (\ref{H+}) and
(\ref{H}). Note that~:
$$
X^{*}Z + T^{*}x = \left(\begin{array}{cc}
      T_{31}^{*}x_{31} & T_{41}^{*}x_{42} \\
      X_{12}^{*}Z_{11} + T_{32}^{*}x_{31} & X_{12}^{*}Z_{12} +
      T_{42}^{*}x_{42}
      \end{array} \right),
$$
and that $ x^{*}Z + Z^{*}x - X^{*}T - T^{*}X $ equals~:
$$
\left(\begin{array}{cc}
x_{31}^{*}Z_{31} + Z_{31}^{*}x_{31}  & x_{31}^{*}Z_{32} +
Z_{41}^{*}x_{42} - T_{11}^{*}X_{12}\\
x_{42}^{*}Z_{41} + Z_{32}^{*}x_{31} - X_{12}^{*}T_{11}  &
x_{42}^{*}Z_{42} + Z_{42}^{*}x_{42} - T_{12}^{*}X_{12} - X_{12}^{*}T_{12}
\end{array} \right).
$$

\. To show that the differential of $\mathcal{F}$ is onto,
consider an element  $(U, V)$  of $L^{1}(H_{+}) \times
\mathcal{S}^{1}(H_{+})$ and denote be $(U_{ij})_{1 \leq i,j \leq
2}$ and $(V_{ij})_{1 \leq i, j \leq 2}$ the block decompositions
of $U$ and $V$ with respect to the direct sum $H_{+} = \ker X
\oplus \left(\ker
  X\right)^{\perp}$. A preimage of $(U, V)$ by $d_{(x, X)}\mathcal{F}$ is
given by the following ordered pair $(Z, T)$:
$$
\begin{array}{ll}
Z = \left(\begin{array}{cc} 0 & 0
     \\
     0 & 0 \\
     \frac{1}{2}x_{31}^{-1*}V_{11} & \frac{1}{2}x_{31}^{-1*}V_{12} \\
     \frac{1}{2}x_{42}^{-1*}V_{21} & \frac{1}{2}x_{42}^{-1*}V_{22}
\end{array} \right)
&
T = \left(\begin{array}{cc} 0 &
0 \\
0 & 0 \\ x_{31}^{-1*}U_{11}^{*} & x_{31}^{-1*}U_{21}^{*}\\
 x_{42}^{-1*}U_{12}^{*} & x_{42}^{-1*}U_{22}^{*}
\end{array} \right).
\end{array}
$$

\. Let us now show that for every  $(x, X)$ in $\W$, the kernel of
the differential $d_{(x, X)}\mathcal{F}$ splits. It is  given by
the following subspace of $T_{(x, X)}\mathcal{M}_{k}$~:
$$
T_{(x, X)}\W ~:=~ \{~(Z, T)~ \in~ T_{(x, X)}T\m ~~|~~ X^{*}Z +
T^{*}x = 0,~
 x^{*}Z + Z^{*}x = X^{*}T + T^{*}X ~\}.
$$
Consider the space $ L^{2}(H_{+}, H) \times L^{2}(H_{+}, H) $
endowed with the complex structure $I(Z, T) = (iZ, -iT)$ and the
strong Riemannian  metric $\bar{\gm}$ given by the real part of
the natural Hermitian  product. Let $E$ be the closure of  $T_{(x,
X)}\W$ in  $L^{2}(H_{+}, H) \times L^{2}(H_{+}, H)$ and
${E}^{\perp_{\bar{\gm}}}$ its orthogonal in $L^{2}(H_{+}, H)
\times L^{2}(H_{+}, H)$. By Lemma  \ref{technical}, to show that
$T_{(x, X)}\W$ is complemented in $T_{(x, X)}(T\m)$, it is
sufficient to show that the orthogonal projection of $L^{2}(H_{+},
H) \times L^{2}(H_{+}, H)$ onto $E$ maps $T_{(x, X)}(T\m)$ onto
 $T_{(x, X)}\W$.
An ordered pair $(Z, T) \in  L^{2}(H_{+}, H) \times L^{2}(H_{+},
H)$ is in $E$ if and only if $Z$ and $T$ are of the following
form:
$$
\begin{array}{l}
Z = \left(\begin{array}{cc} Z_{11} & Z_{12}\\ Z_{21} & Z_{22} \\
    x_{31}^{-1*} \mathfrak{a}_{1} &  Z_{32} \\
x_{42}^{-1*}(X_{12}^{*}T_{11} - Z_{32}^{*}x_{31}) &
x_{42}^{-1*} (\frac{1}{2}(X_{12}^{*}T_{12} +
 T_{12}^{*}X_{12})
 + \mathfrak{a}_{2})
\end{array}\right) \\
\\
T = \left(\begin{array}{cc} T_{11}
& T_{12} \\
T_{21} & T_{22}\\ 0 &
    -x^{-1*}_{31}Z^{*}_{11}X_{12}\\ 0 & -x^{-1*}_{42}Z^{*}_{12}X_{12}
\end{array}\right),
\end{array}
$$
where $\mathfrak{a}_{1}$ (resp. $\mathfrak{a}_{2}$) is an element
of the space $\mathcal{A}^{2}(\ker X)$ (resp.
$\mathcal{A}^{2}\left((\ker X)^{\perp}\right)$ of skew-Hermitian
Hilbert-Schmidt operators on $\ker X$ (resp. on $(\ker
X)^{\perp}$). An ordered pair $(Z, T) \in  L^{2}(H_{+}, H) \times
L^{2}(H_{+}, H)$ is in ${E}^{\perp_{\bar{\gm}}}$ if and only if
$Z$ and $T$ have the following form:
$$
\begin{array}{ll}
Z = \left(\begin{array}{cc} X_{12}T_{32}^{*}x_{31}^{-1*} &
    X_{12}T_{42}^{*}x_{42}^{-1*} \\ 0 & 0 \\
    x_{31}\mathfrak{s}_{1} &  x_{31}Z_{41}^{*}x_{42}^{-1*} \\
    Z_{41} & x_{42} \mathfrak{s}_{2}
\end{array}\right)
&
T = \left(\begin{array}{cc} -X_{12}x_{42}^{-1}Z_{41}  &
-X_{12}\mathfrak{s}_{2}\\
0 & 0\\ T_{31} & T_{32} \\ T_{41} & T_{42}
\end{array}\right),
\end{array}
$$
where  $\mathfrak{s}_{1}$ (resp. $\mathfrak{s}_{2}$) is an element
of the space  $\mathcal{S}^{2}(\ker X)$ (resp.
$\mathcal{S}^{2}\left((\ker X)^{\perp}\right)$ of self-adjoint
Hilbert-Schmidt operators on $\ker X$ (resp. $(\ker X)^{\perp}$).
We will denote by $(p_{E}(Z), ~p_{E}(T))$ the orthogonal
projection of an element $(Z, T) \in L^{2}(H_{+}, H) \times
L^{2}(H_{+}, H)$ to $E$, and  $(p_{E}(Z_{ij}))_{ 1 \leq i \leq
4,\, 1 \leq j \leq 2 }$ (resp. $(p_{E}(T_{ij}))_{ 1 \leq i \leq
4,\, 1 \leq j \leq 2 }$ ) the block decomposition of $p_{E}(Z)$
(resp. $p_{E}(T)$) with respect to the direct sums (\ref{H+}) and
(\ref{H}). Then $p_{E}(Z)$ and $p_{E}(T)$ are given by~:
$$
\begin{array}{l}
p_{E}(Z) = \left(\begin{array}{cc} p_{E}(Z_{11}) & p_{E}(Z_{12})\\
   Z_{21} & Z_{22} \\
     \frac{1}{2}(Z_{31} - x_{31}^{-1*}Z_{31}^{*}x_{31})
&   p_{E}(Z_{32}) \\ p_{E}(Z_{41}) &
     \frac{1}{2}x_{42}^{-1*}\left(\frac{1}{2}\left(X_{12}^{*}p_{E}(T_{12}) +
p_{E}(T_{12})^{*}X_{12}\right)  +
     \mathfrak{a}_{1}\right)
\end{array}\right) \\
\\
p_{E}(T) = \left(\begin{array}{cc} p_{E}(T_{11})
  &
p_{E}(T_{12}) \\
T_{21} & T_{22}\\ 0 &
    -x^{-1*}_{31}(p_{E}(Z_{11}))^{*}X_{12} \\
0 & -x^{-1*}_{42}(p_{E}(Z_{12}))^{*}X_{12}
\end{array}\right).
\end{array}
$$
with:
$$
\begin{array}{l}
p_{E}(Z_{41}) = \frac{1}{2}(Z_{41} + x_{42}^{-1*}Z_{32}^{*}x_{31} -
x_{42}^{-1*}X_{12}^{*}T_{12})\\
p_{E}(Z_{32}) = Z_{32} - x_{31}p_{E}(Z_{41})^{*}x_{42}^{-1*}\\
p_{E}(T_{11}) = T_{11} + X_{12}x_{42}^{-1}p_{E}(Z_{41})
\\
\mathfrak{a}_{1} = \frac{1}{2}(x_{42}^{*}Z_{42} -
Z_{42}^{*}x_{42}) + \frac{1}{2}\left(X_{12}^{*}\left(T_{12} -
p_{E}(T_{12})\right) -  \left(T_{12} - p_{E}(T_{12})\right)^{*}
X_{12}\right).
\end{array}
$$
Denoting by $p_{{+}}$ the orthogonal projection of $H$ onto
$H_{+}$, let us prove that  $p_{{+}} \circ p_{E}(Z)$ and $p_{{+}}
\circ p_{E}(T)$ are of trace class whenever $p_{{+}}(Z) $ and
$p_{{+}}(T)$ are of trace class. Let
$$
p_{{+}} = \left(\begin{array}{cccc} p_{11} & p_{12}  & p_{13} &
    p_{14}
\\ p_{21} &  p_{22} &  p_{23}  & p_{24} \end{array}\right)
$$
be the decomposition of $p_{{+}}$ with respect to the
decompositions (\ref{H}) of $H$ and (\ref{H+}) of $H_{+}$. For
$(x, X) \in \W$, the operators $p_{11} ,\, p_{12} ,\, p_{14} ,\,
p_{21},\, p_{22},\, p_{23} $ are of trace class, as well as
$p_{13} - x_{31}^{-1}$ and $p_{24} - x_{42}^{-1}$. The  condition
$p_{{+}}(Z) \in L^{1}(H_{+})$ implies that $Z_{31},\, Z_{32},\,
Z_{41}$ and $Z_{42}$ are of trace class. It follows that
$p_{E}(Z_{32})$ and $p_{E}(Z_{41})$ are of trace class, as well as
$\mathfrak{a}_{1}$. Hence $p_{{+}}\circ p_{E}(Z)$ is of trace
class. On the other hand, $p_{11},\, p_{12},\, p_{21}$ and $p_{22}
$ being of trace class,  the condition $p_{{+}}(T) \in
L^{1}(H_{+})$ implies that $p_{{+}}\circ p_{E}(T)$ is of trace
class. Thus we obtain the following  direct sum of closed
subspaces~:
$$
 T_{(x, X)}(T\m) =   T_{(x, X)}\W \oplus \left({E}^{\perp_{\bar{\gm}}}
\cap T_{(x, X)}T\m\right).
$$
and the Theorem  follows. \cqfdt

\subsection{Hyperk\"ahler quotient of $T\m$ by $G$}

In this Subsection we will prove that the assumptions (D) and (S)
of Subsection   \ref{hypsub} are satisfied, as well as the
following Theorem.

\begin{thm}\label{qu}
The hyperk\"ahler quotient of the weak hyperk\"ahler space
$T\mathcal{M}_{k}$ with respect to the tri-symplectic action of $G$
 provides $\W / G$ with a structure of strong
hyperk\"ahler smooth Hilbert manifold.
\end{thm}
For the remainder of the paper, we will denote by
$(\mathrm{g}^{\textrm{red}}, I_{1}^{red}, I_{2}^{red},
I_{3}^{red}, \omega_{1}^{red}, \omega_{2}^{red}, \omega_{3}^{red})
$ the hyperk\"ahler structure induced by $T\mathcal{M}_{k}$ on the
reduced space $\W / G$.\\

\prooft \ref{qu}:

\. Let us first show that the quotient space $\W/G$ endowed with
the quotient topology is Hausdorff. Since $G$ acts freely on $\W$,
by Propositions \ref{haus} and \ref{propre}, it it sufficient to
prove that  the graph $\mathcal{C}$ of the equivalence relation
defined by $G$ is closed in $\W \times \W$ and that the canonical
application from $\mathcal{C}$ to $G$ is continuous.

Note that, for all $(y, Y)$ in $\W$, the operator $y$ is an
injective Fredholm operator from $H_{+}$ to $H$, thus establishes
an isomorphism between $H_{+}$ and its (closed) range. We will
denote by $y^{-1}$ the inverse operator of $y$, and extend it to
an operator from $H$ to $H$ by demanding that the restriction of
$y^{-1}$ to the orthogonal complement of $\textrm{ Ran }y$
vanishes.

Let us remark that  the canonical map from the graph $\mathcal{C}$
to $G$ assigns to an ordered pair $\left((x, X), ~(y, Y)\right)$
the element $g = y^{-1}x$. It is continuous with respect to the
topology of $\mathcal{C}$ inherited from the topology of $\W
\times \W$, and  the topology of $G$ induced by the $L^{1}$-norm
on $\Id + L^{1}(H_{+}).$

Let $ \{~(x_{n}\,,\, X_{n})~;~(x_{n}\, \circ\,
g_{n}^{-1},\,X_{n}\, \circ\, g_{n}^{-1})~\} $ be a sequence in
$\mathcal{C}$ which converges to an element $\left( (
x,\,X);~(y,\,Y) \right) $ in $\W \times \W$. The sequence
$\{~g_{n}^{-1} = x_{n}^{-1} \circ (x_{n} \circ g_{n}^{-1})~\}$ is
a sequence of elements of $G$ which converges to the element
$x^{-1}y$ in $\Id + L^{1}(H_{+})$. Since $G$ is closed in $\Id +
L^{1}(H_{+})$, it follows that $x^{-1}y$ is in $G$ and that
$\mathcal{C}$ is closed in $\W \times \W$.

\. By Theorem  \ref{hyper}, for $\W/G$ to be hyperk\"ahler, it is
sufficient to   have the following topological direct sum,
referred to as (D)~:
$$
{\rm (D) }\hspace{40pt}  T_{(x,\, X)}G\!\cdot\!(x, \,X) ~\oplus~
\left(T_{(x,\, X)}G\!\cdot\!(x,\, X)\right)^{\perp_{\textrm{g}}}
~=~ T_{(x,\, X)}\W.
$$
For this purpose, we will use Lemma  \ref{technical}. Let us again
consider  the space
$
L^{2}(H_{+}, H) \times L^{2}(H_{+}, H)
$
endowed with the complex structure $I(Z, T) = (iZ, -iT)$ and the
strong Riemannian  metric $\bar{\gm}$ given by the real part of
the natural Hermitian  product. Let $E$ (resp. $F$) be the closure
of ${T_{(x, X)}\W}$  (resp. ${T_{(x, X)}G\!\cdot\!(x, X)}$) in
$L^{2}(H_{+}, H) \times L^{2}(H_{+}, H)$.   We will show that the
$\bar{\gm}$-orthogonal projection   of  $E$  onto $F$ maps
${T_{(x, X)}\W}$ onto ${T_{(x, X)}G\!\cdot\!(x, X)}$. The
orthogonal of $F$  in $E$ is the set of ordered pairs $(Z, T) \in
L^{2}(H_{+}, H) \times L^{2}(H_{+}, H)$ of the form~:
$$
\begin{array}{ll}
Z = \left(\begin{array}{cc}
    Z_{11} & Z_{12} \\
    Z_{21} & Z_{22}  \\
    0 & x_{31}^{-1*}T_{11}^{*}X_{12} \\
    0 & x_{42}^{-1*}T_{12}^{*}X_{12}
     \end{array}\right);
&

T = \left(\begin{array}{cc}
     T_{11}   & T_{12} \\
     T_{21} & T_{22} \\
     0 & -x_{31}^{-1*}Z_{11}^{*}X_{12}\\
     0 & -x_{42}^{*-1}Z_{12}^{*}X_{12}
 \end{array}\right).
\end{array}
$$
The orthogonal projection of $E$ onto $F$  maps an ordered pair
 $(Z,\, T) \in E$ having the following block decompositions
 with respect to the direct sums (\ref{H+}) and (\ref{H})
$$
\begin{array}{l}
Z = \left(\begin{array}{cc} Z_{11} & Z_{12}\\ Z_{21} & Z_{22} \\
    x_{31}^{-1*} \mathfrak{a}_{1} &  Z_{32} \\
x_{42}^{-1*}(X_{12}^{*}T_{11} - Z_{32}^{*}x_{31}) &
x_{42}^{-1*} (\frac{1}{2}(X_{12}^{*}T_{12} + T_{12}^{*}X_{12}) +
    \mathfrak{a}_{2})
\end{array}\right) \\
\\
T = \left(\begin{array}{cc} T_{11}
& T_{12}\\
T_{21} & T_{22}\\ 0 &
    -x^{-1*}_{31}Z^{*}_{11}X_{12}\\ 0 & -x^{-1*}_{42}Z^{*}_{12}X_{12}
\end{array}\right)
\end{array}
$$
to an ordered pair $(p_{G}(Z), p_{G}(T))$ with decompositions
$$
\begin{array}{l}
p_{G}(Z) = \left(\begin{array}{cc}
           0 & 0 \\ 0 & 0 \\
          x_{31}^{*-1}\mathfrak{a}_{1} &
           -x_{31}(T_{11}^{*}X_{12}-
           x_{31}^{*}Z_{32})x_{42}^{-1}x_{42}^{-1*}\\
            x_{42}^{-1*}(X_{12}^{*}T_{11} - Z_{32}^{*}x_{31})
           & - x_{42} \mathfrak{a}
        \end{array} \right)

\\

p_{G}(T) =  \left(\begin{array}{cc}
            X_{12}x_{42}^{-1}x_{42}^{-1*}(X_{12}^{*}T_{11} -
            Z_{32}^{*}x_{31})
          & -X_{12} \mathfrak{a}\\
            0 & 0 \\0 & 0 \\ 0 & 0
            \end{array} \right),
\end{array}
$$
where $\mathfrak{a}$ satisfies~:
\begin{equation}\label{aaaa}
\mathfrak{a}_{2} + \frac{1}{2}(X_{12}^{*}T_{12} - T_{12}^{*}X_{12})
= k^{2}\mathfrak{a} - (x_{42}^{*}x_{42}\mathfrak{a} +
\mathfrak{a}x_{42}^{*}x_{42}).
\end{equation}
Let us show that the projection $p_{G}$ from $E$ to $F$ maps an
element of $T_{(x, X)}\W$ into $T_{(x, X)}G\!\cdot\!(x, X)$. Let
$$
p_{{+}} = \left(\begin{array}{cccc} p_{11} & p_{12}  & p_{13} &
    p_{14}
\\ p_{21} &  p_{22} &  p_{23}  & p_{24} \end{array}\right)
$$
be the expression of the orthogonal projection onto $H_{+}$ with
respect to the direct sums (\ref{H}) and (\ref{H+}). For $(x, X)
\in \W$, the operators $p_{11} ,\,\, p_{12} ,\,\, p_{14} ,\,
\,p_{21},\,\, p_{22},\,\, p_{23}$ are of trace class as well as
$p_{13} - x_{31}^{-1}$ and $p_{24} - x_{42}^{-1}$. The condition
$p_{{+}}(Z) \in L^{1}(H_{+})$ implies in  particular that
$\mathfrak{a}_{1}$,\, $Z_{32}$ and $\mathfrak{a}_{2}$ are of trace
class. To conclude that $p_{G}$ maps $T_{(x, X)}\W$ into $T_{(x,
X)}G\!\cdot\!(x, X)$, it remains to show that $\a$ defined by
Equation (\ref{aaaa}) is of trace class. This will follow by lemma
\ref{cici} below. As a consequence, $p_{{+}} \circ \,p_{G}(Z)$ and
$p_{{+}} \circ\, p_{G}(T)$ are of trace class.
 Hence,  $F^{\perp_{\bar{\textrm{g}}}} ~\cap~ T_{(x,
X)}\W $ is a closed complement to $T_{(x, X)}G\!\cdot\!(x, X)$.
Since $\textrm{g}$ is the restriction of $\bar{\textrm{g}}$ to
$T_{(x, X)}\W $, it follows that~:
$$
T_{(x, X)}G\!\cdot\!(x, X) \oplus \left(T_{(x, X)}G\!\cdot\!(x,
X)\right)^{{\perp}_{\textrm{g}}} ~=~ T_{(x, X)}\W.
$$
\cqfdt

\begin{lem}\label{cici}
The map $\mathcal{E}_{x_{42}}$ defined by
$$
\begin{array}{llll}
\mathcal{E}_{x_{42}}~:& L^{2}\left((\ker X)^{\perp}\right) &
\longrightarrow &
L^{2}((\ker X)^{\perp})\\
& \mathfrak{a} & \mapsto & k^{2}\mathfrak{a} - (x_{42}^{*}x_{42}\mathfrak{a} +
\mathfrak{a}x_{42}^{*}x_{42}).
\end{array}
$$
is an isomorphism which restricts to an isomorphism of
$L^{1}\left((\ker X)^{\perp}\right)$.
\end{lem}

\proofl \ref{cici}~:\\
  Indeed, $x_{42}^{*}x_{42}$
is a self-adjoint positive definite operator on the space $(\ker
X)^{\perp}$ satisfying
$$
 x_{42}^{*}x_{42} = k^{2} \textrm{ Id } + X_{12}^{*}X_{12},
$$
where $X_{12}^{*}X_{12}$ is a compact operator. It follows that
there exists a diagonal operator $D$ with respect to an
  orthogonal basis
$\{f_{i}\}_{ i \in J }$ of $(\ker X)^{\perp}$, and a unitary
operator  $u \in \U\left((\ker X)^{\perp}\right)$ such that
$x_{42}^{*}x_{42} = u~ D~ u^{*}$. Denote by $D_{i}$ the
eigenvalues of $D$. Remark that $D_{i} > k^{2}$. It follows that
the equation $$ k^{2} \mathfrak{a} = x_{42}^{*}x_{42}\mathfrak{a}
+ \mathfrak{a}x_{42}^{*}x_{42}
$$
is equivalent to
$$
k^{2} u^{*}~\mathfrak{a}~u = u^{*}~\mathfrak{a}~u~ D + D~
u^{*}~\mathfrak{a}~u,
$$
and implies in particular that, for every $i$ and $j$ in $J$, one
has~:
$$
k^{2} \langle ~(u^{*}~\mathfrak{a}~u)(f_{j}),~ f_{i}~\rangle =
(D_{j} + D_{i})~ \langle ~(u^{*}~\mathfrak{a}~u)(f_{j}),~
f_{i}~\rangle.
$$
From $(D_{j} + D_{i}) - k^{2} > k^{2}$, we get that $\langle~
(u^{*}~\mathfrak{a}~u)(f_{j}),~ f_{i}~\rangle = 0$ for every $i$,
$j \in J$. Thus $\ker \mathcal{E}_{x_{42}} = 0$. To see that the
map $\mathcal{E}_{x_{42}}$ is surjective, consider an element $V$
in $L^{2}((\ker X)^{\perp})$, and define an operator
$\tilde{\mathfrak{a}} \in L^{2}((\ker X)^{\perp})$ by~:
$$
\langle ~\tilde{\mathfrak{a}}(f_{j}),~ f_{i}~\rangle =
\frac{1}{(k^{2} -
  (D_{i} + D_{j}))} ~ \langle ~(u^{*}~ V ~u) (f_{j}),~ f_{i}~\rangle.
$$
The operator $u~\tilde{\mathfrak{a}}~u^{*}$ is a preimage of $V$
by $\mathcal{E}_{x_{42}}$. Moreover if $V \in L^{1}((\ker
X)^{\perp})$, then $u~\tilde{\mathfrak{a}}~u^{*} \in L^{1}((\ker
X)^{\perp})$. Since $\mathcal{E}_{x_{42}}$ is clearly continuous,
it follows that $\mathcal{E}_{x_{42}}$ is an isomorphism of
$L^{2}((\ker X)^{\perp})$ that restricts to an isomorphism of
$L^{1}((\ker X)^{\perp})$. \cq

\begin{prop}\label{Sprop}
For every $(x, \,X)$ in $\W$ one has~:
$$
\begin{array}{lll}
T_{(x, X)}\mathcal{M}_{k} &=&
 T_{(x, X)}G\!\cdot\!(x, X) \oplus H_{(x, X)}
\oplus I_{1}\left(T_{(x, X)}G\!\cdot\!(x, X)\right) \\
&& \\ & & \qquad\oplus I_{2}\left(T_{(x, X)}G\!\cdot\!(x,
X)\right) \oplus I_{3}\left(T_{(x, X)}G\!\cdot\!(x, X)\right),
 \end{array}
$$
where $H_{(x,\,X)}$ is the orthogonal of $T_{(x, X)}G\!\cdot\!(x,
X)$ in $T_{(x,\,X)}\W$.
\end{prop}

\proofp \ref{Sprop}:\\
With the previous notation, it follows  from the proof of Theorem
\ref{w} that the orthogonal projection from $L^{2}(H_{+}, H)
  \times L^{2}(H_{+}, H)$ to $\overline{T_{(x, X)}\mathcal{W}_{k}}$ with respect
   to the strong Riemannian  metric $\bar{\gm}$
takes $T_{(x, X)}\mathcal{M}_{k}$ to $T_{(x, X)}\mathcal{W}_{k}$.
It also follows  from the proof of Theorem  \ref{qu} that the
orthogonal projection from $\overline{T_{(x, X)}\mathcal{W}_{k}}$
to $\overline{T_{(x, X)}G\!\cdot\!(x, X)}$ with respect to
$\bar{\textrm{g}}$ takes $T_{(x, X)}\mathcal{W}_{k}$ onto  $T_{(x,
X)}G\!\cdot\!(x, X)$. Now let us remark that the complex
structures $I_j$, $j = 1,\,2,\,3$, extend to complex structures of
$L^{2}(H_{+}, H)
  \times L^{2}(H_{+}, H)$ by the same formulas, making $L^{2}(H_{+}, H)
  \times L^{2}(H_{+}, H)$ into a hyperk\"ahler space.
  Since, for $j = 1,\, 2,\, 3$, the complex
structure $I_{j}$ fixes $T_{(x, X)}\mathcal{M}_{k}$ and is
orthogonal with respect to $\bar{\textrm{g}}$, it follows that for
$j = 1,\,2,\,3$, the orthogonal projection from $L^{2}(H_{+}, H)
  \times L^{2}(H_{+}, H)$ onto $I_{j}\left(\overline{T_{(x, X)}G\!\cdot\!(x, X)}\right)$
  takes $T_{(x, X)}\mathcal{M}_{k}$ onto $I_j\left(T_{(x, X)}G\!\cdot\!(x, X)\right)$.
Hence, from the orthogonal sum
$$
\begin{array}{lll}
L^{2}(H_{+}, H)
  \times L^{2}(H_{+}, H)~ &=& ~~~\overline{T_{(x, X)}G\!\cdot\!(x, X)} \oplus
  \overline{H_{(x, \,X)}} \oplus I_1\left(\overline{T_{(x, X)}G\!\cdot\!(x,
  X)}\right)\\&&\\ & & \qquad
  \oplus I_2\left(\overline{T_{(x, X)}G\!\cdot\!(x, X)}\right) \oplus I_3\left(\overline{T_{(x, X)}G\!\cdot\!(x,
  X)}\right),
  \end{array}
$$
it follows that
$$
\begin{array}{lll}
T_{(x, X)}\mathcal{M}_{k} &=&
 T_{(x, X)}G\!\cdot\!(x, X) \oplus H_{(x, X)}
\oplus I_{1}\left(T_{(x, X)}G\!\cdot\!(x, X)\right)\\&&\\ & &
\qquad\oplus I_{2}\left(T_{(x, X)}G\!\cdot\!(x, X)\right) \oplus
I_{3}\left(T_{(x, X)}G\!\cdot\!(x, X)\right).
 \end{array}
$$
\cqfd

\section{A 1-parameter family of hyperk\"ahler structures on the
  cotangent bundle of the restricted Grassmannian }\label{p5}

\subsection{The stable manifold $\Ws$ associated with the complex
  structure $I_{1}$}

Recall that the complex Banach Lie group $G^{\C}~:= GL(H_{+})
~\cap~\{~\Id + L^{1}(H_{+})~ \}$ acts $I_{1}$-holomorphically on
$T\mathcal{M}_{k}$ by
$$
g\!\cdot\!((x, X)) = (x \circ g^{-1}, X \circ g^{*}),
$$
for all $g$  in $G^{\C}$, and for all $(x, X)$ in $T\m$. Let $\Ws$
be the stable manifold associated with $\W$ with respect to the
complex structure $I_{1}$, i.e. the union of $G^{\C}$-orbits (for
the above action) intersecting $\W$. Since by the proof of Theorem
$3.5$ and Proposition \ref{Sprop}, assumptions (D) and (S) are
satisfied, one has the following Corollary  of Lemma \ref{decs}
and Theorem \ref{kiot}:

\begin{cor}\label{ws}
The space $\Ws$ is a $I_{1}$-complex submanifold of $T\m$ and the
quotient space $\Ws/G^{\C}$ is a smooth complex manifold. The map
from $\mathcal{W}_{k}/G$ to  $\Ws/G^{\C}$ induced by the natural
injection of $\mathcal{W}_{k}$ into $\Ws$ is an
$I_{1}$-holomorphic diffeomorphism. \hfill $\Box$
\end{cor}
In the following Proposition  we give an explicit characterization
of the stable manifold $\Ws$ and we compute the projection~:
$$
\begin{array}{llll}
q_1 :& \Ws & \rightarrow & \W\\
   & (x, X) & \mapsto & q_{1}((x, X)) = g_{(x, X)}\cdot(x, X).
\end{array}
$$
defined by Proposition  \ref{toctoc}.

\begin{prop}\label{Ws}
The stable manifold  $\Ws$ is the set
$$
\{ (x,X) \in T\m \textrm{ such that } X^{*}x = 0 \textrm{ and } x
\textrm{ is  one-to-one}
 \},
$$ and, for all $(x, X)$ in $\Ws$, the unique element $~g_{(x, X)}~$
of $~\exp i\g~$ such that $~ g_{(x, X)}\cdot(x, X)$ belongs to
$~\W~ $ is defined by
$$
g_{(x, X)}^{-1}~:= \left(\frac{k^2}{2}(x^{*}x)^{-1} +
\frac{k^2}{2} (x^{*}x)^{-\frac{1}{2}}\left(Id_{H_{+}} +
\frac{4}{k^{4}}
(x^{*}x)^{\frac{1}{2}}X^{*}X(x^{*}x)^{\frac{1}{2}}\right)^{\frac{1}{2}}
(x^{*}x)^{-\frac{1}{2}}\right)^{\frac{1}{2}}.
$$
\end{prop}

\proofp \ref{Ws}~:\\
Let $\mathcal{A}$ be the set of elements $(x, X)$  in $T\m$  such
that $~ X^{*}x = 0~$  and  $x$ is one-to-one.
 Let us show  that $\Ws \subset \mathcal{A}$.
Consider  $(x, X) \in \Ws$ and $g \in G^{\C}$  such that $g \cdot
(x, X) \in \W$. We have:
$$
(X \circ g^{*})^{*}(x \circ g^{-1}) = gX^{*}xg^{-1} = 0 .
$$
If $g^{-1} = |g|^{-1}.u^{-1}$ denotes the polar decomposition of
$g^{-1}$, the equality:
$$
(x \circ g^{-1})^{*}(x \circ g^{-1}) -
(X \circ g^{*})^{*}(X \circ g^{*}) = k^{2} Id_{H_{+}},
$$
reads:
$$
|g|^{-1} x^{*}x |g|^{-1} - |g| X^{*}X |g| = k^{2} Id_{H_{+}}.
$$
Thus $|g|^{-1} x^{*}x |g|^{-1} = k^{2} Id_{H_{+}} + |g| X^{*}X
|g|$ is a positive definite self-adjoint operator. Since
$|g|^{-1}$ is an isomorphism, the same holds true for $x^{*}x$.
It follows that $x$ is one-to-one.

 To see  that $\mathcal{A} \subset \Ws$,
consider an element  $(x, X)$  of $\mathcal{A}$. We are looking
for a positive definite self-adjoint operator
 $g_{(x, X)}$  such that~:
$$
g_{(x, X)}^{-1} x^{*}x g_{(x, X)}^{-1} - g_{(x, X)} X^{*}X g_{(x,
X)} = k^{2} Id_{H_{+}}.$$ The operator $x$ being one-to-one,
$x^{*}x$ is positive definite and its square root
$(x^{*}x)^{\frac{1}{2}}$ is an invertible  operator on $H_{+}$.
Hence it is sufficient to find a positive definite operator
$\gamma~:= (x^{*}x)^{\frac{1}{2}}g_{(x, X)}^{-1}$ such that:
$$
{\gamma}^{*}\gamma -
{\gamma}^{-1}(x^{*}x)^{\frac{1}{2}}X^{*}X(x^{*}x)^{\frac{1}{2}}{\gamma}^{-1*}
= k^{2} Id_{H_{+}}
$$
$$
\Leftrightarrow (\gamma {\gamma}^{*})^{2} - k^{2}(\gamma {\gamma}^{*})
- (x^{*}x)^{\frac{1}{2}}X^{*}X(x^{*}x)^{\frac{1}{2}} = 0
$$
The unique positive definite solution of the latter equation is~:
$$
\gamma {\gamma}^{*} = \frac{k^{2}}{2}\left(Id_{H_{+}} +
\left(Id_{H_{+}} + \frac{4}{k^{4}}(x^{*}x)^{\frac{1}{2}}
X^{*}X(x^{*}x)^{\frac{1}{2}}\right)^{\frac{1}{2}}\right).
$$
Therefore
\begin{equation}\label{ggamma}
g_{(x, X)}^{-2}~:=(x^{*}x)^{-\frac{1}{2}}\gamma
{\gamma}^{*}(x^{*}x)^{-\frac{1}{2}}
\end{equation}
is positive definite and self-adjoint, and its square root
satisfies the required condition. \cqfd
\\

\subsection{Identification of $\Ws/G^{\C}$ with the cotangent
  space $T'Gr_{res}^{0}$ of the restricted Grassmannian  }

In this Subsection  , we will use the following Theorem  to
identify the complex quotient space $\Ws/G^{\C}$ with the
cotangent bundle $T'Gr_{res}^{0}$ of the connected component
$Gr_{res}^{0}$ of the restricted Grassmannian. Recall that
$\textrm{Ran } x$ denotes the range of an operator $x$.

\begin{thm}\label{subco}
The map $\Psi$ defined by
$$
\begin{array}{lcll}
\Psi~: & \Ws & \longrightarrow & T'Gr_{res}^{0}\\
       & (x, X) & \longmapsto & ( \Ran \,x,\,
       \frac{1}{k^{2}} x \circ X^{*} )
\end{array}
$$
is an $I_1$-holomorphic submersion whose fibers are the orbits
under the $I_1$-holomorphic action of the complex group $G^{\C}$
on $\Ws \subset T\M_k$.
\end{thm}

\prooft \ref{subco}:

 \. For $(x,\,X)$ in $\Ws$, the range $P$ of
$x$ is an element of $Gr_{res}^{0}$ since $p_{+} \circ x$ belongs
to $ \{ Id_{H_{+}} + L^{1}(H_{+}) \}$, thus is a Fredholm operator
with  index $0$, and $p_{-} \circ x$ is a Hilbert-Schmidt
operator. Furthermore the condition $X^{*}x = 0$ implies that the
restriction of $\eta~:= \frac{1}{k^{2}} x \circ X^{*}$ to $P$
vanishes. Thus $\eta$ can be identified with an  element of
$L^{2}(P^{\perp}, P)$ which is the cotangent space of
$Gr_{res}^{0}$ at $P$.

\. Let us check that $\Psi$ is onto. For $P$ in $Gr_{res}^{0}$,
denote by $x_{P}$  the operator from $H_{+}$ to $H$ whose columns
are the vectors of  the canonical basis of $P$ as defined in
\cite{PS}. Then $k \, x_{P}$ is in $\m$ (see \cite{Tum} for the
details of this affirmation). On the other hand, for every $V \in
L^{2}(P^{\perp}, P)$, the operator $X$ defined by $X~:= k^{2}~
V^{*} \circ x_{P}^{*-1}$ (where $x_{P}$ is viewed as an
isomorphism between $H_{+}$ and $P$) satisfies $\frac{1}{k^{2}}
x_P \circ X^{*} = V$ and is an element of $L^{2}(H_{+},
P^{\perp})$. Moreover, since $p_{-}~: P \rightarrow H_{-}$ is
Hilbert-Schmidt, $p_{+}~: P^{\perp} \rightarrow H_{+}$ is also
Hilbert-Schmidt and it follows that $p_{+} \circ X \in
L^{1}(H_{+})$. Thus $\Psi\left( (k x_P,~k^{2}~ V^{*} \circ
x_{P}^{*-1})\right) = (P, V)$.

\. Let us show that two elements $(x_{1}, X_{1})$ and $(x_{2},
X_{2})$ in $\Ws$ have the same image by $\Psi$ if and only if they
are in the same orbit under $G^{\C}$. We have:
$$
\textrm{ Ran }x_{1} = \textrm{ Ran }x_{2} \Leftrightarrow x_{2} =
x_{1} \circ g^{-1} \textrm{ for some }  g^{-1} \in G^{\C},
$$
thus~:
$$
x_{2} \circ X_{2}^{*} = x_{1} \circ X_{1}^{*} = x_{2} \circ g \circ
X_{1}^{*},
$$
which is equivalent to~: $X_{2} = X_{1} \circ g^{*}$ since $x_{2}$
is one-to-one.

\. Let us explicit the differential of $\Psi$ at $(x, \,X)$.
Denote by $P$ the range of $x$, by
 $\mathcal{U}_{P} \subset
Gr_{res}^{0}$ the open subset of elements $P' \in Gr_{res}^{0}$ such
 that the orthogonal projection of $P'$ onto  $P$ is an isomorphism
 and by
 $\varphi_{P}$ the chart from $\mathcal{U}_{P}$ onto $L^{2}(P,
 P^{\perp})$ which maps $P'$ to the unique element $U$  in $ L^{2}(P,
 P^{\perp})$ whose graph is $P'$. Let
$$(Z, T) \in T_{(x, X)}\Ws
$$
and
$$
(x(t), X(t)) \in \mathcal{C}^{1}\left((-\epsilon, \epsilon),
\Ws\right)
$$
be such that~:
$$\dot{x}(0) = Z \textrm{ and } \dot{X}(0) = T.
$$
Denote by $(U(t), V(t))$ the curve $ \varphi \circ \Psi\left(x(t),
X(t)\right)$. Since
$$ \textrm{ Ran }x(t) = \textrm{ Ran }(\textrm{ Id}_{P}+ U(t))
$$
 and since $U(0) = 0$, there exists  $g(t) \in
\mathcal{C}^{1}((-\epsilon, \epsilon), G^{\C})$ such that
$$x(t) \circ g(t)^{-1} = \textrm{ Id}_{P}+ U(t) \textrm{ and  } x(0) \circ g(0)^{-1}
 =  \textrm{ Id}_{P}.
$$
Considering the decomposition of $H$ into  $P \oplus P^{\perp}$,
 one has~: $g(t) = pr_{P} \circ x(t)$, where $pr_{P}$ denotes the orthogonal
 projection onto $P$. Thus $U(t) = pr_{P^{\perp}} \circ x(t)\circ
 (pr_{P} \circ x(t))^{-1}$ and
$$
\frac{d}{dt}_{|t = 0}U(t) =  pr_{P^{\perp}}\circ  Z \circ
x(0)^{-1}.
$$
Moreover one has $V(t) = \frac{1}{k^{2}} pr_{P} \circ x(t) \circ
X(t)^{*}_{|P^{\perp}}$ and
$$
\frac{d}{dt}_{|t = 0}V(t) = \frac{1}{k^{2}} \left( pr_{P}(Z)\circ
X^{*} +  x \circ pr_{P^{\perp}}(T)^{*} \right).
$$
Therefore
$$
d\varphi_{P}\circ d\Psi_{(x, X)}((Z, T)) =
\left(pr_{P^{\perp}}\circ~ Z ~\circ x^{-1},~ \frac{1}{k^{2}}
\left( pr_{P}(Z) ~\circ~ X^{*} + x ~\circ~
pr_{P^{\perp}}(T)^{*}\right) \right).
$$
It follows that~:
$$d\varphi_{P}\circ d\Psi_{(x, X)}(I_{1}(Z, T)) = i
d\varphi_{P}\circ d\Psi_{(x, X)}((Z, T))
$$ thus $\Psi$ is
$I_1$-holomorphic. Furthermore $d\varphi_{P}\circ d\Psi_{(x, X)}$
is surjective, a preimage of $(U, V) \in L^{2}(P, P^{\perp})
\times L^{2}(P^{\perp}, P)$ being given by $(U \circ x, k^{2}
V^{*} x^{*-1})$. At last, from the above considerations it follows
that the kernel of $d\Psi_{(x, \,X)}$ is the tangent space of the
$G^{\C}$-orbit $G^{\C}\!\cdot\!(x,\,X)$, which splits by
Proposition \ref{Sprop}. \cqfdt

\begin{cor}\label{identik}
The quotient space  $\Ws / G^{\C}$ is isomorphic as a smooth
complex manifold to the cotangent space $T'Gr_{res}^{0}$ endowed
with its natural complex structure via the following isomorphism
$$
\begin{array}{lcll}
\tilde{\Psi}~: & \Ws /G^{\C} & \longrightarrow & T'Gr_{res}^{0}\\
       & [(x, X)] & \longmapsto & ( \Ran \,x,
       \frac{1}{k^{2}} x \circ X^{*}).
\end{array}
$$
Hence $T'Gr_{res}^{0}$ carries a 1-parameter family of
hyperk\"ahler structures indexed by $k \in \R^*$.
\end{cor}

\begin{rem}{\rm
By exchanging $H_{+}$ with a subspace of another connected
component of $Gr_{res}$, we obtain the cotangent space of every
connected component of $Gr_{res}$ as a hyperk\"ahler quotient.
}\end{rem}

By restriction to the zero section of the tangent space $T\m$ one
deduces from the previous Theorem  the following result, which has
been, as already mentioned in the Introduction, partially obtained
by T. Wurzbacher (cf \cite{Wur2})~:

\begin{cor}
For every $k \in \R^{*}$, the connected component $Gr_{res}^{0}$
of the restricted Grassmannian  is diffeomorphic to the K\"ahler
quotient of the space $\m$ by the Hamiltonian  action of the
unitary group $G$, with level set $$\{~ x \in \m \,,\, x^{*}x ~=
~k^{2} \textrm{ Id} ~\}.$$
\end{cor}

\subsection{The K\"ahler potential $K_{1}$ of $T'Gr_{res}^{0}$ }

The hyperk\"ahler manifold  $T\mathcal{M}_{k}$ admits a globally
defined hyperk\"ahler potential, i.e. a K\"ahler potential with
respect to all complex structures, which has the following
expression~:
$$
\begin{array}{llll}
K~: & T\m & \rightarrow & \R\\
    & (x, X) & \mapsto & \frac{1}{4} \Tr ( x^{*}x + X^{*}X - k^{2}Id).
\end{array}
$$
The theory of Subsection   \ref{potquo} applied to the particular
case of $T'Gr_{res}^{0}$ yields  the following Theorem~:
\begin{thm}\label{k}
For all  $\frac{k^{2}}{2} \in \N^{*}$, the  2-form
$\Psi^{*}\omega_{1}^{red}$ on $\Ws$ satisfies $~
i\Psi^{*}\omega_{1}^{red} = d d^{c_{1}} {K}_{1},~ $ where   for
all $(x, X)$ in $\Ws$,
$$
{K}_{1}\left((x,\, X)\right) = \frac{k^{2}}{4} \log \det
\left(\frac{x^{*}x}{k^{2}}\right) + \frac{k^{2}}{2}
\Tr\left(\frac{\gamma {\gamma}^{*}}{k^{2}} - \textrm{ Id }\right)
- \frac{k^{2}}{4}\Tr\left(\log \frac{\gamma
{\gamma}^{*}}{k^{2}}\right),
$$
with $ \gamma {\gamma}^{*} := \frac{k^{2}}{2}\left(Id_{H_{+}} +
\left(Id_{H_{+}} + \frac{4}{k^{4}}(x^{*}x)^{\frac{1}{2}}
X^{*}X(x^{*}x)^{\frac{1}{2}}\right)^{\frac{1}{2}}\right). $
\end{thm}

\prooft \ref{k}:\\
By Theorem  \ref{potentiel}, one has~:
$$
{K}_{1}((x, X))~:= K\left(g_{(x, X)}\!\cdot\!(x, X)\right) +
\frac{1}{2} \log |\,\chi_{\frac{k^{2}}{2}}(g_{(x, X)})\,|^{2}.
$$
Since $ g_{(x, X)} X^{*}X g_{(x, X)} = g_{(x, X)}^{-1} x^{*}x
g_{(x, X)}^{-1}   - k^{2} \textrm{ Id}$, one has~:
$$
\begin{array}{ll}
K\left(g_{(x, X)}\!\cdot\!(x, X)\right) & := \frac{1}{4} \Tr
\left(g_{(x, X)}^{-1} x^{*}x g_{(x, X)}^{-1} + g_{(x, X)} X^{*}X
g_{(x, X)} -
k^{2} \textrm{ Id}\right)\\
          & = \frac{1}{2} \Tr\left( g_{(x, X)}^{-1} x^{*}x g_{(x, X)}^{-1}
          - k^{2}\textrm{ Id}\right) .
\end{array}
$$
Thus, after conjugation by $g_{(x,\, X)}^{-1}$~:
$$
K\left(g_{(x, X)}\!\cdot\!(x, X)\right) = \frac{k^{2}}{2}
\Tr\left( g_{(x, X)}^{-2} \frac{x^{*}x}{k^{2}} - \textrm{
Id}\right),
$$
and by \eqref{ggamma}
$$
g_{(x, X)}^{-2} x^{*}x =
(x^{*}x)^{-\frac{1}{2}}\gamma {\gamma}^{*}(x^{*}x)^{\frac{1}{2}}.
$$
After conjugation by $(x^{*}x)^{-\frac{1}{2}}$, we have~:
$$
K\left(g_{(x, X)}.(x, X)\right) = \frac{k^{2}}{2}
\Tr\left(\frac{\gamma
  {\gamma}^{*}}{k^{2}}
- \textrm{  Id}\right).
$$

On the other hand,
$$
\begin{array}{ll}
 \frac{1}{2}\log |\chi_{\frac{k^{2}}{2}}\left(g_{(x, X)}\right)|^{2} & = -\frac{1}{2}
\log \left(\det \left(g_{(x, X)}^{-2}\right)\right)^{\frac{k^{2}}{2}}\\
& = -\frac{k^{2}}{4} \log \det \left(x^{*}x\right)^{-\frac{1}{2}}
\gamma {\gamma}^{*}(x^{*}x)^{-\frac{1}{2}}\\
& = \frac{k^{2}}{4} \log \det \left(\frac{x^{*}x}{k^{2}}\right)
    - \frac{k^{2}}{4} \log \det \left(\frac{\gamma {\gamma}^{*}}{k^{2}}\right).
\end{array}
$$
Furthermore, the operator
$$A~:= \frac{\gamma {\gamma}^{*}}{k^{2}} - \textrm{ Id }
  = \frac{1}{2}\left( \left(\textrm{ Id } + \frac{4}{k^{4}}(x^{*}x)^{\frac{1}{2}}
X^{*}X(x^{*}x)^{\frac{1}{2}}\right)^{\frac{1}{2}} -  \textrm{ Id }
\right)$$ is self-adjoint positive and of trace class. Thus~:
$$ \log \det \left(\frac{\gamma {\gamma}^{*}}{k^{2}}\right) =
\Tr \log \left(\frac{\gamma {\gamma}^{*}}{k^{2}}\right).$$ \cqfdt

\begin{prop}\label{k1hat}
For every  $\frac{k^{2}}{2} \in \N^{*}$, the $2$-form
$\Psi^{*}\omega_{1}^{red}$ on $\Ws$ satisfies
$~i\Psi^{*}\omega_{1}^{red} = d d^{c_{1}} {K}_{1},~$ with~:
$$
\begin{array}{lll}
{K}_{1}\left( (x, X) \right) & = & \frac{k^{2}}{4} \log \det
 \left(\frac{x^{*}x}{k^{2}}\right)
 +
\frac{k^{2}}{4} \Tr \left( \left(\textrm{ Id } + 4V^{*}V
 \right)^{\frac{1}{2}}  -
  \textrm{ Id }\right) \\ & &
- \frac{k^{2}}{4} \Tr \log \frac{1}{2} \left(\textrm{ Id } +
    \left(\textrm{ Id } + 4V^{*}V \right)^{\frac{1}{2}} \right)
\end{array}
$$
where  $ V^* = \frac{1}{k^{2}} x \circ X^{*}$ is the image of the
class  $[(x, X)]$ under the identification $\tilde{\Psi}~:\Ws /
G^{\C} \rightarrow T'Gr_{res}$ given by Corollary  \ref{identik} .
\end{prop}

\dem \ref{k1hat}:\\
Since~:
$$
\frac{\gamma \gamma^{*}}{k^{2}} = \frac{1}{2}\left( \textrm{Id} +
\left(\textrm{Id} + \frac{4}{k^{4}}
|x|X^{*}X|x|\right)^{\frac{1}{2}}\right),
$$
the operator $\frac{\gamma \gamma^{*}}{k^{2}}$ is conjugate to~:
$$
\frac{1}{2}\left( \textrm{Id} + \left(\textrm{Id} +
\frac{4}{k^{4}} x X^{*} X x^{*}\right)^{\frac{1}{2}}\right).
$$
Hence one has~:
$$
\begin{array}{ll}
\Tr\left(\frac{\gamma {\gamma}^{*}}{k^{2}} - \textrm{ Id }\right)
- \frac{1}{2}\Tr \left(\log \frac{\gamma
{\gamma}^{*}}{k^{2}}\right)  & = \Tr \left( \left(\textrm{ Id } +
4V^{*}V \right)^{\frac{1}{2}} -
  \textrm{ Id }\right) \\ & - \frac{1}{2} \Tr \log \frac{1}{2}
\left(\textrm{ Id } +
    \left(\textrm{ Id } + 4V^{*}V \right)^{\frac{1}{2}}\right).
\end{array}
$$
\cqfd

\subsection{The K\"ahler potential $K_{1}$ as a function of the
  curvature of $Gr_{res}^{0}$}

\begin{thm}\label{koko}
The potential ${K}_{1}$ has the following expression in terms of
the curvature~:
$$
{K}_{1}((x, X)) = \frac{k^{2}}{4} \log \det
\left(\frac{x^{*}x}{k^{2}}\right) +
 k^{2} \gm_{Gr} \left( f(I_{1}R_{I_{1}V, V}) V, V\right),$$ with
$V = \frac{1}{k^{2}} X \circ x^{*}$ and $f(u) =\frac{1}{u}\left(
\sqrt{1 + u} - 1 - \log \frac{1 + \sqrt{1 + u}}{2}\right).$
\end{thm}

\demthm \ref{koko}:\\
The Grassmannian  $Gr_{res}^{0}$ is a Hermitian-symmetric orbit of
the connected component $\mathcal{U}_{res}^{0}$ of the restricted
unitary group. Its curvature is therefore given by (see
\cite{SpWu})~:
$$
R_{X, Y} Z = Y X^{*} Z - Z Y^{*}X + Z X^{*}Y - X Y^{*}Z,
$$
for all $X, Y, Z \in T_{P}Gr_{res}^{0}$. The operator $R_{I_{1}V,
V}$ acts on $T_{P}Gr_{res}^{0}$ by~:
$$
R_{I_{1}V, V}Y = -2i(VV^{*}Y + Y V^{*}V).
$$
It follows that~:
$$
\begin{array}{ll}
\gm_{Gr}(I_{1}R_{I_{1}V, V}V, V) & = 2 \Re \Tr(V^{*}VV^{*}V +
V^{*}VV^{*}V) = 4 \Re \Tr((V^{*}V)^{2}) \\& = \frac{1}{4}\Re
\Tr((4V^{*}V)^{2}),
\end{array}
$$
and~:
$$
\begin{array}{ll}
\gm_{Gr}\left((I_{1}R_{I_{1}V, V})^{j}V, V \right) & =  \Re \Tr
\left( 4^{j} (V^{*}V)^{j+1}\right)
\\ &  =  \frac{1}{4} \Re \Tr \left( (4V^{*}V)^{j+1}\right).
\end{array}
$$
Therefore one has~:
$$
\begin{array}{c}
\frac{1}{4}\Tr \left( \left( \textrm{ Id } + 4V^{*}V
  \right)^{\frac{1}{2}} -
  \textrm{ Id } \right) - \frac{1}{2} \Tr \log \frac{1}{2} \left(
\textrm{ Id } +
    (\textrm{ Id } + 4 V^{*}V)^{\frac{1}{2}} \right) \\ =
\gm_{Gr}\left(f(I_{1}R_{I_{1}V, V}) V, V \right),
\end{array}
$$ with
$$
f(u) = \frac{1}{u}\left( \sqrt{1 + u} - 1 - \log \frac{1 + \sqrt{1
+ u}}{2}\right).
$$
\cqfdt

\begin{rem} {\rm
The first summand in the expression of $K_{1}$ is the pull-back to
$\Ws$ of the K\"ahler potential of the restricted Grassmannian  (
defined on the stable manifold $\mathcal{M}_{k}^{s}$ of
$Gr_{res}^{0}$) via the canonical injection $\mathcal{M}_{k}^{s}
\hookrightarrow \Ws$. Note that the K\"ahler potential of
$Gr_{res}^{0}$ is the pull-back of  the K\"ahler potential of the
complex projective space of a separable Hilbert space by
Pl\"ucker's embedding. The second summand is expressed as a
function of the curvature of the restricted Grassmannian  applied
to the image of an  element $(x, X)$ in $\Ws$ by the
identification $\Ws/G^{\C} = T'Gr_{res}^{0}$ given by Corollary
\ref{identik}. }
\end{rem}

\section{A 1-parameter family of
hyperk\"ahler structures on a natural complexification of the
restricted Grassmannian }\label{p6}

\subsection{Definition of the complexified orbit $\mathcal{O}^{\C}$
of   $Gr_{res}^{0}$}

Let $\U_{2}(H)$ be the Banach Lie group $\U(H) \cap \{
\textrm{Id}_{H} + L^{2}(H) \}$.  An element $P$ of $Gr^{0}_{res}$
can be identified with $ik^{2} pr_{P}$, $k \neq 0$, where $pr_{P}$
denotes the orthogonal projection
  of $H$ onto $P$. Via this identification, the natural action of
  $\U_{2}(H)$ on $Gr^{0}_{res}$ is given by the conjugation.
  The complexified orbit
$\mathcal{O}^{\C}$ of $Gr_{res}^{0}$ is the orbit of an element $P
\in Gr_{res}^{0}$ under the action  of the complex Lie group
$GL_{2}(H) := GL(H) \cap  \{ \textrm{Id}_{H} + L^{2}(H) \}$. It is
the set of operators $z \in B(H)$ whose spectrum is the pair $\{
ik^{2}\,,\, 0 \}$ with $k \neq 0$, and such that the eigenspace
associated with $ik^{2}$ (resp. $0$) is an element of
$Gr_{res}^{0}$ (resp. of the Grassmannian  $Gr_{res}^{0*}$
obtained from the definition of $Gr_{res}^{0}$ by exchanging the
roles of $H_{+}$ and $H_{-}$). This complexified orbit has been
introduced in particular by J. Mickelsson in \cite{Mic}.

\begin{prop}\label{oc}
The complexified orbit $\mathcal{O}^{\C}$ of the connected
component $Gr_{res}^{0}$ defined as the homogeneous space~:
$$\mathcal{O}^{\C} := GL_{2}(H)/\left(GL_{2}(H_{+}) \times
GL_{2}(H_{-})\right)$$ is a Hilbert manifold modelled over the
Hilbert space $L^{2}(H_{+}, H_{-}) \times L^{2}(H_{-}, H_{+})$,
diffeomorphic to the open set of $Gr_{res}^{0} \times
Gr_{res}^{0*}$ consisting of all ordered pairs
 $(P, Q) \in Gr_{res}^{0} \times Gr_{res}^{0*}$ such that $P
\cap Q = \{ 0 \}$.
\end{prop}

\dem \ref{oc}:\\
Let us denote by $\varepsilon$ the operator $ik^{2} p_{+}$. The
stabilizer of $\varepsilon$ under the action of $GL_{2}(H)$ by
conjugation is $GL_{2}(H_{+}) \times GL_{2}(H_{-})$. The tangent
space at
 $\varepsilon$ of the homogeneous space
$GL_{2}(H) \cdot \varepsilon$ is isomorphic to
$\mathfrak{gl}_{2}(H)/\left(\mathfrak{gl}_{2}(H_{+}) \times
\mathfrak{gl}_{2}(H_{-})\right)$ which can be identified with
$L^{2}(H_{+}, H_{-}) \times L^{2}(H_{-}, H_{+})$. For $g \in
GL_{2}(H)$, $g \varepsilon g^{-1} = i k^{2} pr_{g.H_{+}}$, where
$pr_{g.H_{+}}$ denotes the projection on $g.H_{+}$ parallel to
$g.H_{-}$. Since $g$ belongs to $ GL_{2}(H)$, the orthogonal
projection of $g.H_{+}$ to $H_{-}$ is an Hilbert-Schmidt operator,
and the orthogonal   projection of $g.H_{+}$ to $H_{+}$ is a
Fredholm operator of  index $0$. Similarly, the orthogonal
projection of $g.H_{-}$ to $H_{+}$ is Hilbert-Schmidt, and the
orthogonal projection of $g.H_{-}$ to $H_{-}$ is a  Fredholm
operator of index $0$. Thus $g.H_{+}$ belongs to $ Gr^{0}_{res}$
and $g.H_{-}$ to $  Gr_{res}^{*0}$. Moreover $g.H_{+} \cap g.H_{-}
= \{ 0 \}$. \cqfd

\subsection{The stable manifold $\wss$ associated with the complex
  structure  $I_{3}$}

Recall that  $I_{3}(Z, T) = (iT,\, iZ)$ for $(Z, T) \in T_{(x,
X)}T\mathcal{M}_{k}$, and that the action of the Lie algebra $\g$
on $(x, X) \in T\mathcal{M}_{k}$ is given by
$$
\a.(x, X) = (- x \circ \a, - X \circ \a),
$$
for all $\a$ in $\g$.  The action $\cdot_{3}$ of  the
complexification $\g^{\C}$ of the Lie algebra $\g$ on
$T\mathcal{M}_{k}$ compatible with $I_3$ is defined  by
$$
i\a \cdot_{3}(x, X) := I_{3}\left(\a.(x, X)\right) = (-i X\circ
\a, -i x \circ \a) = (x, X) \left( \begin{array}{cc} 0 & -i\a \\
-i\a & 0
\end{array} \right),
$$
for all $\a$ in $\g$. This action integrates into an
$I_{3}$-holomorphic action  of $G^{\C} = \exp i\g.~G$ on
$T\mathcal{M}_{k}$, also denoted by $\cdot_{3}$,  and given by
$$
 \exp (i\a) u \cdot_{3}(x, X)
:= (x \circ u^{-1}\,,\, X \circ u^{-1}) \left( \begin{array}{cc} \cosh i\a & -\sinh i\a \\
-\sinh  i\a & \cosh i\a \end{array} \right),
$$
for all $\a$ in $\g$ and for all $u$ in $G$.  By Lemma~\ref{decs},
the stable manifold associated with $I_{3}$ is the $I_{3}$-complex
submanifold of $T\mathcal{M}_{k}$, contained in
${\mu}_{1}^{-1}\left(-\frac{i}{2}k^2
\Tr\right)\cap{\mu}_{2}^{-1}\left(0 \right)$,  defined as
$$
\wss:= \left\{~ (x, X) \in T\mathcal{M}_{k},~~ \exists ~\a \in
\g,~ \exp i\a \cdot_{3}(x, X) \in \W~ \right\}.
$$

\subsection{Identification of $\wss/G^{\C}$ with $\mathcal{O}^{\C}$}

\begin{thm}\label{orbus}
The map $\psi$ defined by~:
$$
\begin{array}{llll}
\psi: & \wss & \rightarrow & \mathcal{O}^{\C}\\
      & (x, X) & \mapsto & z = i(x + X)(x^{*} - X^{*})
\end{array}
$$
is an $I_{3}$-holomorphic submersion whose fibers are the orbits
of
 the $I_{3}$-holomorphic action of $G^{\C}$ on $T\M$.
\end{thm}

\prooft \ref{orbus}:

 \. Let us show that $\psi$ takes all its values in
$\mathcal{O}^{\C}$. Recall that, for $(x, X) \in \wss$,\, one
has~:
\begin{center}
\begin{tabular}{ccc}
$x^{*}x - X^{*}X = k^{2}\textrm{ Id} $ &and &$X^{*}x = x^{*}X.$
\end{tabular}
\end{center} Thus~:
\begin{equation}\label{rap}
\begin{array}{l}
(x^{*} - X^{*})(x + X) = k^{2} \textrm{ Id},\\ (x^{*} + X^{*})(x -
X) = k^{2} \textrm{ Id}.
\end{array}
\end{equation}
It follows that $\ker(x + X) = \{0\}$ and $\ker(x - X) = \{0\}$.
The kernel of  $z$ is therefore~:
$$\ker z ~= ~\ker(x^{*} - X^{*}) ~=~ \left(\textrm{Ran}\,(x -
X)\right)^{\perp}.$$ Moreover,  since for all $v\in H$~:
$$
z\left((x + X)v\right) = i(x + X)(x^{*} - X^{*})(x + X)v =
ik^{2}(x + X)v,
$$
the subspace $\textrm{ Ran}\,(x + X)$ is contained in the
eigenspace of $z$ corresponding  to the eigenvalue $ik^{2}$.
Hence~:
$$\textrm{ Ran}\,(x + X)~~ \cap~~ \ker(x^{*} - X^{*}) = \{0\}.$$
Further, the projection of $H$ onto $\textrm{ Ran}\,(x + X)$ is
given by
$$
\begin{array}{llll}
p_{1}:& H & \rightarrow & \textrm{ Ran}(x + X)\\
      & v & \mapsto & \frac{1}{k^{2}}(x + X)(x^{*} - X^{*}) v
\end{array}
$$
and is continuous. Since  $\textrm{ Id}_{H} - p_{1}$ takes its
values in
 $\ker(x^{*} - X^{*}) $,
one has
$$\textrm{ Ran}(x + X) \oplus \ker(x^{*} - X^{*}) = H$$
 as
a direct topological  sum. Moreover, for $(x, X) \in
T\mathcal{M}_{k}$, $\textrm{ Ran}(x + X)$ and $\textrm{ Ran}(x -
X)$ are elements of $Gr_{res}^{0}$, thus $\ker(x^{*} - X^{*}) =
\left(\textrm{ Ran}(x - X)\right)^{\perp}$ is an element of
$Gr_{res}^{0*}$. It follows that $\psi$ takes values in
$\mathcal{O}^{\C}$.

\. Let us show that the fibers of $\psi$ are the orbits under the
$I_3$-holomorphic action of $G^{\C}$. Suppose that
$\psi\left((x_{1}, X_{1})\right) = \psi\left((x_{2},
X_{2})\right)$ where $(x_{1}, X_{1})$ and $(x_{2}, X_{2})$ are in
$\wss$. It follows that~:
\begin{center}
\begin{tabular}{ccc}
$\textrm{ Ran}(x_{1} + X_{1}) = \textrm{ Ran}(x_{2} + X_{2})$& and
& $\textrm{ Ran}(x_{1} - X_{1}) = \textrm{ Ran}(x_{2} - X_{2})$.
\end{tabular}
\end{center}
 Therefore there exists $g \in GL(H_{+})$ such that
$(x_{2} + X_{2}) = (x_{1} + X_{1}) \circ g$ and $g' \in GL(H_{+})$
such that $(x_{2} - X_{2}) = (x_{1} - X_{1}) \circ g'$. This
implies that:
$$
\begin{array}{ll}
2x_{2} & = x_{1}(g + g') + X_{1}(g - g')\\
2X_{2} & = x_{1}(g - g') + X_{1}(g + g')
\end{array}
$$
Recall that for $i = 1,\, 2$, $x_{i}^{*}x_{i} - X_{i}^{*}X_{i} =
k^{2}\textrm{ Id}$ and $X^{*}_{i}x_{i} = x_{i}^{*}X_{i}$. We have:
$$
\begin{array}{ll}
4(x_{2}^{*}x_{2}  - X_{2}^{*}X_{2}) &  = (g^{*}+g'^{*})(x_{1}^{*}x_{1}
- X_{1}^{*}X_{1})(g+g') \\ &  + (g^{*}-g'^{*})(X_{1}^{*}X_{1} -
x_{1}^{*}x_{1})(g-g')\\
& + (g^{*}+g'^{*})(x_{1}^{*}X_{1} - X_{1}^{*}x_{1})(g-g') \\ & +
(g^{*}-g'^{*}) (X_{1}^{*}x_{1} - x_{1}^{*}X_{1})(g+g'),
\end{array}
$$
i.e.
$$
g^{*}g' + g'^{*}g = 2\textrm{ Id},
$$
and
$$
\begin{array}{ll}
4(X_{2}^{*}x_{2} - x_{2}^{*}X_{2}) & = (g^{*} - g'^{*})(x_{1}^{*}x_{1}
- X_{1}^{*}X_{1})(g + g') \\ & + (g^{*} + g'^{*})(X_{1}^{*}X_{1} -
x_{1}^{*}x_{1})(g - g')\\
& + (g^{*} - g'^{*})(x_{1}^{*}X_{1} - X_{1}^{*}x_{1})(g - g') \\ &
+ (g^{*} + g'^{*})(X_{1}^{*}x_{1} - x_{1}^{*}X_{1})(g + g'),
\end{array}
$$
that is ~:
$$
g^{*}g' = g'^{*}g.
$$
Thus $g' = g^{*-1}$. Denoting by $\exp(-i\a).u^{-1} = g^{-1}$ the
polar decomposition of $g^{-1} $, with $u \in \U(H_{+})$ and $\a
\in \mathfrak{u}(H_{+})$, it follows that:
$$
\begin{array}{ll}
x_{2} & = x_{1}  \cosh (i\a)  u + X_{1} \sinh (i\a)  u \\
X_{2} & = -x_{1}  \sinh (i\a)  u + X_{1}  \cosh (i\a) u.
\end{array}
$$
Consequently~: $(x_{2}, X_{2}) = \exp(-i\a)u^{-1}\cdot(x_{1},
X_{1})$, in other words $(x_{1}, X_{1})$ and $(x_{2}, X_{2})$
belong to the same $G^{\C}$-orbit.

\. Let us show that $\psi$ is onto. Let $P \in Gr_{res}^{0}$ and
$Q \in Gr_{res}^{0*}$ be such that $P \cap Q = \{ 0 \}$.
$Q^{\perp}$ is the graph of a Hilbert-Schmidt operator $A: P
\rightarrow P^{\perp}$ and $Q$ is the graph of $-A^{*}: P^{\perp}
\rightarrow P$. Let $f$  be the map that takes an orthonormal
basis  $\{ e_{i} \}_{ i \in \N }$ of $H_{+}$ to the associated
canonical basis of $P$ and that takes an orthonormal basis $\{
e_{-i} \}_{ i \in \N^{*} }$ of $H_{-}$ to the associated canonical
basis of $P^{\perp}$. Denote by $g$  the unitary element $f \circ
|f|^{-1}$. Let us remark that $g$ belongs to $\U_{2}(H)$ and
satisfies~: $p_+ \circ g_{|H_+} \in \Id_{H_+} + L^{1}(H_{+})$ as
well as $p_- \circ g_{|H_{-}} \in \Id_{H_-} + L^{1}(H_{-})$.
Define~:
$$
\left\{
\begin{array}{ll}
x & ~=~ k (\textrm{Id}_{P} ~+~ \frac{1}{2}A) \circ g_{|H_{+}}\\
X & ~=~ -\frac{k}{2} A \circ g_{|H_{+}}.
\end{array}\right.
$$
One has: $\textrm{Ran }(x + X) = \textrm{Ran }(g_{|H_{+}}) = P$
and $\textrm{Ran }(x - X) = \textrm{Ran }\left(\textrm{Id}_{P}
\,+\, A\right) \circ g_{|H_{+}} = Q^{\perp}$. Let us check that
$(x\,,\, X)$ is an element of $T\M$. Denote by~:
$$
\begin{array}{ll}
\Id_H = & \left( \begin{array}{cc} a & b \\ c & d \end{array}
\right)
\end{array}
$$
the block decomposition of the identity operator with respect to
the direct sums $H = P \oplus P^{\perp}$ and $H = H_+ \oplus H_-$,
where $a$ (resp.  $d$) belongs to $\textrm{Fred}(P, H_{+})$ (resp.
$\textrm{Fred}(P^{\perp}, H_{-})$ )   and where $b$ (resp. $c$)
belongs to $L^{2}(P^{\perp}, H_{+})$ (resp. $L^{2}(P, H_{-})$).
Further, denote by~:
$$
\begin{array}{ll}
g = & \left( \begin{array}{cc} u_1 & 0 \\ 0 & u_2 \end{array}
\right)
\end{array}
$$
the block decomposition of $g$ with respect to the directs sums $H
= H_+ \oplus H_-$ and $H = P \oplus P^{\perp}$. The block
decomposition of $g$ with respect to $H = H_+ \oplus H_-$ is~:
$$
\begin{array}{ll}
g = & \left( \begin{array}{cc} a u_1 & b u_2 \\ c u_1 & d u_2
\end{array} \right).
\end{array}
$$
As mentioned above $p_+ \circ g_{|H_+} = a u_1$ belongs to  $
\Id_{H_+} + L^{1}(H_{+})$, and $p_- \circ g_{|H_{-}} = d u_2$
belongs to $\Id_{H_-} + L^{1}(H_{-})$. It follows that with
respect to the direct sum $H = H_+ \oplus H_-$, the operator $x$
has the following expression~:
$$
\begin{array}{lll}
x = & \left( \begin{array}{cc} a & b \\ c & d \end{array} \right)
\left( \begin{array}{c} k\Id_P \\ \frac{k}{2}A \end{array} \right)
\circ u_1 = \left( \begin{array}{c}k a u_1 + \frac{k}{2} b A u_1 \\
k c u_1 + \frac{k}{2} d A u_1
\end{array} \right).
\end{array}
$$
It follows that:
$$
p_{+} \circ x =  k a u_1 + \frac{k}{2} b A u_1 \in k
\textrm{Id}_{H_{+}} + L^{1}(H_{+})
$$
and
$$
p_{-} \circ x =  k c u_1 + \frac{k}{2} d A u_1 \in L^{2}(H_{+},
H_{-}).
$$
Similarly,
$$p_{+} \circ X = -\frac{k}{2} b A u_1 \in L^{1}(H_{+})
$$
and
$$p_{-} \circ X = -\frac{k}{2} d A u_1 \in L^{2}(H_{+},
H_{-}).
$$
Hence the  ordered pair $(x, X)$ is in $T\mathcal{M}_{k}$.
Besides, $x^{*}x - X^{*}X = k^{2} \textrm{Id}_{H_{+}}$ and $X^{*}x
- x^{*}X = 0$. It remains to prove that $(x, X) \in \wss$. For
this purpose, observe that:
$$
\begin{array}{l}
x^{*}x + X^{*}X = k^2\textrm{Id}_{H_{+}} + \frac{k^2}{2}u_1^{*} A^{*}A u_1\\
X^{*}x + x^{*}X = - \frac{k^2}{2} u_{1}^{*} A^{*}A u_{1}.
\end{array}
$$
The condition $\exp i\a\cdot_{3}(x\,,\, X) \in \W$ is  equivalent
to the following  equation~:
$$
\begin{array}{rl}
\cosh i\a \circ \left(\textrm{Id}_{H_{+}} + \frac{1}{2}u_1^{*}
A^{*}A u_{1}\right) \circ \sinh i\a + \sinh i\a \circ
\left(\textrm{Id}_{H_{+}} + \frac{1}{2} u_{1}^{*} A^{*}A u_{1}\right) \circ \cosh i\a  & \\
 + \cosh i\a  \circ \left(\frac{1}{2} u_{1}^{*} A^{*}A u_{1}\right) \circ  \cosh i\a
+ \sinh i\a \circ \left(\frac{1}{2} u_{1}^{*} A^{*}A u_{1}\right)
\circ \sinh i\a & = 0,
\end{array}
$$
whose solution is:
$$
\a = \frac{i}{4} \log \left(\textrm{Id}_{H_{+}} +  u_{1}^{*}
A^{*}A u_1 \right),
$$
which belongs to $\mathcal{A}^{1}(H_{+})$.

\. The differential of  $\psi$ at an element $(x, X) \in \wss$
maps the ordered pair $(Z, T) \in T_{(x, X)}\wss$ to:
$$
d\psi_{(x, X)}\left((Z, T)\right) = i (Z + T)(x^{*} - X^{*}) + i
(x + X)(Z^{*} - T^{*}).
$$
One has $d\psi_{(x, X)}(I_{3}\left((Z\,,\, T)\right) = i
d\psi_{(x, X)}\left((Z\,, \,T)\right)$, thus  $\psi$ is
holomorphic. Let $z$ be in $\mathcal{O}^{\C}$, and let $P$ (resp.
$Q$) be the eigenspace of $z$ with respect to the eigenvalue $i
k^2$ (resp. $0$). Let $(U\,,\, V)$ be an element of $L^{2}(P\,,\,
P^{\perp}) \times L^{2}(Q\,,\, Q^{\perp})$. A preimage of $(U\,,\,
V)$ by $d\psi_{(x, X)}$ is given by the ordered pair $(Z\,,\, T)
\in L^{2}(H_{+}, H) \times L^{2}(H_{+}, H)$ defined by
$$
\begin{array}{l}
U = i (Z + T)(x^{*} - X^{*})\\
V =  i (x + X)(Z^{*} - T^{*}).
\end{array}
$$
Using equations (\ref{rap}) page \pageref{rap}, one gets~:
$$
\begin{array}{l}
(Z + T) = -\frac{i}{k^{2}} U (x + X)\\
(Z - T) = \frac{i}{k^{2}} V^{*} (x - X).
\end{array}
$$
Hence~:
$$
\begin{array}{ll}
Z & = -\frac{i}{2k^2}\left(  U (x + X) - V^* (x - X)\right)\\
T & = -\frac{i}{2k^2}\left(  U (x + X) + V^* (x - X)\right).
\end{array}
$$
Moreover, for $(x\,,\, X) \in \wss$, $p_{+} \circ Z $ and $p_{+}
\circ T $ are trace class operators. It follows that the
differential $d\psi_{(x, X)}$ is onto. \cqfdt

\begin{cor}
The quotient space $\wss/G^{\C}$ is  diffeomorphic to the
complexified orbit $\mathcal{O}^{\C}$ via the isomorphism
$$
\begin{array}{llll}
\tilde{\psi}: & \wss/G^{\C} & \rightarrow & \mathcal{O}^{\C}\\
      & [(x, X)] & \mapsto & z =  i(x + X)(x^{*} - X^{*})
\end{array}
$$\hfill $\Box$
\end{cor}

\subsection{The K\"ahler potential
$\hat{K}_{3}$ of $T'Gr_{res}^{0}$}\label{q10}

From the general theory it follows that:
$$
\psi^{*}\omega_{3}^{red}((x, X)) = d d^{c_{3}} K(q_{3}(x, X)),
$$
where: $K((x, X)) = \frac{1}{4}\Tr (x^{*}x + X^{*}X -
k^{2}\textrm{Id})$, and where $q_{3}$ is the projection from
$\wss$ to $\W$. This Subsection is devoted to the computation of
the K\"ahler potential $K_{3} := K \circ q_{3}$ associated with
the complex structure $I_{3}$ at a point $(x\,,\,X)$ of the stable
manifold $\wss$ by the use of  an invariant of the
$G^{\C}$-orbits. We will use the following notation. The injection
of $\g$ into $\g'$ given by the trace allows to identify the
moment map $\mu_{3}$ with the map (still denoted by ${\mu}_{3}$)
defined by
$${\mu}_{3}\left((x, X)\right) = \frac{i}{2}(X^{*}x + x^{*}X).$$
Define a function ${\mu}_4$ by~:
$$
{\mu}_{4}\left((x, X)\right):= \frac{i}{2}(x^{*}x + X^{*}X).
$$
We have the following Lemma~:

\begin{lem}\label{dx}
For every $(x\,,\, X)$ in $\wss$, one has
$$
K_{3}\left((x, X)\right) = \frac{1}{4} \Tr
\left(\left({\mu}_{4}^{2}\left((x, X)\right) -
{\mu}_{3}^{2}\left((x, X)\right)\right)^{\frac{1}{2}} -
k^{2}\textrm{Id}\right).
$$
\end{lem}

\proofl \ref{dx}:\\
One has~:
$$
\begin{array}{lll}
{\mu}_{3}\left(\exp i\a\cdot_3(x, X)\right) & =& \cosh i\a \circ
{\mu}_{4}\left((x, X)\right)
\circ \sinh i\a + \sinh i\a \circ {\mu}_{4}\left((x, X)\right) \circ \cosh i\a \\
& &+ \cosh i\a \circ {\mu}_{3}\left((x, X)\right) \circ \cosh i\a
+ \sinh i\a \circ
{\mu}_{3}\left((x, X)\right) \circ \sinh i\a,\\
{\mu}_{4}\left(\exp i\a\cdot_3(x, X)\right) &  = & \cosh i\a \circ
{\mu}_{4}\left((x, X)\right)
\circ \cosh i\a + \sinh i\a \circ {\mu}_{4}\left((x, X)\right) \circ \sinh i\a \\
& &+ \cosh i\a \circ {\mu}_{3}\left((x, X)\right) \circ \sinh i\a
+ \sinh i\a \circ {\mu}_{3}\left((x, X)\right) \circ \cosh i\a.
\end{array}
$$
Therefore:
$$
\begin{array}{ll}
\left({\mu}_{3} + {\mu}_{4}\right)\left(\exp i\a\cdot_3(x,
X)\right) & = \exp i\a \circ \left({\mu}_{3} + {\mu}_{4}\right)
\circ
\exp i\a,\\
\left({\mu}_{3} - {\mu}_{4}\right)\left(\exp i\a\cdot_3(x,
X)\right) & = \exp(-i\a) \circ \left({\mu}_{3} - {\mu}_{4}\right)
\circ \exp(-i\a),
\end{array}
$$
and
$$
\left({\mu}_{4}^{2} - {\mu}_{3}^{2}\right)\left(\exp i\a\cdot_3(x,
X)\right) = \exp (i\a) \circ \left({\mu}_{4}^{2}\left((x,
X)\right) - {\mu}_{3}^{2}\left((x, X)\right)\right) \circ \exp
(-i\a).
$$
For $\a\in \g$ such that $\exp i\a\cdot_3\left((x, X)\right) =
q_{3}\left((x, X)\right)$, it follows that
$$
{\mu}_{4}\left(q_{3}\left((x, X)\right)\right) = \exp (i\a) \circ
\left({\mu}_{4}^{2}\left((x, X)\right) - {\mu}_{3}^{2}\left((x,
X)\right)\right)^{\frac{1}{2}} \circ \exp (-i\a),
$$
and
$$
K_{3}\left((x, X)\right) = \frac{1}{4} \Tr
\left(\left({\mu}_{4}^{2}\left((x, X)\right) -
{\mu}_{3}^{2}\left((x, X)\right)\right)^{\frac{1}{2}} -
k^{2}\textrm{Id}\right).
$$
\cq

\begin{prop}\label{calculpo2}:\\
For every $(x,\, X)$ in  $ \wss$, one has
$$
K_{3}\left((x, X)\right) = \frac{k^{2}}{4} \Tr\left(\left(\textrm{
Id } + \frac{4}{k^{4}}\left( xX^{*}Xx^{*} -
Xx^{*}Xx^{*}\right)\right)^{\frac{1}{2}} - \textrm{ Id }\right).
$$
\end{prop}

\proofp \ref{calculpo2}:\\
Since  $\wss \subset
{\mu}_{1}^{-1}\left(-\frac{i}{2}k^{2}\Tr\right)\,\cap\,
{\mu}_{2}^{-1}(0)$, for all $(x, X)$ in  $ \wss$, one has $x^{*}x
- X^{*}X = k^{2}\textrm{ Id }$ and $X^{*}x = x^{*}X$. Hence~:
$$
\begin{array}{lll}
\left({\mu}_{4}^{2}\left((x, X)\right) - {\mu}_{3}^{2}\left((x,
X)\right)\right) & = &
x^{*}xx^{*}x + x^{*}xX^{*}X + X^{*}Xx^{*}x + X^{*}XX^{*}X\\
& =&  x^{*}x(X^{*}X + k^{2}) + x^{*}xX^{*}X + X^{*}Xx^{*}x \\ & &
+ (x^{*}x - k^{2})X^{*}X
  - x^{*}Xx^{*}X \\ & & - x^{*}XX^{*}x - X^{*} x x^{*} X - X^{*}xX^{*}x\\
& = & k^{4} + 4 x^{*}xX^{*}X - 4 x^{*}Xx^{*}X.
\end{array}
$$
The result then follows after conjugation by $x^{*-1}$ viewed as
an operator of $H_{+}$ onto $\textrm{Ran}\,x$. \cqfd

\begin{thm}\label{fifi}:\\
The symplectic form $\omega_{3}^{red}$ on the cotangent space
$T'Gr_{res}^{0}$ admits a globally defined K\"ahler potential
$\hat{K}_{3}$ on $T'Gr_{res}$, whose expression at $(P,\, V^{*})
\in T'Gr_{res}^{0}$ is   given by
$$
\begin{array}{ll}
\hat{K}_{3}\left((P, V^{*})\right) & =~ \frac{k^2}{4} \Tr
\left(\left( \Id
+ 4 V^{*}V \right)^{\frac{1}{2}} - \Id \right)\\
& =~
 \gm_{Gr}\left(h\left(
I_{1}R_{I_{1}V, V}\right) V, V\right),
\end{array}
$$
where~: $$h(u):= \frac{1}{u}(\sqrt{1 + u} - 1).$$
\end{thm}

\prooft \ref{fifi}:\\
When  $(x, X)$ belongs to the level set, $x^{*}X = 0$ and
$$
\begin{array}{ll}
K_{3}\left((x, X)\right) & = \frac{k^{2}}{4}
\Tr\left(\left(\textrm{ Id } + \frac{4}{k^{4}}(
xX^{*}Xx^{*})\right)^{\frac{1}{2}}
- \textrm{ Id }\right)\\
& = \frac{k^{2}}{4} \Tr\left(\left(\textrm{ Id } +
4V^{*}V\right)^{\frac{1}{2}} - \textrm{ Id }\right),
\end{array}
$$
where  $V = \frac{1}{k^{2}} X \circ x^{*}$. The theorem then
follows from the identities:
$$
\gm_{Gr}\left(I_{1}R_{I_{1}V, V}V, V\right) = \frac{1}{4}\Re
\Tr\left((4V^{*}V)^{2}\right)
$$
and
$$
\gm_{Gr}\left((I_{1}R_{I_{1}V, V})^{j}V, V\right) = \frac{1}{4}
\Re \Tr\left( (4V^{*}V)^{j+1}\right),
$$
and from the fact that $p_{3}^{*}\omega_{3}^{red} = dd^{c_3}K_{3}
= dd^{c_3} p_{3}^{*}\hat{K}_3 = p_3^{*}dd^{c_3}\hat{K}_3$ since
$p_3$ is holomorphic with respect to the complex structure $I_3$.
\cqfdt

\subsection{The K\"ahler potential  $\hat{K}_{3}$ of
  $\mathcal{O}^{\C}$ as a function of the characteristic angles}

In this Subsection, we show that the general formulas of
\cite{BG3} giving $K_{3}$  in terms of the characteristic angles
have an analogue in the infinite-dimensional setting. For this
purpose, we use a section of the application  $\psi$ (defined in
Theorem \ref{orbus}), which has been already used in the proof of
Theorem \ref{orbus}.

 Theorem \ref{orbus} states that every ordered pair  $(P,\, Q)$ belonging to
$Gr_{res}^{0} \times Gr_{res}^{*0}$ with $P \cap Q = \{0\}$
represents an element of the complexified orbit. A preimage $(x,\,
X)$ of $(P,\, Q)$ by the application $\psi:  \wss \rightarrow
\mathcal{O}^{\C}$ is given by~:
\begin{equation}\label{xX}
\left\{
\begin{array}{l}
x = k( \textrm{Id}_{P} +\frac{1}{2}A) g_{|H_{+}}\\
X = -\frac{k}{2} A g_{|H_{+}},
\end{array}\right.
\end{equation}
where $A$ is a Hilbert-Schmidt operator from $P$ to $P^{\perp}$
whose graph is $Q^{\perp}$ (determined modulo the right action of
$GL(P)$), and where $g$ is a unitary operator uniquely defined if
$P$ and $Q$ are endowed with their canonical bases. Note that the
eigenvalues  $\{ a^2_{i}\}_{ i \in \N }$ of $A^{*}A$ are
independent of the operator  $A$ chosen to represent  the ordered
pair $(P,\, Q)$. If $A$ is generic, i.e. if all the eigenvalues
$a^2_{i}$ are distinct, it is possible to define pairs of
characteristic lines $\{l_{i}, l_{i}'\}$, $i \in \N$, as follows.
The complex line $l_{i}$ is the eigenspace in $P$ of the operator
$A^{*}A$ with respect to the eigenvalue $a_{i}^{2}$, and $l_{i}'$
is the complex line in $Q^{\perp}$ which is the image of $l_{i}$
under the operator $\textrm{
  Id}_{P} + A$. The angle $\theta_{i}$ between the two complex
  line
 $l_{i}$ and $l_{i}'$ is defined by
$$
\cos \theta_{i} = |\langle e_{i}, e_{i}' \rangle|,
$$
where $e_{i}$ is a unitary generator of  $l_{i}$ and where~:
$$
e_{i}' := \frac{e_{i} + A(e_{i})}{|e_{i} + A(e_{i})|}.
$$
The angle $\theta_{i}$ is related to the eigenvalue  $a_{i}^2$ by
the following formula~:
$$
\cos \theta_{i} = \frac{1}{\sqrt{1 + a_{i}^{2}}}.
$$
The latter expression makes sense even in the non-generic case,
and allows one to uniquely define the set of  characteristic
angles $\theta_{i} \in (-\frac{\pi}{2}, +\frac{\pi}{2})$, $i \in
\N$.

\begin{rem}{\rm
The orbit of an ordered pair  $(P, \,Q)$ in $Gr_{res}^{0} \times
Gr_{res}^{0*}$ under the natural action of  $GL_{2}(H)$ is
characterized by the dimension of  $P \,\cap \,Q$. The orbit of
 $(P,\, Q)$ under the action of
$\U_{2}(H)$ on $Gr_{res}^{0} \times Gr_{res}^{0*}$ is
characterized by the set of characteristic angles $\theta_{i}$.}
\end{rem}

Proposition \ref{calculpo2} allows to express the K\"ahler
potential  $K_{3}$ on the complexified orbit either in terms of
the eigenvalues  $a_{i}^2$ of $A^{*}A$ or in terms of the
characteristic angles  $\theta_{i}$~:

\begin{thm}\label{aa}
The form $\omega_{3}^{red}$ defined on the complexified orbit
$\mathcal{O}^{\C}$ and associated with the natural complex
structure of $\mathcal{O}^{\C}$ satisfies $\omega_{3}^{red} =
dd^{c_{3}}\hat{K}_{3}$ with~:
$$
\hat{K}_{3}\left((P\,,\, Q)\right) = k^{2} \Tr \left(
\left(\textrm{ Id}_{P} + A^{*}A \right)^{\frac{1}{2}} - \textrm{
Id}_{P}\right),
$$
for $(P\,,\,Q)$ in $\mathcal{O}^{\C}$, where $A$ is such that
$\textrm{ Ran}(\textrm{Id}_{P} + A) = Q^{\perp}$. Denoting by
$a_{i}^2$ the eigenvalues of the operator $A^{*}A$, and by
$\theta_{i}$ the characteristic angles of the pair $(P\,,\, Q)$,
one has~:
$$
\begin{array}{ll}
\hat{K}_{3}\left((P\,,\, Q)\right) & = k^{2} \sum_{i \in \N} \left(\sqrt{1 + a_{i}^{2}} - 1\right)\\
              & = k^{2} \sum_{i \in \N}
\left(\frac{1}{\cos \theta_{i}} - 1\right).
\end{array}
$$
\end{thm}

\demthm \ref{aa}:\\
From the proof of Theorem  \ref{calculpo2} it follows that  the
potential ${K}_{3}$ is given at an element $(x, X)$ of the stable
manifold  $\wss$ by~:
$$
K_{3}((x, X)) = \Tr \left( \left( k^{4} + 4 x^{*}xX^{*}X - 4
x^{*}Xx^{*}X \right)^{\frac{1}{2}} - k^{2} \textrm{ Id}\right).
$$
To proceed, let us recall the element  $(x, X)$ of $\wss$ defined
in the proof of Theorem  \ref{orbus}  by~:
$$
\begin{array}{l}
x = k( \textrm{Id}_{P} +\frac{1}{2}A) \circ u_1\\
X = -\frac{k}{2} A \circ u_1,
\end{array}
$$
where $u_1$ is a unitary operator from $H_{+}$ to $P$.  One has
$\psi\left((x,\,X)\right) = (P,\,Q)$ and
$$
\hat{K}_{3}\left((P\,,\,Q)\right) = K_{3}((x,\, X)) = k^{2} \Tr
\left( \left(\textrm{ Id}_{P} + u_1^{*} A^{*}A u_1
\right)^{\frac{1}{2}} - \textrm{ Id}_{P}\right),
$$
which, after conjugation by $u_1$, gives the result. \cqfdt


\begin{thebibliography}{10}


\bibitem{Bel} D.\,Belti\c{t}\u{a},  {\it Integrability of analytic
  almost complex structures on Banach manifolds}, Annals of Global
  Analysis and Geometry \textbf{28}, (\textbf{2005}), 59-73.

\bibitem{BP} D.\,Belti\c{t}\u{a}, B.\,Prunaru,   {\it
Amenability, completely bounded projections, Dynamical systems and
smooth orbits}, to appear in Integral Equations and Operator
Theory, arXiv:math.OA/0504313 v2, (20 Apr \textbf{2005}).


\bibitem{Biq} O.\,Biquard,  {\it Sur les {\'e}quations de Nahm et la
    structure de Poisson des alg{\`e}bres de Lie semi-simples complexes},
    Math. Ann. \textbf{304}, $n^{o}$ 2, (\textbf{1996}), 253-276.

\bibitem{BG1} O.\,Biquard, P.\,Gauduchon,  {\it Hyperk{\"a}hler metrics
    on cotangent bundles of Hermitian  Symmetric spaces}, Geometry and
    Physics, {\it Lect. notes Pure Appl. Math.} Serie \textbf{184}, Marcel
    Dekker (\textbf{1996}), 287-298.

\bibitem{BG2} O.\,Biquard, P.\,Gauduchon,  {\it La m{\'e}trique
  hyperk{\"a}hl{\'e}rienne des orbites coadjointes de type sym{\'e}trique
  d'un groupe de Lie complexe semi-simple}, C. R. Acad. Sci. Paris,
    t. {\bf 323}, s{\'e}rie {\bf I} (\textbf{1996}), 1259-1264.




\bibitem{BG3} O.\,Biquard, P.\,Gauduchon,  {\it G{\'e}om{\'e}trie
    hyperk{\"a}hl{\'e}rienne des espaces hermitiens sym{\'e}triques
complexifi{\'e}s},
    S{\'e}minaire de th{\'e}orie spectrale et g{\'e}om{\'e}trie, Grenoble,
    Vol {\bf 16} (\textbf{1998}), 127-173.


\bibitem{Bou1} N.\,Bourbaki,  {\it Topologie G\'en\'erale}
  \'El\'ements de Math\'ematiques, chap 1-4, Masson, (\textbf{1990}).

\bibitem{Bou2} N.\,Bourbaki,  {\it Groupes et Alg{\`e}bres de Lie}
{\'E}l{\'e}ments de Math{\'e}matiques, chap 1-6, Masson
(\textbf{1981}).

\bibitem{Bou3}  N.\,Bourbaki,  {\it Vari\'et\'es diff\'erentielles
  et analytiques } \'El\'ements de Math\'ematiques, Fascicule de
  r\'esultats, paragraphes 1 \`a 7, Hermann, (\textbf{1967}).


\bibitem{Bry} J-L.\,Brylinsky,  {\it Loop Spaces, Characteristic
  Classes and Geometric Quantization}, Progress in Mathematics, \textbf{107},
  Birkh\"auser (\textbf{1992}).


\bibitem{Don} S.K.\,Donaldson,  {\it Remarks on gauge theory, complex geometry and
$4$-manifold topology}, Fields Medallists' lectures,  384--403,
World Sci. Ser. 20th Century Math., \textbf{5}, World Sci.
Publishing, River Edge, NJ, (\textbf{1997}).



\bibitem{Fei} B.\,Feix,  {\it Hyperk\"{a}hler Metrics on cotangent Bundles},
  PhD dissertation, University of Cambridge, (\textbf{1999}).

\bibitem{HH} G.\,Helminck, A.\,Helminck,  {\it The
structure of Hilbert flag varieties}, Publ. Res. Inst. Math. Sci.
\textbf{30}, $n^{o}$ \textbf{3}, (\textbf{1994}), 401-441.

\bibitem{HKLR} N.J.\,Hitchin, A.\,Karlhede, U.\,Lindstr{\"o}m,
M.\,Ro\v{c}ek, {\it Hyperk{\"a}hler Metrics and Supersymmetry},
  Commun. Math. phys. \textbf{108}, (\textbf{1987}), 535-589.


\bibitem{Kac}  V.G.\,Kac,  {\it Infinite-dimensional Lie Algebras},
  Progress in Mathematics, Edited by J. Coates and S. Helgason,
  Birkh{\"a}user, (\textbf{1983}).

\bibitem{Kal}  D.\,Kaledin,  {\it Hyperk{\"a}hler structures on total
    spaces of holomorphic cotangent bundles}, Quaternionic structures
    in mathematics and physics (Rome, \textbf{1999}), 195-230.

\bibitem{Kirw1} F.\,Kirwan,  {\it Momentum maps and reduction in
  algebraic geometry}, J. Diff. Geom. Appl. \textbf{9}, $n^{o}$ \textbf{1,2},
  (\textbf{1998}), 135-171.

\bibitem{Kirw2} F.\,Kirwan,  {\it Cohomology of moduli spaces},
Li, Ta Tsien (ed.) et al., Proceedings of the international
congress of mathematicians, ICM 2002, Beijing, China, August
20-28, 2002. Vol. I: Plenary lectures and ceremonies. Beijing:
Higher Education Press; Singapore: World Scientific/distributor.
363-382 (\textbf{2002}).


\bibitem{Kirw3} F.\,Kirwan,  {\it  Cohomology of quotients in
  symplectic and algebraic geometry}, Princeton University Press,
  (\textbf{1984}), 210p.


\bibitem{KS} P.Z.\,Kobak, A.\,Swann,  {\it Quaternionic geometry of a
    nilpotent variety}, Math. Ann. \textbf{297}, (\textbf{1993}), 747-764.


\bibitem{Kov} A.G.\,Kovalev,  {\it Nahm's  equation   and complex
  adjoint orbits}, Quart. J. Math., \textbf{47}, 41-58, (\textbf{1993}).


\bibitem{Kro1} P.B.\,Kronheimer,  {\it A hyper-K{\"a}hlerian structure on
    coadjoint orbits of a semisimple complex group}, J. London
    Math. Soc. (2) \textbf{42}, (\textbf{1990}), 193-208.

\bibitem{Kro2} P.B.\,Kronheimer,  {\it Instantons and the geometry of the
    nilpotent variety}, J. Differential Geometry \textbf{32},
    (\textbf{1990}), 473-490.

\bibitem{Kro3}  P.B.\,Kronheimer,  {\it A hyperk\"ahler structure
  on the cotangent bundle of a complex Lie group},
  Archiv:math.DG/0409253 (june \textbf{1988}).


\bibitem{Lem} L.\,Lempert,  {\it Loop spaces as complex
    manifolds}, Journal of Differential Geometry \textbf{38},
(\textbf{1993}),  519-543.

\bibitem{LT} J.\,Lindenstrauss,  L.\,Tzafriri,  {\it On the
  complemented subspaces problem} Israel Journal Math. \textbf{9},
  (\textbf{1971}), 263-269.




\bibitem{MR} J.E.\,Marsden, T.\,Ratiu,  {\it Introduction to Mechanics
    and Symmetry}, Texts in Applied Mathematics, Springer-Verlag,
    (\textbf{1999})

\bibitem{MW1} J.E.\,Marsden, A.\,Weinstein,  {\it Reduction of
    symplectic manifolds with symmetry}, Rep. Math. Phys. \textbf{5}
    (\textbf{1974}), 121-130.

\bibitem{MW2} J.E.\,Marsden, A.\,Weinstein,  {\it Comments on the
    history, theory, and applications of the symplectic reduction},
    Progress. Math \textbf{198} (\textbf{2001}) Springer.

\bibitem{MW3} J.E.\,Marsden, A.\,Weinstein,  {\it The Hamiltonian
    Structure of the Maxwell-Vlasov equations}, Physica \textbf{4D}, (\textbf{1982}), 394-406.

\bibitem{MW4} J.E.\,Marsden, A.\,Weinstein,  {\it Coadjoint
orbits, vortices, Clebsch variables for incompressible fluids}
Physica \textbf{ 7 D}, (\textbf{1983}), 305-323.

\bibitem{Mic} J.\,Mickelsson,  {\it Current algebras and groups},
  New York: Plenum Press, (\textbf{1989}).


\bibitem{MFK} D.\,Mumford, J.\,Fogarty, F.\,Kirwan,  {\it
Geometric Invariant Theory}, (3rd. ed.), Springer,
(\textbf{1994}).


\bibitem{NN} A.\,Newlander,
L.\,Nirenberg,  {\it Complex analytic
  coordinates in almost complex manifolds}, Annals of Math. Vol. \textbf{65},
  $n^{o}$ \textbf{3}, (May \textbf{1957}), 391--404.

\bibitem{Pat} I.\,Patyi, {\it On the $\bar{\partial}$-equation  in a
  Banach space}, Bull. Soc. math. France, \textbf{128}, (\textbf{2002}), 391-406.


\bibitem{Pen} J-P.\,Penot,  {\it Sur le th\'eor\`eme de
  Frobenius}, Bull. Soc. math. France, \textbf{98}, (\textbf{1970}),  47-80.


\bibitem{Pre} A.\,Pressley, {\it Loop Groups, Grassmannians and KdV
     equations}, Infinite-dimensional groups with applications,
    Publ. Math. Sci. Res. Inst. \textbf{4}, 285-306 (\textbf{1985}).


\bibitem{PS} A.\,Pressley, G.\,Segal,  {\it Loop Groups},
Oxford Mathematical Monographs. Oxford (UK): Clarendon Press.
viii, 318 p. (\textbf{1988})

\bibitem{Sat} M.\,Sato, Y.\,Sato,  {\it Soliton  equations as dynamical
systems on infinite-dimensional Grassmann manifold}, Nonlinear
partial differential  equations in applied science, Proc. U.S. -
Jap. Semin., Tokyo 1982, North-Holland Math. Stud. \textbf{81},
 (\textbf{1983}), 259-271.


\bibitem{Seg1} G.\,Segal,  {\it The geometry of the KdV equation },
Int. J. Mod. Phys. \textbf{A 6}, $n^{o}$ \textbf{16},
(\textbf{1991}), 2859-2869.

\bibitem{Seg2} G.\,Segal,  {\it Loop groups and harmonic maps},
Advances in homotopy theory, Proc. Conf. in Honour of I. M. James,
Cortona/Italy 1988, Lond. Math. Soc. Lect. Note Ser. \textbf{139},
(\textbf{1989}), 153-164.


\bibitem{Seg3} G.\,Segal,  {\it Unitary Representations of some Infinite
    Dimensional Groups}, Comm. Math. Phys \textbf{80}, $n^{o}$ \textbf{3},
    (\textbf{1981}), 301-342.


\bibitem{SW} G.\,Segal, G\,Wilson,  {\it Loop Groups and equations of
    KdV type}, Terng, Chuu Lian (ed.) et al., Surveys in differential
    geometry, Vol. IV. A supplement to the Journal of Differential
    Geometry. Integral systems (integrable systems).
Lectures on geometry and topology. Cambridge, MA: International
Press,  (\textbf{1998}), 403-466.



\bibitem{Sim} B.\,Simon,  {\it Trace ideals and their applications}, Cambridge
  University Press, Cambridge, (\textbf{1979}).



\bibitem{SpVa} M.\,Spera, G.\,Valli,  {\it Pl\"ucker embedding of
  the Hilbert space Grassmannian and the CAR algebra}, Russian
  J. Math. Phys. \textbf{2}, $n^{o}$ \textbf{3}, (\textbf{1994}), 383-392.


\bibitem{SpWu} M.\,Spera, T.\,Wurzbacher,  {\it Differential
  geometry of Grassmannian embeddings of based loop groups},
  Differential Geometry and its Applications \textbf{13}, (\textbf{2000}),
43-75, North-Holland.

\bibitem{Tum} A.B.\,Tumpach,  {\it Vari\'et\'es k\"ahl\'eriennes et
hyperk\"ahl\'eriennes de dimension infinie}, Ph.D Thesis, \'Ecole
Polytechnique, Palaiseau, France, (july \textbf{2005}).

\bibitem{Tum2} A.B.\,Tumpach, {\it Mostow Decomposition Theorem
for a $L^*$-group and Applications to affine coadjoint orbits and
stable manifolds}, preprint arXiv:math-ph/0605039, (May
\textbf{2006}).


\bibitem{Tum3} A.B.\,Tumpach,  {\it Infinite-dimensional hyperk\"ahler manifolds associated with
Hermitian-symmetric affine coadjoint orbits}, preprint
arXiv:math-ph/0605032, (May \textbf{2006}).


\bibitem{Wur1} T.\,Wurzbacher,  {\it Fermionic Second Quantization and the
    Geometry of the Restricted Grassmannian}, in Infinite-Dimensional
    K{\"a}hler Manifolds, DMV Seminar, Band \textbf{31}, Birkh{\"a}user, (\textbf{2001}).

\bibitem{Wur2} T.\,Wurzbacher,  {\it La grassmannienne d'un espace
de Hilbert comme r\'eduction symplectique}, talk given at the
South-Rhodanian seminar of geometry  ``Autour de la r\'eduction
symplectique'', CIRM, Luminy, (\textbf{1-5/12/97}).
\end{thebibliography}
\end{document}